\documentclass[11pt]{article}
\input tcilatex
\def\bldmth#1{%
\mathchoice
{{\hbox{\boldmath$\displaystyle#1$\unboldmath}}}%
{{\hbox{\boldmath$\textstyle#1$\unboldmath}}}%
{{\hbox{\boldmath$\scriptstyle#1$\unboldmath}}}%
{{\hbox{\boldmath$\scriptscriptstyle#1$\unboldmath}}}%
}
\def\vec#1{\bldmth{#1}}

\title{Finite temperature Thirring model: from linearization through canonical 
transformations to correct normal form of thermofield solution}
\author{V.V. Semenov and S.E. Korenblit 
{ \small (E-mail: korenb@ic.isu.ru) }
}
\begin{document}
\maketitle
\begin{abstract}
It is shown that exact solvability of the finite temperature massless Thirring
model, as well as of its zero temperature case, in canonical quantization scheme 
originates from the intrinsic hidden exact linearizability of Heisenberg equations 
by means of dynamical mapping onto the Schr\"odinger physical fields. The normal forms of  
different one- and two- parametric (thermo) field's solutions are obtained. They are 
connected with each other by making use of generalized conformal shift transformations. 
The sequential use of bosonic canonical transformations provides a correct 
renormalization, anticommutation and symmetry properties of these solutions. 
The dynamical role of inequivalent representations of 1+1-D free massless Dirac fields, 
that are induced by inequivalent representations of 1+1-D free massless (pseudo) scalar 
field, and the appearance of Schwinger terms are elucidated. The inequivalent vacuum is 
established as coherent state for SU(1,1) group. A new alternative sources of 
superselection rules are shown. A generalization of Ojima tilde conjugation rules is 
suggested, which reveals the properties of coherent state for SU(2) group for the 
fermionic thermal vacuum state and is useful for the thermofield bosonization. 
The notions of ``hot'' and ``cold'' thermofields are introduced to distinguish different 
thermofield representations giving the correct normal form of thermofield solution. 
The weak sense of definition of zero and finite temperature operator bosonization rules 
in the framework of thermofield dynamics is demonstrated. 

\end{abstract}
\section{Introduction}

Despite a considerable age the two-dimensional Thirring model
\cite{thi}--\cite{s_w} is still remained as important
touchstone for non-perturbative methods of quantum field theory
\cite{Leut}--\cite{man} revealing new features both in the
well-known \cite{nak}--\cite{fab-iva_2} and in newly obtained
solutions \cite{raja}--\cite{fujita_2}, \cite{alv-gom}. At the same time the 
methods of integration of such two-dimensional models provide a clue for
understanding some non-linear theories of higher dimensions  
\cite{fab-iva}. In particular the Thirring model turns out to
be a two-dimensional analog of the well-known
Nambu-Jona-Lasinio model \cite{fab-iva}, \cite{fujita,fujita_2}
and together with the Schwinger model provides an important
example of using the well-known bosonization procedure (BP)
\cite{col}--\cite{blot}, \cite{oksak}.

In the present work the BP for Thirring model is considered as a particular case of
dynamical mapping (DM) \cite{mtu}, \cite{green_1}, what for Schwinger model was
previously done in Greenberg's works \cite{green}. In the framework of canonical
quantization scheme \cite{hep} the DM method \cite{mtu} consists of the construction of 
Heisenberg field (HF) $\Psi(x)$ as a solution of Heisenberg equations of motion (HEq) in 
the form of Haag expansion built as a sum of normal products \cite{gldj} of free 
physical fields $\psi(x)$, whose representation space accords with unknown 
{\it a priori} physical states of the given field theory \cite{mtu}. The DM
$\Psi(x)\stackrel{\rm w}{=}\Upsilon[\psi(x)]$, being generally
speaking a weak equality, implies the choice of appropriate
initial conditions for the HEq. For example \cite{blot}, \cite{mtu}, if both sets of 
fields are complete, irreducible and coincide asymptotically as $t\to -\infty$, the 
HF will tend in a weak sense to appropriate asymptotic physical field $\psi_{in}(x)$: 
$\lim\limits_{t\to-\infty}\Psi(x^1,t)\stackrel{\rm w}{=}
\Upsilon[\psi_{in}(x^1,-\infty)]$. 
However the (asymptotic) completeness and irreducibility are absent in
the presence of bound states \cite{mtu}, \cite{green_1}. In particular for the exactly
solvable two-dimensional Thirring and Schwinger models \cite{fab-iva}, \cite{blot}
the physical asymptotic states of propagated physical particles have nothing
to do with massless free Dirac asymptotic fields.

As was shown in the works \cite{vklt}--\cite{ks}, in general, it is more natural and 
convenient, in the spirit of Ref. \cite{fadd, shir}, to make DM onto the 
Schr\"odinger physical field $\psi_s(x)$, associated with the HF at $t\to 0$: 
$\lim\limits_{t\to 0}\Psi(x^1,t)\stackrel{\rm w}{=}\Upsilon[\psi_s(x^1,0)]$, 
which is a generalization \cite{kt}, \cite{ks} of the well-known interaction
representation and is closely related to the procedure of canonical quantization 
\cite{blot}, \cite{hep}. In this representation the corresponding time-dependent 
coefficient functions of DM \cite{vklt}, \cite{kt} contain all the information about 
bound states and scattering, and exactly solvable Federbush model \cite{wai} leads to the 
exactly linearizable HEq \cite{ks}. Such kind of exact linearization of HEq with 
so-generalized initial conditions in a weak sense allows to overcome the restrictions of 
Haag theorem \cite{blot,hep}, removing them into the representation construction of the 
free physical field $\psi_s(x)$ \cite{vklt,kt,ks,ks_tt}.  Arisen as Schr\"odinger 
physical field, the field $\psi_s(x)$ in fact plays the role of asymptotic one 
\cite{schwarz}. 

The present paper shows that HEq of the Thirring model admits a similar operator 
linearization and that the choice of free massless (pseudo) scalar fields as the 
physical ones is a consequence of reducibility of the free massless Dirac field 
\cite{blot} in the space of these fields.
The problem of Schwinger terms in the currents commutator \cite{Leut}, being closely
related to BP \cite{nak}--\cite{blot}, also finds here a natural solution \cite{ks} in
fact borrowed from QED \cite{sok}, where it is also sufficient to define this commutator
only for the free fields in the corresponding ``interaction representation''.

The presence of finite temperature: ${\rm k}_B T=1/\varsigma>0$, aggravates the above 
mentioned problems. The introduction of the temperature in quantum field theory 
may be achieved in the real-time formalism of Thermofield Dynamics \cite{mtu}, 
\cite{ojima} by means of reordering with respect to a new thermal vacuum 
$|0(\varsigma)\rangle$ the 
above DM expression of the HF $\Psi(x)\stackrel{\rm w}{=}\Upsilon[\psi(x)]$ given at 
first in terms of the free physical fields $\psi(x)$. 
The latter ones are defined now with respect to the ``cold'' vacuum 
$|0\widetilde{0}\rangle$ at $T=0$, 
which is connected with the new thermal vacuum $|0(\varsigma)\rangle$ by the thermal 
Bogoliubov transformation 
\cite{mtu}, \cite{ojima}: ${\cal V}_\vartheta|0(\varsigma)\rangle=|0\widetilde{0}\rangle$.  
Here for 1+1-D space-time: $\vartheta=\vartheta(k^1,\varsigma)$. Thus, for bosonic $(B)$ 
or fermionic $(F)$ fields with the energy $k^0=\omega(k^1)$ one has, respectively:  
\begin{eqnarray}
&&\!\!\!\!\!\!\!\!\!\!\!\!\!\!\!\!\!\!
|0(\varsigma)\rangle_{(B,F)}=
{\cal V}^{-1}_{\vartheta(B,F)}|0\widetilde{0}\rangle_{(B,F)},\;\mbox{ where for: }\;
g(k^1,\varsigma)\{\mp\} f(k^1,\varsigma)=1,
\label{K_0--} \\
&&\!\!\!\!\!\!\!\!\!\!\!\!\!\!\!\!\!\!
\left. \begin{array}{cc}(B)\{-\}:\;\;&
\cosh^2\vartheta(k^1,\varsigma) \\ (F)\{+\}:\;\;& \cos^2\vartheta(k^1,\varsigma)
\end{array}\right\}=g(k^1,\varsigma)=\exp\{\varsigma\omega(k^1)\}f(k^1,\varsigma). 
\label{K_00} 
\end{eqnarray}
If the corresponding physical thermofields $\psi(x,\varsigma)$ and interpolating them 
at $t\rightarrow\pm\infty$ thermofields $\Phi(x,\varsigma)$, as well as their 
tilde - conjugate ones $\widetilde{\psi}(x,\varsigma)$ and 
$\widetilde{\Phi}(x,\varsigma)$, are determined for $\sigma=\pm 1$ by the relations 
\cite{mtu}, (see however section 3):
\begin{eqnarray}
\psi(x,\varsigma)=g^{1/2}(i\partial_1,\varsigma)\psi(x)-
\sigma f^{1/2}(i\partial_1,\varsigma)\widetilde{\psi}(x),
\label{K_2} \\
\Phi(x,\varsigma)=g^{1/2}(i\partial_1,\varsigma)\Psi(x)- \sigma
f^{1/2}(i\partial_1,\varsigma)\widetilde{\Psi}(x),
\label{K_1} 
\end{eqnarray}
this means in general the absence of any equations of motion for these thermofields, 
since already the free fields $\psi(x)$ and $\widetilde{\psi}(x)$ in Eq. (\ref{K_2}) obey 
different free field equations \cite{mtu}. 
The happy exceptions of this rule are at least the free Hermitian (pseudo) scalar 
field, $(\partial_\mu\partial^\mu+m^2)\varphi(x,\varsigma)=0$, free massless Majorana 
field and the 1+1-D free massless Dirac field: 
$i(\gamma^\mu\partial_\mu)\chi(x,\varsigma)=0$, with not yet important factor $i$. 

However, the interaction contributions in the HEqs will inevitably destroy the 
superposition principle for the interpolating thermofield (\ref{K_1}). 
Therefore, both for the self-interacting (pseudo) scalar field, with coinciding HEqs for 
the tilde HF $\widetilde{\Psi}(x)$ and non-tilde HF $\Psi(x)$, and for a self-interacting 
Dirac field, where the above factor $i$ now changes the relative sign of the interaction 
in the 
HEq for $\widetilde{\Psi}(x)$ (see (\ref{tT_45bn6l5}) below), the left hand side of the 
Eq. (\ref{K_1}) is not a solution of the HEq for the fields in the right hand side. 

However, assuming again both sets of fields are complete, irreducible 
\cite{blot} and asymptotically coincide at $t\to\pm\infty$, the Heisenberg fields 
$\Psi(x)$ and interpolating them thermofields (\ref{K_1}) should tend in a weak sense 
\cite{mtu} to the asymptotic physical fields $\psi(x)$ and physical thermofields 
(\ref{K_2}) respectively, what is not the case for fermionic solutions of exactly 
solvable two-dimensional Thirring and Schwinger models \cite{blot}. 

Here, by using the methods of thermofield dynamics \cite{mtu}, \cite{ojima}, the answer 
to the question of \cite{gom_ste} about the fate of operator bosonization relations for 
the Thirring model at finite temperature is given constructively. It is shown that  
HEq of Thirring model \cite{thi} admits exact solutions in terms of free (pseudo) 
scalar ``cold'' physical thermofields similar (but not the same!) to (\ref{K_2}), as 
fermionic Heisenberg thermofields $\Psi(x,\varsigma)$, for which the relations like 
(\ref{K_1}) are meaningless even asymptotically at $t\to\pm\infty$ in the weak sense 
\cite{mtu}, with respect to the both thermal (``hot'') $|0(\varsigma)\rangle_{(B,F)}$ 
or ``cold'' $|0\widetilde{0}\rangle_{(B,F)}$ vacua. To this end the DM onto the 
Schr\"odinger physical fields with above generalized weak initial condition at 
$t\to 0$ becomes especially relevant. 

Unlike the recent works \cite{abr}, \cite{abr2}, \cite{abr3}, by using direct step by 
step integration of exactly linearized HEq, we obtain the normal form of HF and 
Heisenberg thermofields, which includes all the necessary Klein factors providing a 
correct doubling of the number of degrees of freedom and consistent with all necessary 
anticommutation relations, renormalization, and symmetry conditions for the both free 
and Thirring thermofields. 

To explain our method we start with zero temperature case in the next section, which 
extends and elaborates our previous results \cite{ks_tt}.  
Definition of the model in canonical quantization scheme is given in the first 
subsection. Then the linearization procedure with corresponding definition of Heisenberg 
currents is advocated. The bosonization rules that we need for the free fields only are 
discussed in the subsection 3 with the appropriate choice of representation space for 
(pseudo) scalar fields. That all is used in subsection 4 
for direct integration of HEq with chosen initial condition. 
The connection to different one- and two- parametric solutions, the values of Schwinger 
terms, and $n$- point fermionic Wightman functions are considered in subsection 5, where 
new sources of superselection rules are shown.

The third section is devoted to the finite temperature case and extends and elaborates 
our previous preliminary results \cite{ks_ttt}, what was announced also in \cite{ks_tttt}. 
We start with qualitative thermodynamic manifestations of bosonization for the free 
massless 1D gases. Then we discuss an extension of Ojima rules. The notion of ``hot'' and 
``cold'' thermofields is introduced in subsections 3.3, and in subsections 3.4 -- 3.6 the 
thermal Thirring HEqs are obtained and solved by the 
same way as before, and the thermal $n$- point fermionic Wightman functions are obtained. 
The final remarks are made in Conclusion. Some details are explained, and useful formulas 
are collected in the Appendixes A-E. 

\section{Thirring model without temperature}
\subsection{Definition of model}

Following to the canonical quantization procedure \cite{hep} we
start with the formal Hamiltonian of the Thirring model
\cite{thi}, which in two-dimensional space-time\footnote{Here:
$x^\mu=\left(x^0,x^1\right)$; $x^0=t$; $\hbar=c=1$;
$\partial_\mu=\left(\partial_0,\partial_1\right)$; for
$g^{\mu\nu}$: $g^{00}=-g^{11}=1$; for $\epsilon^{\mu\nu}$:
$\epsilon^{01}=-\epsilon^{10}=1$;
$\overline{\Psi}(x)=\Psi^\dagger(x)\gamma^0$;
$\gamma^0=\sigma_1$, $\gamma^1=-i\sigma_2$,
$\gamma^5=\gamma^0\gamma^1=\sigma_3$, $\gamma^\mu\gamma^5=
-\epsilon^{\mu\nu}\gamma_\nu$, where $\sigma_i$ -- Pauli
matrices, and $I$ -- unit matrix; $(px)=p^0x^0-p^1x^1$, $x^\xi = x^0 + \xi x^1$, 
$P^1=-i\partial_1=-i{\partial}/{\partial x^1}$,
$2\partial_\xi =2{\partial}/{\partial x^\xi}=\partial_0+\xi\partial_1$; summation
over repeated $\xi=\pm$ is nowhere implied.} defines a Fermi
self-interaction with fixed (and further unrenormalizable)
dimensionless coupling constant $g$ for spinor field with spin
$1/2$ and zero mass:
\begin{eqnarray}
&&
H[\Psi]=H_{0[\Psi]}(x^0)+H_{I[\Psi]}(x^0),
\label{K_0} \\
&&
H_{I[\Psi]}(x^0)=\frac{g}{2}\int\limits_{-\infty}^\infty dx^1
J_{(\Psi)\,\mu}(x) J_{(\Psi)}^\mu (x),
\label{vbnmei1} \\
&&
H_{0[\Psi]}(x^0)=\int\limits_{-\infty}^\infty
dx^1\Psi^\dagger(x)E(P^1)\Psi(x), \quad E(P^1)=\gamma^5P^1,
\label{K_3}
\end{eqnarray}
satisfying the equal-time canonical anticommutation relations (CAR) and locality  
condition:
\begin{eqnarray}
&&\!\!\!\!\!\!\!\!\!\!\!\!\!\!\!\!\!\!\!\!
\left\{\Psi_\xi(x), \Psi_{\xi'}^\dagger(y)\right\}\Bigr|_{x^0= y^0} =
\delta_{\xi, \xi'}\delta\left(x^1-y^1\right),
\label{vbnmei2} \\
&&\!\!\!\!\!\!\!\!\!\!\!\!\!\!\!\!\!\!\!\!
\left\{\Psi_\xi(x),\Psi_{\xi'}(y)\right\}\Bigr|_{x^0=y^0}=0,
\label{vbnmei3} \\
&&\!\!\!\!\!\!\!\!\!\!\!\!\!\!\!\!\!\!\!\!
\left\{\Psi_\xi(x),\Psi^{\#}_{\xi'}(y)\right\}\Bigr|_{(x-y)^2<0}=0,\;\,\mbox{ with: }\; 
\Psi^{\#}_\xi(y)=\Psi_\xi(y),\;\Psi^{\dagger}_\xi(y). 
\label{vbnmei4}
\end{eqnarray}
The indices $\xi, \xi' = \pm $, as well as for $x^\xi$, enumerate here the components 
of HF by the rule:
\begin{eqnarray}
\Psi (x) = \left(\begin{array}{c}
\Psi_1(x) \\ \Psi_2(x) \end{array}\right)=
\left(\begin{array}{c}
\Psi_{+}(x) \\ \Psi_{-}(x)\end{array}\right)\mapsto \Psi_\xi(x),\quad 
\left(\overline{\Psi}(x)\right)_\xi=\Psi^\dagger_{-\xi}(x), 
\label{vbnmei6}
\end{eqnarray}
and the vector current $J_{(\Psi)}^\mu(x)$, together with the pseudovector current
$J_{(\Psi)}^{5\mu}(x)$, for $\mu,\nu = 0,1$, is their yet formal local bilinear
functional of the form:
\begin{eqnarray}
&&\!\!\!\!\!\!\!\!\!\!\!\!\!\!\!\!\!\!\!\!
J_{(\Psi)}^\mu (x)\longmapsto \overline{\Psi}(x)\gamma^\mu \Psi(x),
\label{vbnmei5} \\
&&\!\!\!\!\!\!\!\!\!\!\!\!\!\!\!\!\!\!\!\!
J_{(\Psi)}^{5\mu}(x)\longmapsto\overline{\Psi}(x)\gamma^\mu\gamma^5\Psi(x)=
-\epsilon^{\mu\nu}J_{(\Psi)\nu}(x),
\nonumber
\end{eqnarray}
which due to (\ref{K_0})--(\ref{vbnmei6}) formally appears also in the canonical 
equations of motion\footnote{Contribution to (\ref{45bn6l4}) due to non-commutativity of 
$J_{(\Psi)}^\nu(x)$ and $\Psi(x)$ is formally proportional to 
$\delta(0)\gamma^0\gamma_\nu\gamma^0\gamma^\nu=0$.} \cite{s_w}--\cite{wai}:
\begin{eqnarray}
&&
i\partial_0\Psi(x)=\left[\Psi(x), H[\Psi]\,\right]=
\left[E(P^1)+g\gamma^0\gamma_\nu J_{(\Psi)}^\nu(x)\right]\Psi(x),
\label{45bn6l4} \\
&&
\mbox{or: }\;
2\partial_\xi \Psi_\xi(x)=-igJ^{-\xi}_{(\Psi)}(x)\Psi_\xi(x), \quad  \xi=\pm ,
\label{45bn6l5}
\end{eqnarray}
-- for each $\xi$-component of the field (\ref{vbnmei6}), also are related formally to 
the corresponding current components:
\begin{eqnarray}
J_{(\Psi)}^\xi (x)=J_{(\Psi)}^0 (x)+\xi J_{(\Psi)}^1(x)\longmapsto
2\Psi_\xi^\dagger(x)\Psi_\xi (x), \quad \xi=\pm .
\label{vbnmei8}
\end{eqnarray}
The correct definitions of these formal operator products will be discussed hereinafter.

\subsection{Linearization of the Heisenberg equation}

An immediate consequence of the field equations of motion
(\ref{45bn6l4}), (\ref{45bn6l5}) are the local conservation
laws \cite{s_w}--\cite{wai} for the currents (\ref{vbnmei5}),
(\ref{vbnmei8}):
\begin{eqnarray}
\partial_\mu J_{(\Psi)}^\mu(x)=0,\quad
\partial_\mu J_{(\Psi)}^{5\mu}(x)=-\epsilon_{\mu\nu}\partial^\mu J^\nu_{(\Psi)}(x)=0,
\;\mbox{ or: }\; \partial_\xi J_{(\Psi)}^\xi(x)=0,\quad \xi=\pm ,
\label{CC_1} 
\end{eqnarray}
fully determine their dynamics as a free one \cite{Leut}, \cite{d_f_z}.
Therefore it is not surprising, that by virtue of the same equations of motion
(\ref{45bn6l4}), (\ref{45bn6l5}), as well as by means of the anti-commutation
relations (\ref{vbnmei2}) for HF, it is a simple matter to show \cite{ks} that:
\begin{eqnarray}
i\partial_0\gamma^0\gamma_\nu J_{(\Psi)}^\nu(x)-
\left[\gamma^0\gamma_\nu J_{(\Psi)}^\nu (x), H_{0[\Psi]}(x^0)\right]
= iI\,\partial_\mu J_{(\Psi)}^\mu(x)+
i\gamma^5\,\epsilon_{\mu\nu}\partial^\mu J^\nu_{(\Psi)}(x)\equiv 0,
\label{bnemti4}
\end{eqnarray}
where the first term on the r.h.s. of equality (\ref{bnemti4})
comes evidently from the left terms with $\nu=0$, while the
second term on the r.h.s. comes from the left terms with
$\nu=1$. The canonical equation of motion for this operator of
``total current'' in Eq. (\ref{45bn6l4}), containing of course
its commutator with the total Hamiltonian $H[\Psi]$ (\ref{K_0}), recasts then to the 
following equation:
\begin{eqnarray}
i\partial_0\gamma^0\gamma_\nu J_{(\Psi)}^\nu(x)-
\left[\gamma^0\gamma_\nu J_{(\Psi)}^\nu (x), H_{0[\Psi]}(x^0)\right]
=\left[\gamma^0\gamma_\nu J_{(\Psi)}^\nu (x), H_{I[\Psi]}(x^0)\right] = 0,
\label{bnemti5}
\end{eqnarray}
which thus cannot contain a contribution from the commutator
with the interaction Hamiltonian $H_{I[\Psi]}(x^0)$ given by Eq. (\ref{vbnmei1}).
Hence, as well as for the Federbush model \cite{ks}, a non-zero contribution of
Schwinger terms in HEq (\ref{bnemti5}) would be premature, because, due to Eq.
(\ref{bnemti4}), it leads to violation of the current conservation laws (\ref{CC_1}).

On the one hand, within the framework of canonical quantization procedure \cite{hep},
the vanishing of expressions (\ref{bnemti4}), (\ref{bnemti5}) means, that temporal
evolution of this ``total current'' is governed by a free Hamiltonian
$H_{0[\chi]}\left(x^0\right)$ of the same form (\ref{K_3}) quadratic on some kind of
free massless trial physical Dirac fields $\chi(x)$, furnished by the same
anti-commutation relations and by the same conservation laws for corresponding currents
$J_{(\chi)}^\nu (x)$, $J_{(\chi)}^{5\nu}(x)$, defined formally by
Eqs. (\ref{vbnmei2})--(\ref{vbnmei5}),  (\ref{vbnmei8}), (\ref{CC_1}) with
$\Psi(x)\mapsto\chi(x)$:
\begin{eqnarray}
i\partial_0\gamma^0\gamma_\nu J_{(\chi)}^\nu (x)-
\left[\gamma^0\gamma_\nu J_{(\chi)}^\nu (x), H_{0[\chi]}(x^0)\right]=
iI\,\partial_\mu J_{(\chi)}^\mu(x)+
i\gamma^5\,\epsilon_{\mu\nu}\partial^\mu J^\nu_{(\chi)}(x)=0.
\label{bnemti6}
\end{eqnarray}
On the other hand, the Heisenberg current operators appearing in (\ref{bnemti4}),
(\ref{bnemti5}) acquire precise operator meaning -- with non-vanishing Schwinger
term -- only after the choice of the representation space \cite{Leut}, \cite{hep},
\cite{Vlad} for anticommutation relations (\ref{vbnmei2})--(\ref{vbnmei4}) and
subsequent reduction in this representation to the normal-ordered form by means of
renormalization, for example, via point-splitting and subtraction of the vacuum
expectation value \cite{blot}:
\begin{eqnarray}
&&\!\!\!\!\!\!\!\!\!\!\!\!\!\!\!\!\!\!
J^0_{(\Psi)} (x) \longmapsto
\lim\limits_{\widetilde{\varepsilon} \rightarrow 0}
\widehat{J}^0_{(\Psi)}(x;\widetilde{\varepsilon})=\widehat{J}^0_{(\Psi)}(x),
\quad J^1_{(\Psi)} (x) \longmapsto
\lim\limits_{\varepsilon \rightarrow 0}
\widehat{J}^1_{(\Psi)}(x;\varepsilon)=\widehat{J}^1_{(\Psi)}(x),
\label{bos-111}\\
&&\!\!\!\!\!\!\!\!\!\!\!\!\!\!\!\!\!\!
\mbox{where for: }\; \widetilde{\varepsilon}^\mu=-\epsilon^{\mu\nu}\varepsilon_\nu,\; 
\mbox{ at first: }\; \widetilde{\varepsilon}^0 = \varepsilon^1\rightarrow 0,\;
\mbox{ with fixed: }\; \widetilde{\varepsilon}^1 =\varepsilon^0,\quad 
\varepsilon^2=-\widetilde{\varepsilon}^2>0,
\label{K_E} \\
&&\!\!\!\!\!\!\!\!\!\!\!\!\!\!\!\!\!\!
\mbox{for: }\;
\widehat{J}^\nu_{(\Psi)}(x;a)=
Z^{-1}_{(\Psi)}(a)\left[\overline{\Psi}(x + a)\gamma^\nu \Psi (x)-
\langle 0|\overline{\Psi}(x + a)\gamma^\nu\Psi(x)|0\rangle\right],
\label{K_Z}
\end{eqnarray}
and accordingly for every $\xi$- component (\ref{vbnmei8}). 
The renormalization 
``constant'' $Z_{(\Psi)}(a)$ is defined below in (\ref{K_Za}). The definition of 
renormalized current (\ref{bos-111})--(\ref{K_Z}) used here corresponds to the 
well-known Mandelstam-Schwinger prescription \cite{man}, \cite{sok} specified in the 
work \cite{ot} and, unlike Johnson definition \cite{Jon}, \cite{wai}, 
directly depends on the choice of representation space via the vacuum expectation value 
\cite{blot} in Eq. (\ref{K_Z}) like the very meaning of Schwinger term 
\cite{Leut,ot,fab-iva}. One can show \cite{ot}, these different current definitions 
for the massless case coincide only for the free Dirac fields (cf. Eqs. (\ref{nweyi19}) 
and (\ref{J_L}) below).

The comments given above jointly with the foregoing arguments deduced from Eq.
(\ref{CC_1})--(\ref{bnemti6}) allow to identify in HEq (\ref{45bn6l4}), at least in a
weak sense, the Heisenberg operator of ``total current'', defined by Eqs.
(\ref{vbnmei5}), (\ref{bnemti4}), with that operator, defined by Eqs. 
(\ref{vbnmei5}), (\ref{bnemti6}) for the free massless trial physical Dirac fields 
$\chi(x)$ and renormalized in the sense of normal form (\ref{bos-111})--(\ref{K_Z}) up 
to an unknown yet constant $\beta$:
\begin{eqnarray}
&&\!\!\!\!\!\!\!\!\!\!\!\!\!\!\!\!\!\!
\gamma^0\gamma_\nu J_{(\Psi)}^\nu(x) \stackrel{\rm w}{\longmapsto}
\frac{\beta}{2\sqrt{\pi}}\gamma^0\gamma_\nu\widehat{J}_{(\chi)}^\nu(x),\;\;
\mbox{ where:}
\label{ns94m61} \\
&&\!\!\!\!\!\!\!\!\!\!\!\!\!\!\!\!\!\!
\widehat{J}_{(\chi)}^\nu(x)=
\lim\limits_{\varepsilon,(\widetilde{\varepsilon})\rightarrow 0}
\widehat{J}_{(\chi)}^\nu\left(x;\varepsilon (\widetilde{\varepsilon})\right)\,
\equiv\, :J_{(\chi)}^\nu(x):\,, \;\mbox{ for: }\;Z_{(\chi)}(a)=1. 
\label{ns94m62}
\end{eqnarray}
Here the symbol $:\ldots:$ means for $Z_{(\chi)}(a)=1$ the usual normal form 
\cite{blot,gldj} with respect to free fermionic annihilation/creation operators of 
the field $\chi(x)$ \cite{fab-iva}, defined in Appendix E. 

The identification (\ref{ns94m61}), (\ref{ns94m62}) leads to linearization of both 
Eqs. (\ref{45bn6l4}), (\ref{45bn6l5}) in the representation of these trial physical 
fields $\chi(x)$. Of course, the Eq. (\ref{45bn6l4}) is linearized with respect to $x^0$, 
while the Eq. (\ref{45bn6l5}) -- with respect to $x^\xi$. However, the latter equation 
is the preference of two-dimensional world with initial condition being far from 
evidence. Whereas the former equation admits the above-mentioned in the Introduction 
physically reasonable generalized initial condition at $t=x^0=0$: 
$\lim\limits_{t\to 0}\Psi(x^1,t)\stackrel{\rm w}{=}\Upsilon[\psi_s(x^1,0)]$. 
The observed weak linearization (\ref{ns94m61}) of HEq with so-generalized initial 
conditions in a weak sense allows below to overcome the restrictions of Haag theorem 
\cite{blot,hep}, removing them into the representation construction of the massless free 
Dirac field $\chi(x)$. Since unlike \cite{vklt,kt,ks} the massless free Dirac field does 
not defines here the true asymptotic states of the model, this initial condition does not 
fix here the constant $\beta$, which will be defined dynamically in subsequent 
subsections contrary to \cite{Leut,blot}. 

\subsection{Bosonization and scalar fields}

As was shown in \cite{ks} such kind of linearization of HEq for the Federbush model
directly leads to its solution in the form of DM 
$\Psi(x)=\Upsilon[\psi_{1}(x), \psi_{2}(x)]$ onto the free massive Dirac fields
$\psi_{1,2}(x)$ with different non-zero masses $m_{1,2}$, or in the form of DM 
$\Psi(x)=\Upsilon[\Phi_{1}(x), \Phi_{2}(x)]$ onto the pseudoscalar sine-Gordon fields 
$\Phi_{1,2}(x)$ furnished by appropriately defined normal ordering prescription 
\cite{wai}. Unlike the massive case, the components 
$\chi_\xi(x)$ of two-dimensional free massless field become completely decoupled, 
$\partial_\xi \chi_\xi(x)=0$. As a consequence, this field turns out to be
essentially non-uniquely defined or reducible and equipped by many inequivalent
representations both in the spaces of a free massless (pseudo) scalar field ($\phi(x)$), 
$\varphi(x)$ \cite{blot} and free massive pseudoscalar field $\phi_m(x)$ \cite{raja}. 
Of course, the DM is physically meaningful only onto the complete, irreducible sets of 
fields: $\Psi(x)=\Upsilon[\varphi(x),\phi(x)]$, or $\Psi(x)=\Upsilon[\phi_m(x)]$, or
$\Psi(x)=\Upsilon[\psi_M(x)]$, -- for the phase with spontaneously broken chiral
symmetry \cite{fab-iva}--\cite{fab-iva-09}, \cite{fujita_2}, so further we consider here 
only the first possibility.
The corresponding DM is known also as BP, allows to operate with functionals of boson 
fields instead of fermionic operators and forms a powerful tool for obtaining 
non-perturbative solutions in various two-dimensional models \cite{man}, \cite{nak}, 
\cite{blot}, \cite{fab-iva}. Its use also simplifies integration of the linearized HEq 
like (\ref{45bn6l4}) \cite{ks}.

Being a formal consequence of the current conservation conditions (\ref{CC_1}) only,
the bosonization rules have, generally speaking, the sense of weak equalities only for
the current operator in the normal-ordered form (\ref{bos-111})--(\ref{K_Z}), that
already implies a choice of certain representations of (anti-) commutation relations 
(\ref{vbnmei2}) and (\ref{K_8_1}) below.
However, for the free massless fields $\chi(x)$, $\varphi(x)$, $\phi(x)$, this choice 
(\ref{chi_xi_0}) and (\ref{phi_+_-}) below is carried out automatically. This, due to the 
linearization condition (\ref{ns94m61}), 
(\ref{ns94m62}), becomes enough for our purposes, since for the free fields these 
relationships appear as strong operator equalities \cite{blot, tvash}:
\begin{eqnarray}
&&\!\!\!\!\!\!\!\!\!\!\!\!\!\!\!\!\!\!
\widehat{J}_{(\chi)}^\mu(x)=\frac{1}{\sqrt{\pi}}\,\partial^\mu\varphi(x)=
-\,\frac{1}{\sqrt{\pi}}\,\epsilon^{\mu\nu}\partial_\nu \phi (x), \quad
\widehat{J}_{(\chi)}^{-\xi}(x)= 
2:\chi_{-\xi}^\dagger (x)\chi_{-\xi}(x): =
\frac{2}{\sqrt{\pi}}\,\partial_{\xi} \varphi^{\xi}\left(x^{\xi}\right)\,,
\label{nweyi19} \\
&&\!\!\!\!\!\!\!\!\!\!\!\!\!\!\!\!\!\!
\mbox{where: }\;a^\mu\partial_\mu \varphi^\xi(x^\xi)=a^\xi\partial_\xi\varphi^\xi(x^\xi)=
a^\xi\partial_0\varphi^\xi(x^\xi)=\xi a^\xi\partial_1\varphi^\xi(x^\xi),\;\mbox 
{ for: }\; x^\xi=x^0+\xi x^1, 
\label{a_d_xi_phi} \\
&&\!\!\!\!\!\!\!\!\!\!\!\!\!\!\!\!\!\!
\mbox{and for $\nu=0,1$: }\;
\sum\limits_{\xi=\pm }(\xi 1)^\nu\partial_\xi\varphi^\xi(x^\xi)=
\partial_\nu\sum\limits_{\xi=\pm}\varphi^\xi(x^\xi)=
\epsilon_{\nu\mu}\partial^\mu\sum\limits_{\xi=\pm}\xi\varphi^\xi(x^\xi).
\label{d_xi_phi}
\end{eqnarray}
Unlike \cite{nak} the free massless scalar and pseudoscalar fields $\varphi(x)$ 
and $\phi(x)$, $\partial_\mu\partial^\mu\varphi(x)=0=\partial_\mu\partial^\mu\phi(x)$, 
are taken mutually dual and coupled by symmetric integral relations with the step 
function $\varepsilon(x^1)={\rm sgn}(x^1)$:
\begin{eqnarray}
&&\!\!\!\!\!\!\!\!\!\!\!\!\!\!\!\!\!\!
\left. 
\begin{array}{c} \phi(x) \\ \varphi(x) \end{array}
\right\}
=-\frac{1}{2}\int\limits_{-\infty}^\infty dy^1
\varepsilon \left(x^1-y^1\right)\partial_0 \left\{
\begin{array}{c}
\varphi\left(y^1,x^0\right), \\ \phi\left(y^1,x^0\right), \end{array} \right. 
\label{K_5} \\
&&\!\!\!\!\!\!\!\!\!\!\!\!\!\!\!\!\!\!
\mbox{where: }\;
\varepsilon(x^1)= 
\frac{1}{i\pi}\int\limits_{-\infty}^\infty dk^1\,{\rm P}
\frac 1{k^1}\,e^{ik^1x^1}=\left\{
\begin{array}{c}
1,\;\mbox{ for }\; x^1>0, \\
-1,\;\mbox{ for }\; x^1<0, 
\end{array}  \right.
\label{K_5_ep}
\end{eqnarray}
imposing the following asymptotical boundary conditions of solitonic type: 
\begin{eqnarray}
&&\!\!\!\!\!\!\!\!\!\!\!\!\!\!\!\!\!\!
\varphi(-\infty,x^0)+\varphi(\infty,x^0)=0,\quad 
\phi(-\infty,x^0)+\phi(\infty,x^0)=0,  
\label{as_con}
\end{eqnarray}
that admit also nonzero behavior on both space infinities $x^1\to\pm\infty$, providing 
for the conserved charges of these fields, similar to \cite{nak,fab-iva,blot,oksak},  
usual and topological definitions (also with 
$\phi(x)\longmapsto\widehat{\phi}(x)+\upsilon_{cl}(x)$):
\begin{eqnarray}
&&\!\!\!\!\!\!\!\!\!\!\!\!\!\!\!\!\!\!
\left. \begin{array}{c} O \\ O_5\end{array}\right\}=
\lim_{L\to\infty}
\int\limits_{-\infty}^\infty dy^1\Delta\left(\frac{y^1}L\right)\partial_0
\left\{\begin{array}{c}\varphi (y^1,x^0) \\
\phi(y^1,x^0) \end{array}\right\}
\stackunder{\Delta=1}{\Longrightarrow}
\left\{\begin{array}{c}\phi(-\infty, x^0)-\phi(\infty,x^0)\longmapsto\widehat{O}+o_{cl},\\
\varphi(-\infty,x^0)-\varphi(\infty,x^0)\longmapsto\widehat{O}_{5}+o_{5\,cl}.\end{array}
\right. 
\label{K_O} 
\end{eqnarray}
The latter ones correspond to the possible c-number contribution of classical fields 
$\upsilon_{cl}(x)$, marking the various unitarily inequivalent vacua \cite{blot,oksak} 
by various values of respective charges $o_{cl}, o_{5\,cl}\neq 0$ (\ref{K_O}). 

The above right ($\xi=-$) and left ($\xi=+$) moving fields $\varphi^{\xi}(x^{\xi})$ and 
their charges $Q^\xi$ for $\xi=\pm $ are defined by linear combinations \cite{blot}:
\begin{eqnarray}
&&\!\!\!\!\!\!\!\!\!\!\!\!\!\!\!\!\!\!
\varphi^\xi(x^\xi)=\frac{1}{2}\left[\varphi(x)-\xi\phi(x)\right],
\quad 
Q^\xi=\frac{1}{2}\left[O-\xi{O}_5\right]\Longrightarrow
\xi\varphi^\xi(x^0\!+\xi\infty)-\xi\varphi^\xi(x^0\!-\xi\infty)
=\pm 2\varphi^\xi(x^0\pm\infty), 
\label{K_7} \\
&&\!\!\!\!\!\!\!\!\!\!\!\!\!\!\!\!\!\!
\varphi(x)=\sum\limits_{\xi=\pm }\varphi^{\xi}\left(x^{\xi}\right),\quad 
\phi(x)=- \sum\limits_{\xi=\pm }\xi\varphi^{\xi}\left(x^{\xi}\right),
\quad O=\sum\limits_{\xi=\pm }Q^\xi, \quad 
O_5=- \sum\limits_{\xi=\pm }\xi Q^\xi.
\label{sum_xi} 
\end{eqnarray} 
The last expressions (\ref{K_7}) for $Q^\xi$ explicitly demonstrate the   
$x^0$ - independence of all charges. 
All commutation relations \cite{nak,fab-iva_2,blot} for the quantum fields 
$\varphi(x)$, $\phi(x)$, $\varphi^\xi(x^\xi)$, and the quantum charges $O,O_5, Q^\xi$:
\begin{eqnarray}
&&\!\!\!\!\!\!\!\!\!\!\!\!\!\!\!\!\!\!
\left[\varphi(x),\partial_0 \varphi (y)\right] \bigr|_{x^0=y^0}=
\left[\phi(x),\partial_0\phi(y)\right]\bigr|_{x^0=y^0}=i\delta(x^1-y^1),
\label{K_8_1} \\
&&\!\!\!\!\!\!\!\!\!\!\!\!\!\!\!\!\!\!
\left[\varphi(x), \varphi (y)\right]=
\left[\phi(x),\phi(y)\right]=
-\,\frac{i}{2}\,\varepsilon(x^0-y^0)\,\theta\left((x-y)^2\right)=
\frac 1i {\rm D}_0(x-y),
\label{K_8} \\
&&\!\!\!\!\!\!\!\!\!\!\!\!\!\!\!\!\!\!
\left[\varphi(x),\phi(y)\right]=
\frac{i}{2}\,\varepsilon(x^1-y^1)\,\theta\left(-(x-y)^2\right),
\label{K_8_8} \\
&&\!\!\!\!\!\!\!\!\!\!\!\!\!\!\!\!\!\!
\left[\varphi(x), O\right]=\left[\phi(x), O_5\right]=i,\quad 
\left[\varphi(x), O_5\right]=\left[\phi(x), O\right]=0,
\label{K_9_9} \\
&&\!\!\!\!\!\!\!\!\!\!\!\!\!\!\!\!\!\!
\left[\varphi^\xi\left(s\right),\varphi^{\xi'}\left(\tau\right)\right]=
-\frac{i}{4}\varepsilon(s - \tau)\delta_{\xi, \xi'}, \quad
\left[\varphi^\xi(s), Q^{\xi'}\right]=\frac{i}{2}\delta_{\xi,\xi'},
\label{K_9}
\end{eqnarray}
with the step function $\theta(s)=\left(1+\varepsilon(s)\right)/2$, are reproduced by 
commutators of their annihilation/creation (frequency) parts (see 
(\ref{D_-0})--(\ref{thet_delt}) of Appendix B) \cite{d_f_z,nak,fab-iva}:
\begin{eqnarray}
&&\!\!\!\!\!\!\!\!\!\!\!\!\!\!\!\!\!\!
\langle 0|\phi(x)\phi(y)|0\rangle=
\langle 0|\varphi(x)\varphi(y)|0\rangle=
\left[\phi^{(+)}(x),\phi^{(-)}(y)\right]=
\left[\varphi^{(+)}(x),\varphi^{(-)}(y)\right]=
\frac 1i {\rm D}^{(-)}(x-y)\equiv 
\label{D_} \\
&&\!\!\!\!\!\!\!\!\!\!\!\!\!\!\!\!\!\!
\equiv
\frac 1i D^{(-)}\left(x^\xi-y^\xi\right)+ \frac 1i D^{(-)}\left(x^{-\xi}-y^{-\xi}\right)=
-\frac 1{4\pi}\ln\biggl(-\overline{\mu}^2(x-y)^2+i0\,\varepsilon(x^0-y^0)\biggr), 
\label{DD_} \\
&&\!\!\!\!\!\!\!\!\!\!\!\!\!\!\!\!\!\!
\left[\varphi^{\xi(\pm)}(s),\varphi^{\xi'(\mp)}(\tau)\right]=
\pm\frac{\delta_{\xi, \xi'}}i D^{(-)}(\pm(s-\tau))\equiv 
\mp\frac{\delta_{\xi,\xi'}}
{4\pi}\ln\biggl(i\overline{\mu}\Bigl\{\pm(s-\tau)-i0\Bigr\}\biggr),
\label{nblaie16} \\
&&\!\!\!\!\!\!\!\!\!\!\!\!\!\!\!\!\!\!
\left[\varphi^{\xi(\pm)}(s),  Q^{\xi'(\mp)}\right]=
\frac{i}{4}\delta_{\xi, \xi'}, \quad
\left[ Q^{\xi(\pm)},  Q^{\xi'(\mp)}\right]=\pm a_0 \delta_{\xi, \xi'}, \quad 
\left[ O^{(\pm)}, O^{(\mp)}\right]= \left[ O^{(\pm)}_5, O^{(\mp)}_5\right]=\pm 2a_0,
\label{nblaie19} \\
&&\!\!\!\!\!\!\!\!\!\!\!\!\!\!\!\!\!\!
a_0=a_0(L)=\pi\int\limits_{0}^\infty d k^1 k^0\left(\delta_L(k^1)\right)^2, \quad 
L\rightarrow\infty, \quad \lim_{L\rightarrow\infty}\delta_L(k^1)=\delta(k^1), 
\label{D_a0} 
\end{eqnarray}
defined here through the annihilation/creation operators $c(k^1),\,c^\dagger(k^1)$ 
of the {\bf pseudoscalar field} $\phi(x)$, integrated with some distributions 
\cite{gshl}: 
\begin{eqnarray}
&&\!\!\!\!\!\!\!\!\!\!\!\!\!\!\!\!\!\!
{\cal P}c(k^1){\cal P}^{-1}=-c(-k^1),\quad 
[c(k^1),c^\dagger(q^1)]=2\pi\, 2k^0\delta(k^1-q^1),\quad 
c(k^1)|0\rangle=0,
\label{cc_k0} \\
&&\!\!\!\!\!\!\!\!\!\!\!\!\!\!\!\!\!\!
\mbox{for: }\;
k^0=|k^1|,\quad {\rm P}\frac 1{k^1}=\frac{\varepsilon\left(k^1\right)}{|k^1|},
\quad \frac 14\left({\rm P}\frac 1{k^1}-\frac{\xi}{k^0}\right)=
\frac{-\xi\theta\left(-\xi k^1\right)}{2 k^0}, \;\mbox{ as:}  
\label{P_e_k0} \\
&&\!\!\!\!\!\!\!\!\!\!\!\!\!\!\!\!\!\!
\phi(x) = \frac{1}{2\pi} \int\limits_{-\infty}^\infty\frac{d k^1}{2 k^0}
\left[c\left(k^1\right)e^{-i(kx)}+c^\dagger\left(k^1\right)e^{i(kx)}\right]\equiv
\phi^{(+)}(x)+\phi^{(-)}(x),
\label{phi_+_-} \\
&&\!\!\!\!\!\!\!\!\!\!\!\!\!\!\!\!\!\!
\varphi(x) = \frac{1}{2\pi}\int\limits_{-\infty}^\infty d k^1\,{\rm P}\frac 1{k^1} 
\left[c\left(k^1\right)e^{-i(kx)}+c^\dagger\left(k^1\right)
e^{i(kx)}\right]\equiv 
\varphi^{(+)}(x)+\varphi^{(-)}(x),
\label{varphi_+_-} \\
&&\!\!\!\!\!\!\!\!\!\!\!\!\!\!\!\!\!\!
\varphi^{\xi(+)}(s)= 
-\,\frac{\xi}{2\pi}\int\limits_{-\infty}^\infty \frac{d k^1}{2 k^0}
\theta\left(-\xi k^1\right)c(k^1) e^{-i k^0 s},
\label{ph_pm} \\
&&\!\!\!\!\!\!\!\!\!\!\!\!\!\!\!\!\!\!
\varphi^{\xi(-)}(s)=\left[\varphi^{\xi(+)}(s)\right]^\dagger,\quad 
\varphi^{\xi}(s)=\varphi^{\xi(+)}(s)+\varphi^{\xi(-)}(s), 
\label{ph_mp} \\
&&\!\!\!\!\!\!\!\!\!\!\!\!\!\!\!\!\!\!
 Q^{\xi(+)}(\widehat{x}^0)=\lim_{L\rightarrow\infty} \frac{i\xi}{2}
\int\limits_{-\infty}^\infty d k^1 \theta\left(-\xi k^1\right) c(k^1) 
e^{-ik^0 \widehat{x}^0}\delta_L (k^1),
\label{Q_pm} \\
&&\!\!\!\!\!\!\!\!\!\!\!\!\!\!\!\!\!\!
 Q^{\xi(-)}=\left[ Q^{\xi(+)}\right]^\dagger,\quad 
 Q^{\xi}= Q^{\xi(+)}+ Q^{\xi(-)}, \quad O_5=\frac i2\left[c^\dagger(0)-c(0)\right]. 
\label{Q_mp} 
\end{eqnarray}
Here the expression (\ref{varphi_+_-}) for the scalar field follows from substitution of 
Eqs. (\ref{phi_+_-}), (\ref{K_5_ep}) to the second line of Eq. (\ref{K_5}), and, 
without loss of generality, here and below we omit the classical parts of fields and 
charges $\upsilon_{cl}(x),o_{cl},o_{5\,cl}$ in the Eq. (\ref{K_O}). The labeled by them 
inequivalent representations of (pseudo) scalar fields have a purely topological nature 
\cite{blot,oksak}, which is not relevant for our further dynamical purposes.
According to Eqs. (\ref{Q_pm}), (\ref{D_a0}), the quantum charges $O, O_5, Q^\xi$ are 
defined by zero mode's contribution only. Nevertheless, the both definitions of Eq. 
(\ref{K_O}) recast $ \widehat{O}_5\Rightarrow O_5 $ to the last formal expression of 
Eq. (\ref{Q_mp}). Whereas for $O,\,Q^\xi$ the distributions 
$\varepsilon(k^1),\theta(k^1)$ appear, inventing the infrared regularization 
(\ref{Q_pm}), (\ref{D_a0}), which is necessary also to obtain all the relations 
(\ref{nblaie19}).

According to \cite{fab-iva}, \cite{fab-iva_2}, the invariance under the parity 
transformation ${\cal P}\{\ldots\}{\cal P}^{-1}$ of generating functional 
${\cal Z}[\phi(x);{\cal J}]$ for the free massless pseudoscalar field $\phi(x)$ 
(\ref{phi_+_-}), (\ref{cc_k0}), unlike that for the theory of scalar field $\varphi(x)$ 
(\ref{varphi_+_-}), implies the following feature of the classical source ${\cal J}(x)$: 
${\cal J}(-x^1,x^0)=-{\cal J}(x^1,x^0)$, $\int d^2x {\cal J}(x)=0$, which automatically 
removes the zero mode contribution of this field into the generating functional, leading 
to its well definiteness and gauge invariance 
under field-translation by arbitrary constant. The appearance of Cauchy main value 
(\ref{P_e_k0}) in the expression (\ref{varphi_+_-}) for the dual scalar field also 
automatically excludes its zero mode contribution \cite{Leut}. 
This relaxes for the chosen representation space of pseudoscalar field (\ref{cc_k0}) 
the problem of non-positivity of inner product \cite{wai,blot} induced by Wightman 
functions ${\rm D}^{(-)}$ (\ref{D_}) or $D^{(-)}$ (\ref{nblaie16}). 


A number of alternative methods are used to solve this problem\footnote{May be the most 
global of them is bosonization (DM) onto the free massive pseudoscalar field $\phi_m(x)$ 
\cite{raja} for $\beta^2<4\pi$, what eliminates this problem together with gauge 
invariance under field-translation.}. 
Some of them use instead of the state $|0\rangle$ of Eq. (\ref{cc_k0}) the vacuum state 
$|\widehat{\upsilon}\rangle$ averaged with respect to above gauge group 
(cf. (\ref{gauge}) below), that provides by GNS construction for non-normalizable vacuum 
functional \cite{blot,oksak} or by quantization in the indefinite metric space 
\cite{mps_1,mps_2,blot}. Another methods use the non degenerate vacuum state 
$|0\rangle$ of Eq. (\ref{cc_k0}), taken or as the unique \cite{col}--\cite{nak} or as the 
only one of degenerate family states  \cite{fab-iva}--\cite{fab-iva_2,fujita_2}, or with 
or without spontaneous breaking of the above 
field-translation symmetry, or with \cite{fab-iva}--\cite{fab-iva_2,fujita_2} or without 
\cite{nak} replacing of the infrared regularization parameter $\overline{\mu}$ 
(\ref{DD_}), (\ref{nblaie16}), (\ref{D-_ln}) to arbitrary fixed finite scale $M$. 

These various representation spaces of massless (pseudo) scalar field result in 
particular in various meanings of non-negative value $a_0$ (\ref{D_a0}) in Eq. 
(\ref{nblaie19}) varying from $0$ for some ones to $\infty$ for another ones. It 
depends in turn on the choice of the volume cut-off regularization function 
$\Delta(y^1/L)$, with the Fourier image $\delta_L (k^1)$ (\ref{DDddDD}) (analysed in 
Appendix C), which should be inserted into the integral (\ref{K_O}) for correct charge 
definition. Unfortunately any of such (continuous or not) cut-off regularization induces 
the fictitious and non-physical $\widehat{x}^0$ - dependence of formally conserved 
charges $O$, $O_5$ (\ref{K_O}) or $Q^\xi$ (\ref{K_7}) and their frequency parts 
(\ref{Q_pm}), and, in general, destroys the last equalities of Eqs. (\ref{K_O}), 
(\ref{K_7}), important in topological sense \cite{fab-iva_2,raja,blot,tvash}. 
Such kind of $\widehat{x}^0$ - dependence, being an artifact of the charge's 
regularization (\ref{K_O}), should be eliminated at the end of calculation. 
Due to (\ref{D_a0}) this is achieved by virtue of thermodynamic limit $L\to\infty$. 
Indeed, because for zero temperature $\widehat{x}^0$ does not appear explicitly in 
commutation relations (\ref{nblaie19}), further without loss of generality it may be 
fixed as $\widehat{x}^0=0$ in accordance with (\ref{K_O}), (\ref{K_7}). 
However, for the nonzero temperature (\ref{pm__ph_Qt}) this dependence will 
require additional care.

The parameter $a_0$ itself makes sense of regularization parameter, which in the 
end of calculation should disappear in physical quantities. This will be especially 
important in the next section, where it acquires a temperature dependence 
(\ref{QQ_QtQt}), 
(\ref{a_1_00}). Besides, this parameter, for any real $\upsilon $, defines the vacuum 
expectation value (VEV) of the operator of the above field-translation gauge 
transformation: 
\begin{eqnarray}
&&\!\!\!\!\!\!\!\!\!\!\!\!\!\!\!\!\!\!
\exp\left\{i\upsilon O_5\right\}\phi(x)\exp\left\{-i\upsilon O_5\right\}=\phi(x)+\upsilon, 
\label{gauge} \\
&&\!\!\!\!\!\!\!\!\!\!\!\!\!\!\!\!\!\!
\langle 0|\exp\left\{i\upsilon O_5\right\}|0\rangle\equiv\langle 0|\upsilon\rangle
=\exp\left\{-a_0\upsilon^2\right\}, 
\label{VEV_0}
\end{eqnarray}
which is well known in quantum theory of free massless (pseudo) scalar field \cite{i_z} 
\S 11.2.2. The state $|\upsilon\rangle$ is a well known coherent state of harmonic 
oscillator, which corresponds to zero mode like (\ref{Q_mp}), $k^1=0$, for $L\to\infty$, 
and if simultaneously $a_0\to\infty$, then $\langle 0|\upsilon\rangle\to 0$, and the 
state $|\upsilon\rangle$ defines another orthogonal vacuum state 
of the above degenerate family \cite{fab-iva_2,mtu,i_z}, for example, for the case of 
usual box of the length $2L$ from the last but one column of the Table given in 
Appendix C. While for any continuous regularization function from another columns of this 
Table the finite $a_0$ means, that the charge definition has nothing to do with previous 
standard thermodynamic limit, and corresponds to another vacuum structure of the 
representation space of (pseudo) scalar field \cite{fab-iva}--\cite{fab-iva_2}. 
The randomization \cite{fab-iva_2} of the function $\Delta(y^1/L)$ 
(or $k^0\delta_L(k^1)$ and the corresponding value of $a_0$) may be 
used for infrared ``stabilization'' of Wightman functions (\ref{D_}), (\ref{nblaie16}), 
which replaces \cite{fab-iva} the infrared regularization parameter $\overline{\mu}$ 
(\ref{DD_}), (\ref{nblaie16}), (\ref{D-_ln}) to an arbitrary finite scale $M$. 

According to \cite{d_f_z,blot,oksak}, in such a (more or less) well-defined space of 
bosonic fields (\ref{K_5})--(\ref{Q_pm}) one can construct the variety of different 
inequivalent representations of the solutions of Dirac equation 
$\partial_\xi\chi_\xi(x)=0$ for the free massless trial physical field $\chi(x)$ in 
the form of local normal-ordered exponentials of right and left moving boson fields 
$\varphi^{-\xi}(x^{-\xi})$ and their charges $Q^\xi$ (\ref{K_7}), (\ref{K_9}). 
The most simple field \cite{blot}, which satisfies to extra free equations of 
\cite{d_f_z}: $\partial_s\chi_\xi(s)=
-i\pi{\cal N}_\varphi\left\{\widehat{J}_{(\chi)}^\xi(s)\chi_\xi(s)\right\}$ with  
$s=x^{-\xi}$, and leads for $Z_{(\chi)}(a)=1$ exactly to the bosonization 
relations (\ref{nweyi19}) for the corresponding currents (\ref{bos-111})--(\ref{K_Z}),  
reads\footnote{With taking into account 
the difference in $-\,\sqrt{2}$ of the field's and charge's normalization, and the losted 
coefficient 1/4 instead 1/2 before the charge in (11.94) of \cite{blot}.}:
\begin{eqnarray}
&&\!\!\!\!\!\!\!\!\!\!\!\!\!\!\!\!\!\!
\chi_\xi\left(x\right)=\chi_\xi\left(x^{-\xi}\right)=
{\cal N}_\varphi\left\{\exp\biggl(-i2\sqrt{\pi}\varphi^{-\xi}(x^{-\xi})\biggr)\right\}
\exp\left(-i\sqrt{\pi}\,\frac{\xi}{2}\,Q^\xi\right)
\left(\frac{\overline{\mu}}{2\pi}\right)^{1/2}e^{i\varpi-i\xi\Theta/4},
\label{nblaie12_0} \\
&&\!\!\!\!\!\!\!\!\!\!\!\!\!\!\!\!\!\! 
\mbox{or: }\;
\chi_\xi\left(x^{-\xi}\right)=
{\cal N}_\varphi\left\{\exp\left(-i2\sqrt{\pi}\left[\varphi^{-\xi}\left(x^{-\xi}\right)+
\frac{\xi}{4}Q^\xi\right]\right)\right\}u_\xi,
\label{nblaie12} \\
&&\!\!\!\!\!\!\!\!\!\!\!\!\!\!\!\!\!\! 
\mbox{where: }\;
u_\xi =\left(\frac{\overline{\mu}}{2\pi}\right)^{1/2}e^{i\varpi-i\xi\Theta/4}
\exp\left\{-a_0\frac \pi 8\right\}\equiv u^{Ok}_\xi
\exp\left\{-a_0\frac \pi 8\right\},
\label{u_xi}
\end{eqnarray}
-- is a two-component c-number spinor containing arbitrary, random, non-observable initial 
overall and relative phases $\varpi$ and $\Theta$ and above mentioned regularization 
parameter $a_0$ from (\ref{nblaie19}), (\ref{D_a0}), and the VEV (\ref{VEV_0}). 
The infrared regularization parameter $\overline{\mu}$ from (\ref{nblaie16}) can 
subsequently tend to zero \cite{blot} or remain to be fixed, $\overline{\mu}\mapsto M$, 
\cite{fab-iva}--\cite{fab-iva_2}, depending on the phase of the model under consideration. 
The total field $\chi(x)$ is the sum of Right and Left fields defined according to their 
chirality, as $\gamma^5\psi_{\rm R}=+\psi_{\rm R}$, $\gamma^5\psi_{\rm L}=-\psi_{\rm L}$: 
\begin{eqnarray}
&&\!\!\!\!\!\!\!\!\!\!\!\!\!\!\!\!\!\! 
\chi(x)=\chi_{\rm R}(x^-)+\chi_{\rm L}(x^+)=\left(\begin{array}{c}
\chi_{+}(x^-) \\ \chi_{-}(x^+) \end{array}\right), \quad 
\chi_{\rm R}(x^-)= \left(\begin{array}{c}
\chi_{+}(x^-) \\ 0 \end{array}\right), \quad 
\chi_{\rm L}(x^+)=\left(\begin{array}{c}
0 \\ \chi_{-}(x^+)\end{array}\right).  
\label{L_R_chi}
\end{eqnarray}
In accordance with (\ref{chi_xi_0}) of Appendix E, the field $\chi_{\rm R}(x^-)$ 
describes the right moving particles, while the $\chi_{\rm L}(x^+)$ describes the 
left moving ones \cite{blot,shif,rub}. 

\subsection{Integration of the Heisenberg equation}

For the chosen representation (\ref{nweyi19})--(\ref{nblaie16})
the operator product in the linearized by means of (\ref{ns94m61}),
(\ref{ns94m62}) HEq (\ref{45bn6l4}) or (\ref{45bn6l5}) is
naturally redefined into the normal-ordered form \cite{blot}
with respect to the fields $\varphi^\xi(x^\xi)$:
\begin{eqnarray}
\partial_0 \Psi_\xi (x)=\left(-\xi \partial_1-i\frac{\beta g}{2\sqrt{\pi}}
\widehat{J}_{(\chi)}^{-\xi(-)}(x)\right)\Psi_\xi (x)-
\Psi_\xi(x)\left(i\frac{\beta g}{2\sqrt{\pi}}\widehat{J}_{(\chi)}^{-\xi(+)}(x)\right).
\label{nweyi13}
\end{eqnarray}
The famous expression for the derivative of function $F\left(x^1\right)$ in terms of the
operator $P^1$: $-i\partial_1 F(x^1)=\left[P^1,F(x^1)\right]$, and its finite-shift
equivalent: $e^{i a P^1}F(x^1)e^{- iaP^1}=F(x^1+a)$, allows to transcribe the equation
(\ref{nweyi13}) for $x^0=t$, $\Psi_\xi(x)\longleftrightarrow Y(t)$, as follows:
\begin{eqnarray}
\frac {d}{dt}Y(t) = A(t)Y(t)-Y(t)B(t),
\label{nweyi15}
\end{eqnarray}
and to obtain then its formal solution in the form of
time-ordered exponentials:
\begin{eqnarray}
Y(t) = {T}_A \left\{\exp\left(\int\limits_0^t d\tau
A(\tau)\right)\right\}Y(0)
\left[{T}_B \left\{\exp\left(\int\limits_0^t d\tau
B(\tau)\right) \right\}\right]^{-1},
\label{nweyi16}
\end{eqnarray}
that are immediately replaced here by the usual ones, recasting the solution already
into the normal form:
\begin{eqnarray}
\Psi_\xi (x)= e^{C^{\xi(-)}(x)}\Psi_\xi\left(x^1-\xi x^0, 0\right) e^{C^{\xi(+)}(x)},
\label{nweyi21}
\end{eqnarray}
where operator bosonization (\ref{nweyi19}) of the vector current of trial field 
$\chi(x)$ (\ref{nblaie12}) gives:
\begin{eqnarray}
&&\!\!\!\!\!\!\!\!\!\!\!\!\!\!\!\!\!\!\!\!
C^{\xi(\pm)} (x)=-i\frac{\beta g}{2\sqrt{\pi}}\int\limits_0^{x^0}d y^0
\widehat{J}_{(\chi)}^{-\xi(\pm)}\left(x^1+\xi y^0 -\xi x^0, y^0\right)=
\label{nweyi22}  \\
&&\!\!\!\!\!\!\!\!\!\!\!\!\!\!\!\!\!\!\!\!
= -i\frac{\beta g}{2\pi}\left[\varphi^{(\pm)}\left(x^1,x^0\right)-
\varphi^{(\pm)}\left(x^1-\xi x^0,0\right)\right]
= -i\frac{\beta g}{2\pi}\left[\varphi^{\xi(\pm)}\left(x^\xi\right)-
\varphi^{\xi(\pm)}\left(-x^{-\xi}\right)\right].
\nonumber
\end{eqnarray}
Remarkably, that the completely unknown ``initial'' HF
$\Psi_\xi(x^1-\xi x^0,0)=\lambda_\xi(x^{-\xi})$ appears here also as a solution of
free massless Dirac equation, $\partial_\xi\lambda_\xi(x^{-\xi})= 0$,
but certainly unitarily inequivalent to the free field $\chi(x)$ (\ref{nblaie12}).
The expressions (\ref{nweyi21}), (\ref{nweyi22}) suggest to choose it also in the
normal-ordered form with respect to the field $\varphi^\xi(s)$ using appropriate
bosonic canonical transformation of this field with constant parameters
$\overline{\alpha}=2\sqrt{\pi}\cosh\eta$ and $\overline{\beta}=2\sqrt{\pi}\sinh\eta$, 
obeying $\overline{\alpha}^2-\overline{\beta}^2 = 4\pi$, which is generated by the
operator $F_\eta$ (for $y^0=x^0$) in the form $U_\eta =\exp F_\eta$:
\begin{eqnarray}
&&\!\!\!\!\!\!\!\!\!\!\!\!\!\!\!\!\!\!\!\!
U^{-1}_\eta\varphi^\xi\left(x^\xi\right)U_\eta=\omega^\xi\left(x^\xi\right)=
\frac{1}{2\sqrt{\pi}}\left[\overline{\alpha}\varphi^\xi \left(x^\xi\right)+
\overline{\beta}\varphi^{-\xi}\left(-x^\xi\right)\right], 
\label{intro_007} \\
&&\!\!\!\!\!\!\!\!\!\!\!\!\!\!\!\!\!\!\!\!
U^{-1}_\eta\varphi\left(x\right)U_\eta=\omega(x)\equiv
\omega^\xi\left(x^\xi\right)+\omega^{-\xi}\left(x^{-\xi}\right)=
\frac{1}{2\sqrt{\pi}}\Bigl[\overline{\alpha}\varphi(x^1,x^0)+
\overline{\beta}\varphi(x^1,-x^0)\Bigr], 
\label{om_U} \\
&&\!\!\!\!\!\!\!\!\!\!\!\!\!\!\!\!\!\!\!\!
U^{-1}_\eta\phi\left(x\right)U_\eta=\Omega(x)\equiv
\xi\left(\omega^{-\xi}\left(x^{-\xi}\right)-\omega^\xi\left(x^\xi\right)\right)=
\frac{1}{2\sqrt{\pi}}\Bigl[\overline{\alpha}\phi(x^1,x^0)-
\overline{\beta}\phi(x^1,-x^0)\Bigr], 
\label{oOm_U} \\
&&\!\!\!\!\!\!\!\!\!\!\!\!\!\!\!\!\!\!\!\!
U^{- 1}_\eta  Q^\xi U_\eta= {\cal W}^\xi= 
\frac{1}{2\sqrt{\pi}}\left[\overline{\alpha} Q^\xi-\overline{\beta}Q^{-\xi}\right], 
\;\mbox{ where for: }\;U_\eta =\exp F_\eta,
\label{intro_008} \\
&&\!\!\!\!\!\!\!\!\!\!\!\!\!\!\!\!\!\!\!\!
F_\eta = 2i\eta\int\limits_{-\infty}^\infty d y^1 \varphi^\xi\left(y^\xi\right)
\partial_0\varphi^{-\xi}\left(-y^\xi\right)=
2i\eta\int\limits_{-\infty}^\infty d y^1 \omega^\xi\left(y^\xi\right)
\partial_0\omega^{-\xi}\left(-y^\xi\right),
\label{intro_009}
\end{eqnarray}
which in fact does not depend on $\xi$ and $x^0$ at all, reading as: 
\begin{eqnarray}
&&\!\!\!\!\!\!\!\!\!\!\!\!\!\!\!\!\!\!\!\!
F_\eta =\frac{i\eta}2\int\limits_{-\infty}^\infty d y^1
\left\{\phi(y^1,-x^0)\partial_0\phi(y^1,x^0)-
\phi(y^1,x^0)\partial_0\phi(y^1,-x^0)\right\}=\vec{\rm F}_\eta[c]\equiv
\label{Fpp_009} \\
&&\!\!\!\!\!\!\!\!\!\!\!\!\!\!\!\!\!\!\!\!
\equiv \lim_{\eta(k^1)\to\eta}
\int\limits_{-\infty}^\infty \frac{d k^1\,\theta(k^1) }{2\pi 2k^0}
\eta(k^1)\left(c(k^1)c(-k^1)-c^\dagger(-k^1)c^\dagger(k^1)\right),\;
\mbox { and with (\ref{cc_k0}), (\ref{ph_pm}), (\ref{Q_pm}),  gives:}
\label{Fcc_009} \\
&&\!\!\!\!\!\!\!\!\!\!\!\!\!\!\!\!\!\!\!\!
\left[\varphi^{\xi(\pm)}(s),F_\eta\right]=\eta\,\varphi^{-\xi(\mp)}(-s), \quad 
\left[Q^{\xi(\pm)}(\widehat{x}^0),F_\eta\right]=-\eta\,Q^{-\xi(\mp)}(-\widehat{x}^0),\;
\mbox{ where: }\; \widehat{x}^0\Rightarrow 0, 
\label{intro_008_}
\end{eqnarray}
to have one and the same result (\ref{intro_008}) by making use of this commutator or 
relations (\ref{K_7}), (\ref{K_9}). 

In what follows, we use transformation formulas evident for any operators  
of the form $A=A^{(+)}+A^{(-)}$ and $B=B^{(+)}+B^{(-)}$, whose transformation 
$U^{-1}_\eta A U_\eta$ for some $U_\eta$ also depends linearly on the annihilation $(+)$ 
and creation $(-)$ operators (\ref{cc_k0}), so that all their mutual commutators are 
c-numbers \cite{man}:  
\begin{eqnarray}
&&\!\!\!\!\!\!\!\!\!\!\!\!\!\!\!\!\!\!\!\!
U^{-1}_\eta A U_\eta=\left(U^{-1}_\eta A U_\eta\right)^{(+)}+
\left(U^{-1}_\eta A U_\eta\right)^{(-)},\quad 
A^{(+)}|0\rangle=\left(U^{-1}_\eta A U_\eta\right)^{(+)}|0\rangle=0,\;\mbox{ and since:}
\label{UAU} \\
&&\!\!\!\!\!\!\!\!\!\!\!\!\!\!\!\!\!\!\!\!
{\cal N}_A \left\{e^A \right\}\Longrightarrow
\exp\left(A^{(-)}\right)\exp\left(A^{(+)}\right)
=e^A\exp\left(-\,\frac 12 \left[A^{(+)}, A^{(-)}\right]\right)\equiv
\frac{e^A}{\langle 0|e^A|0\rangle}, \;\mbox{ then:}
\label{AA+-} \\
&&\!\!\!\!\!\!\!\!\!\!\!\!\!\!\!\!\!\!\!\!
{\cal N}_A\left\{e^A \right\}{\cal N}_B\left\{e^B\right\}=
{\cal N}\left\{e^{A+B}\right\}
\exp\left(\left[A^{(+)}, B^{(-)}\right]\right)=
{\cal N}_B\left\{e^B \right\}{\cal N}_A\left\{e^A\right\}
\exp\left(\left[A, B\right]\right), 
\label{AB+-} \\
&&\!\!\!\!\!\!\!\!\!\!\!\!\!\!\!\!\!\!\!\!
\left[{\cal N}_A\left\{e^A \right\},{\cal N}_B\left\{e^B\right\}\right]_\mp=
\left[\exp\left(\left[A^{(+)}, B^{(-)}\right]\right)\mp
\exp\left(\left[B^{(+)},A^{(-)}\right]\right)\right]{\cal N}\left\{e^{A+B}\right\},
\label{eAeBc+-} \\
&&\!\!\!\!\!\!\!\!\!\!\!\!\!\!\!\!\!\!\!\!
B{\cal N}_A\left\{e^A \right\}=
{\cal N}\left\{\left(B+\left[B^{(+)},A^{(-)}\right]\right)e^A \right\}, \quad 
\left[B,{\cal N}_A\left\{e^A \right\}\right]=
\left[B,A\right]{\cal N}_A\left\{e^A \right\}, \;\;\mbox{ whence:}
\label{BeAc-} \\
&&\!\!\!\!\!\!\!\!\!\!\!\!\!\!\!\!\!\!\!\!
U^{-1}_\eta {\cal N}_A \left\{e^A \right\}U_\eta=
{\cal N}_A\left\{\exp\left(U^{-1}_\eta A U_\eta\right)\right\}
\exp\left(\frac{{\rm w}(\eta)}2\right), \;\mbox{ where:}
\label{UeU} \\
&&\!\!\!\!\!\!\!\!\!\!\!\!\!\!\!\!\!\!\!\!
{\rm w}(\eta)=
\left[\left(U^{-1}_\eta A U_\eta\right)^{(+)},\left(U^{-1}_\eta A U_\eta\right)^{(-)}
\right]-\left[A^{(+)}, A^{(-)}\right]=
\langle 0|\left[U^{-1}_\eta, AA\right]U_\eta|0\rangle, \;\mbox{-- is a c-number}.
\label{w_2}
\end{eqnarray}
To avoid meaningless infinities the field $\varphi^{-\xi}(s)$ is smoothed by function 
$f(s)$ \cite{wai,blot}, which at the end tends to delta-function: 
$f(s)\longrightarrow 2\sqrt{\pi}\,\delta(s-x^{-\xi})$, so that: 
\begin{eqnarray}
&&\!\!\!\!\!\!\!\!\!\!\!\!\!\!\!\!\!\!\!\!
\chi_\xi\left(x^{-\xi}\right)\longmapsto
{\cal N}_\varphi \biggl\{\exp A_\xi [f]\biggr\} u_\xi, \quad 
A_\xi [f]=-i\left[\!\int\!\!ds\,\varphi^{-\xi}(s)f(s)+\xi\frac{\sqrt{\pi}}{2}Q^\xi\right]=
-i\left[\varphi^{-\xi}[f]+\xi\frac{\sqrt{\pi}}{2}Q^\xi\right], 
\label{XphA} \\
&&\!\!\!\!\!\!\!\!\!\!\!\!\!\!\!\!\!\!\!\!
\mbox{leading to convolution: }\;\left[\varphi^{\xi(+)}[f],\varphi^{\xi(-)}[g]\right]=
\frac{1}{i}\int\!\!ds\,f(s)\!\int\!\! d\tau D^{(-)}(s-\tau)g(\tau)\equiv 
\frac{1}{i}\left(f*D^{(-)}*g\right), 
\label{fDg} \\
&&\!\!\!\!\!\!\!\!\!\!\!\!\!\!\!\!\!\!\!\!
\mbox{which appears in }\, {\rm w}(\eta), \mbox{ as: }\;
\frac{1}{i}\left(f*D^{(-)}*f\right)\longrightarrow 
\frac{4\pi}{i}D^{(-)}(0)=\int\limits^\infty_\mu\frac{d\lambda}{\lambda}
e^{-\lambda/\Lambda}=\ln\frac{\Lambda}{\overline{\mu}}, \quad 
\overline{\mu}=\mu e^{C_\ni}. 
\label{Lamu}
\end{eqnarray}
Here $C_\ni$ is Euler-Mascheroni constant and the ultraviolet cut-off $\Lambda$ in 
(\ref{Lamu}) is introduced by means of representations  (\ref{Lam_L}), (\ref{D-_ln}), 
(\ref{e_E_e}), \cite{fab-iva}--\cite{fab-iva_2}, thus appearing in the transformed 
field as: 
\begin{eqnarray}
&&\!\!\!\!\!\!\!\!\!\!\!\!\!\!\!\!\!\!\!\!
U^{-1}_\eta\chi_\xi\left(x^{-\xi}\right)U_\eta=\lambda_\xi\left(x^{-\xi}\right)=
{\cal N}_\varphi \left\{\exp\left(-i2\sqrt{\pi}\left[\omega^{-\xi}\left(x^{-\xi}\right)+
\frac{\xi}{4}{\cal W}^\xi\right]\right)\right\}v_\xi\, , 
\label{M_1} \\
&&\!\!\!\!\!\!\!\!\!\!\!\!\!\!\!\!\!\!\!\!
v_\xi =\exp\left\{-a_0\frac{\overline{\beta}^2}{16}\right\}
\left(\frac{\overline{\mu}}{\Lambda}\right)^{\overline{\beta}^2/{4\pi}} u_\xi=
\left(\frac{\overline{\mu}}{2\pi}\right)^{1/2}
\left(\frac{\overline{\mu}}{\Lambda}\right)^{\overline{\beta}^2/{4\pi}}
\exp\left\{-a_0\frac{\pi}4\left(\frac 12+\frac{\overline{\beta}^2}{4\pi}\right)\right\}
e^{i\varpi-i\xi\Theta/4}.
\label{M_2}
\end{eqnarray}
For the corresponding current $\widehat{J}_{(\lambda)}^\mu(x)$, defined by Eqs.
(\ref{bos-111})--(\ref{K_Z}), or by the Johnson definition \cite{Jon,s_w,wai}, but 
with the same renormalization constant $Z_{(\lambda)}(a)$, we find the 
previous bosonization rules (\ref{nweyi19}): 
\begin{eqnarray}
\widehat{J}_{(\lambda)}^\mu(x)=\frac{1}{\sqrt{\pi}}\,\partial^\mu\omega(x)=
-\,\frac{1}{\sqrt{\pi}}\,\epsilon^{\mu\nu}\partial_\nu\Omega(x),\;\mbox{ for: }\;
Z_{(\lambda)}(a)=(\Lambda^2|a^2|)^{-{\overline{\beta}^2}/{4\pi}},
\label{J_L}
\end{eqnarray}
onto the new (pseudo) scalar fields $\omega(x)$, $\Omega(x)$, 
$\omega^\xi\left(x^\xi\right)$, ${\cal W}^\xi$, (\ref{om_U})--(\ref{intro_008}), obeying 
the same respective commutation relations (\ref{K_8_1})--(\ref{K_9}):
\begin{eqnarray}
&&\!\!\!\!\!\!\!\!\!\!\!\!\!\!\!\!\!\!
\left[\omega(x), \omega(y)\right]=\left[\Omega(x),\Omega(y)\right]=
-\frac{i}{2}\varepsilon(x^0-y^0)\theta\left((x-y)^2\right), 
\label{om_K_8} \\
&&\!\!\!\!\!\!\!\!\!\!\!\!\!\!\!\!\!\!
\left[\omega^\xi\left(s\right),\omega^{\xi'}\left(\tau\right)\right]=
-\frac{i}{4}\varepsilon(s - \tau)\delta_{\xi, \xi'}, \quad
\left[\omega^\xi(s),{\cal W}^{\xi'}\right]=\frac{i}{2}\delta_{\xi,\xi'}.
\label{om_K_9}
\end{eqnarray}
Substituting the normal form (\ref{M_1}) into the solution (\ref{nweyi21}), we
immediately obtain the normal exponential of the DM for Thirring field in the 
form\footnote{With taking again into account the difference in $-\,\sqrt{2}$ of the 
field's and charge's normalization, and the losted coefficient 1/4 instead 1/2 before 
the charges in \cite{blot} (11.129), (11.130).} of Ref. \cite{blot,oksak}:
\begin{eqnarray}
&&\!\!\!\!\!\!\!\!\!\!\!\!\!\!\!\!\!\!\!\!
\Psi_\xi(x)={\cal N}_\varphi
\left\{\exp\left[-i\overline{\alpha}\varphi^{-\xi}\left(x^{-\xi}\right)
-i\frac{\beta g}{2\pi}\varphi^\xi\left(x^\xi\right)
-i\left(\overline{\beta}-\frac{\beta g}{2\pi}\right)\varphi^\xi\left(-x^{-\xi}\right)
-i\overline{\alpha}\frac{\xi}{4} Q^\xi+
i\overline{\beta}\frac{\xi}{4} Q^{-\xi}\right]\right\}v_\xi,
\label{nweyi_29} \\
&&\!\!\!\!\!\!\!\!\!\!\!\!\!\!\!\!\!\!\!\!
\mbox{recasted into: }\; \Psi_\xi(x)={\cal N}_\varphi
\left\{\exp\left[-i\overline{\alpha}\varphi^{-\xi}\left(x^{-\xi}\right)-
i\overline{\beta}\varphi^\xi\left(x^\xi\right)-
i\overline{\alpha}\frac{\xi}{4} Q^\xi+
i\overline{\beta}\frac{\xi}{4} Q^{-\xi}\right]\right\}v_\xi,
\label{nweyi29} \\
&&\!\!\!\!\!\!\!\!\!\!\!\!\!\!\!\!\!\!\!\!
\mbox{or: }\;\Psi_\xi(x)={\cal N}_\varphi\left\{
\exp\left(-i2\sqrt{\pi}\left[\varrho^{-\xi}(x)+
\frac{\xi}{4}{\cal W}^\xi\right]\right)\right\}v_\xi\, ,\quad \varrho^{-\xi}(x)=
\frac{1}{2\sqrt{\pi}}\left[\overline{\alpha}\varphi^{-\xi}\left(x^{-\xi}\right)+
\overline{\beta}\varphi^{\xi}\left(x^\xi\right)\right], 
\label{Psi_varho}
\end{eqnarray}
by imposing the conditions onto the parameters, that are necessary to have a correct
Lorentz -transformation properties corresponding to the spin $1/2$ and correct CAR 
(\ref{vbnmei2})--(\ref{vbnmei4}), respectively:
\begin{eqnarray}
\overline{\alpha}^2-\overline{\beta}^2= 4\pi, \quad
\overline{\beta}-\frac{\beta g}{2\pi} = 0.
\label{nweyi32}
\end{eqnarray}
Straightforward calculation of the vector current operators (\ref{bos-111})--(\ref{K_Z})
for this solution (\ref{nweyi29}), (\ref{Psi_varho}), (\ref{M_2}) by means of Eqs. 
(\ref{nweyi19}), (\ref{d_xi_phi}), (\ref{K_9})--(\ref{nblaie19}) and
(\ref{nweyi32}), under the conditions: 
\begin{eqnarray}
&&\!\!\!\!\!\!\!\!\!\!\!\!\!\!\!\!\!\!\!\!
\overline{\alpha}\equiv \left(\frac{2\pi}{\kappa}+\frac{\kappa}{2}\right),\quad
\overline{\beta}\equiv \left(\frac{2\pi}{\kappa}-\frac{\kappa}{2}\right),\;\mbox{ or: }\;
e^\eta\equiv \frac{2\sqrt{\pi}}{\kappa},
\label{Kab_1}
\end{eqnarray}
reproduces exactly the bosonization and linearization relations (\ref{ns94m61}), 
(\ref{ns94m62}), (\ref{nweyi19}) as following:
\begin{eqnarray}
&&\!\!\!\!\!\!\!\!\!\!\!\!\!\!\!\!\!\!\!\!
\widehat{J}_{(\Psi)}^\mu (x)\stackrel{\rm w}{=}
-\,\frac{\kappa}{2\pi}\,\epsilon^{\mu\nu}\partial_\nu\phi (x)=
\frac{\beta}{2\sqrt{\pi}}\,\widehat{J}_{(\chi)}^\mu (x),\; \mbox{ for: }\;\kappa=\beta, 
\label{K_Za} \\
&&\!\!\!\!\!\!\!\!\!\!\!\!\!\!\!\!\!\!\!\!
Z_{(\chi)}(a)=1,\quad 
Z_{(\Psi)}(a)=(-\Lambda^2 a^2)^{-{\overline{\beta}^2}/{4\pi}}\propto Z_{(\lambda)}(a),
\;\mbox{ whence:}
\label{K_ZaZx} \\
&&\!\!\!\!\!\!\!\!\!\!\!\!\!\!\!\!\!\!\!\!
\overline{\alpha}=\left(\frac{2\pi}{\beta}+\frac{\beta}{2}\right),\quad
\overline{\beta}=\left(\frac{2\pi}{\beta}-\frac{\beta}{2}\right),\;\mbox{ or: }\;
e^\eta=\frac{2\sqrt{\pi}}{\beta}=\sqrt{1+\frac{g}{\pi}}\,,
\label{Kab}
\end{eqnarray}
demonstrating self-consistency of the above calculations. The last equality of
Eq. (\ref{Kab}) is easily recognized as the well-known Coleman identity \cite{col}.
The weak sense of bosonization rules (\ref{K_Za}), unlike (\ref{nweyi19}), is directly
manifested by the difference of renormalization constants $Z_{(\Psi)}(a)$ and 
$Z_{(\chi)}(a)$ defined by Eq. (\ref{K_ZaZx}) for the various fields $\Psi(x)$ and 
$\chi(x)$ respectively. 

Further we also refer to the solution (\ref{nweyi29}), (\ref{Psi_varho}), (\ref{M_2}), as 
Oksak solution $\Psi^{Ok}(x)$, \cite{blot,oksak}. It is worth to note it contains 
all Klein factors also under the normal form, as is demanded for DM. 
By virtue of (\ref{eAeBc+-}) the value of $Z_{(\Psi)}(x-y)$ is fixed also as the wave 
function renormalization constant, which defines dynamical dimension 
${\rm d}_{(\Psi)}$ of the Thirring field \cite{klaib,d_f_z,bip} by 
replacing the CAR (\ref{vbnmei2}) to the following relation: 
\begin{eqnarray}
&&\!\!\!\!\!\!\!\!\!\!\!\!\!\!\!\!\!\!\!\!
\left\{\Psi_\xi(x),\Psi_{\xi'}^\dagger(y)\right\}\Bigr|_{x^0= y^0}=
Z_{(\Psi)}(x-y)\Bigr|_{x^0= y^0}\delta_{\xi,\xi'}\delta\left(x^1-y^1\right),
\label{Ps_d_Z} \\
&&\!\!\!\!\!\!\!\!\!\!\!\!\!\!\!\!\!\!\!\!
Z_{(\Psi)}(x-y)=\frac{2\pi}{\overline{\mu}}
\left(-\overline{\mu}^2(x-y)^2\right)^{-{\overline{\beta}^2}/{4\pi}}
\exp\left\{a_0\frac{\pi}2\left(\frac 12+\frac{\overline{\beta}^2}{4\pi}\right)\right\}
\left(v^*_\xi v_\xi\right), \; \mbox { or:}
\label{ZLa_Z_Z}  \\
&&\!\!\!\!\!\!\!\!\!\!\!\!\!\!\!\!\!\!\!\!
Z_{(\Psi)}(x-y)\Bigr|_{x^0=y^0}=
\left[-\Lambda^2(x-y)^2\right]^{-{\overline{\beta}^2}/{4\pi}}\Bigr|_{x^0= y^0}=
\left[\Lambda^2(x^1-y^1)^2\right]^{-{\overline{\beta}^2}/{4\pi}},\quad 
{\rm d}_{(\Psi)}=\frac 12+ \frac {\overline{\beta}^2}{4\pi}, 
\label{ZLa_Z}
\end{eqnarray}
whence: $Z_{(\Psi)}\bigr|_{x^0= y^0}\rightarrow 1$, for: $x^1-y^1\simeq 1/\Lambda$. For 
the intermediate field $\lambda_\xi(x^{-\xi})$ (\ref{M_1}), (\ref{M_2}) one finds the 
same renormalization constant 
$Z_{(\lambda)}\bigr|_{x^0= y^0}=Z_{(\Psi)}\bigr|_{x^0= y^0}$ (\ref{ZLa_Z}). 

The following comments are in order. 

1. Our way to obtain the Oksak solution for the Thirring field by means of direct 
integration of exactly linearized HEq with initial condition at $t=x^0=0$ gives it as a 
multiplicatively renormalizable operator, whose current's (\ref{K_ZaZx}) and field's 
(\ref{ZLa_Z}) renormalization 
constants depend only on ultraviolet cut-off $\Lambda$ as it should. All the infrared 
regularization parameters $\overline{\mu}$ and $a_0$ are canceled automatically. 
Moreover, the parameter\footnote{which is zero in original Oksak solution 
\cite{blot, oksak} (see Appendix C).} 
$a_0$ fully eliminates from the Oksak-Thirring field (\ref{nweyi29}), 
(\ref{M_2}), as well as from the free one (\ref{u_xi}) by means of the one and the 
same redefinition of parameter $\overline{\mu}$ as:
\begin{eqnarray} 
\overline{\mu}\longmapsto\overline{\overline{\mu}}\exp\left\{a_0\frac \pi 4\right\}. 
\label{mu_a0}
\end{eqnarray}

2. The general form of our solution (\ref{nweyi21}), (\ref{nweyi22}) is very close to 
the Klaiber operator solution \cite{klaib}, but is not the same. The main difference 
manifests in the inequivalent representation for the ``intermediate'' free Dirac field 
arisen as ``initial'' HF: $\lambda_\xi(x^{-\xi})=\Psi_\xi(x^1-\xi x^0,0)$, and induced by 
inequivalent representation of the (pseudo) scalar field. The appearance of ultraviolet 
cut-off $\Lambda$ in (\ref{M_2}) is directly related to well known non-existence of HF 
at fixed point of time \cite{blot,hep}, also leading to necessity to deal with 
inequivalent representations for describing the time evolution of non-trivial interaction 
\cite{blot,schwarz}. 
The fields and the vacuum state $|\widehat{0}\rangle$ for this inequivalent 
representation are defined by the relations (\ref{om_U})--(\ref{intro_008_}) as: 
\begin{eqnarray}
&&\!\!\!\!\!\!\!\!\!\!\!\!\!\!\!\!\!\!\!\!
|\widehat{0}\rangle=U^{-1}_\eta|0\rangle,\quad d(k^1)|\widehat{0}\rangle=0, \quad 
d(k^1)=U^{-1}_\eta c(k^1)U_\eta=
\frac{1}{2\sqrt{\pi}}\left[\overline{\alpha}c(k^1)-\overline{\beta}c^\dagger(-k^1)\right],
\label{N_vac} \\
&&\!\!\!\!\!\!\!\!\!\!\!\!\!\!\!\!\!\!\!\!
d^\dagger(k^1)=U^{-1}_\eta c^\dagger(k^1)U_\eta=
\frac{1}{2\sqrt{\pi}}\left[\overline{\alpha}c^\dagger(k^1)-\overline{\beta}c(-k^1)\right],
\quad [d(k^1),d^\dagger(q^1)]=2\pi\, 2k^0\delta(k^1-q^1),
\label{N_dw}  \\
&&\!\!\!\!\!\!\!\!\!\!\!\!\!\!\!\!\!\!
\widehat{\omega}^{\xi(+)}(s)= 
-\,\frac{\xi}{2\pi} \int\limits_{-\infty}^\infty \!\frac{d k^1}{2 k^0}
\theta \left(-\xi k^1\right)d(k^1) e^{-i k^0 s}=
\frac{1}{2\sqrt{\pi}}\left[\overline{\alpha}\varphi^{\xi(+)}(s)+
\overline{\beta}\varphi^{-\xi(-)}(-s)\right],\quad k^0=|k^1|,
\label{om_pm} \\
&&\!\!\!\!\!\!\!\!\!\!\!\!\!\!\!\!\!\!
\widehat{\omega}^{\xi(-)}(s)=\left[\widehat{\omega}^{\xi(+)}(s)\right]^\dagger,\quad 
\omega^{\xi}(s)=\widehat{\omega}^{\xi(+)}(s)+\widehat{\omega}^{\xi(-)}(s)= 
\omega^{\xi(+)}(s)+\omega^{\xi(-)}(s), 
\label{om_mp} \\
&&\!\!\!\!\!\!\!\!\!\!\!\!\!\!\!\!\!\!
\mbox {while: }\;
\omega^{\xi(+)}(s)=\frac{1}{2\sqrt{\pi}}\left[\overline{\alpha}\varphi^{\xi(+)}(s)+
\overline{\beta}\varphi^{-\xi(+)}(-s)\right], \quad 
\omega^{\xi(-)}(s)=\left[\omega^{\xi(+)}(s)\right]^\dagger,
\label{om_phi_pm} \\
&&\!\!\!\!\!\!\!\!\!\!\!\!\!\!\!\!\!\!
\widehat{\lambda}_\xi\left(x^{-\xi}\right)=
{\cal N}_\omega\left\{\exp\left(-i2\sqrt{\pi}\left[\omega^{-\xi}\left(x^{-\xi}\right)+
\frac{\xi}{4}{\cal W}^\xi\right]\right)\right\}u_\xi\Longleftarrow  
\chi_\xi\left(x^{-\xi}\right)\biggr|
\left[\mbox{for: }\varphi^\xi\mapsto \omega^\xi,Q^\xi\mapsto {\cal W}^\xi\right]. 
\label{lam_om} 
\end{eqnarray}
The latter replacement of (\ref{lam_om}) in (\ref{nweyi19})--(\ref{u_xi}) (see also 
(\ref{om_K_8})--(\ref{om_K_9})) fully determines this representation as ``almost'' 
equivalent to the previous one up to replacing (pseudo) scalar field's representation 
space (\ref{N_vac})--(\ref{om_mp}) and corresponding definition of the free fermionic 
fields (\ref{nblaie12}) $\mapsto$ (\ref{lam_om}) $\mapsto$ 
(\ref{M_1}), (\ref{M_2}), and their currents definitions (\ref{nweyi19}), 
(\ref{J_L}): the current's renormalization constant (\ref{K_Z}),   
$Z_{(\widehat{\lambda})}(a)=Z_{(\chi)}(a)=1$, due to the normal-ordering with respect to 
different spaces, but as well as the current's (\ref{J_L}), as well as the field's 
renormalization constant (\ref{ZLa_Z}) for the one and the same field $\lambda(x)$, 
it becomes representation-dependent. 
The unitarily inequivalent nature (\ref{M_2}) of free Dirac intermediate field 
$\lambda_\xi(x^{-\xi})$ manifests itself namely due to its 
normal-ordering ${\cal N}_\varphi$ in (\ref{M_1}) with respect to the initial vacuum 
$|0\rangle$ of Eq. (\ref{cc_k0}), instead of ${\cal N}_\omega$ (\ref{lam_om}) 
with respect to $|\widehat{0}\rangle$. Other details of this unitarily inequivalent 
transformation are pointed in Appendix A. 

\subsection{Schwinger's terms, other solutions, and superselection rules}

By making use of (\ref{nweyi19}), (\ref{BeAc-}), (\ref{D_eps}), (\ref{D_0_eps}), 
(\ref{K_8_1}), for the Johnson non-equal time and equal time commutators 
\cite{Jon}--\cite{col} of Heisenberg fields (\ref{nweyi29}), (\ref{Psi_varho}) and their 
currents (\ref{K_Za}), understanding in a weak sense, one has:
\begin{eqnarray}
&&\!\!\!\!\!\!\!\!\!\!\!\!\!\!\!\!\!\!
\left[\widehat{J}_{(\Psi)}^\mu (x),\Psi (y)\right]\stackrel{\rm w}{=}
-\left(\underline{a}_{(\Psi)}g^{\mu\nu}-\overline{a}_{(\Psi)}\gamma^5\epsilon^{\mu\nu}
\right)\frac{\partial}{\partial x^\nu}{\rm D}_0(x-y)\Psi(y), 
\label{J_Psi_D} \\
&&\!\!\!\!\!\!\!\!\!\!\!\!\!\!\!\!\!\!
\left[\widehat{J}^0_{(\Psi)}(x),\Psi(y)\right]\Bigr|_{x^0=y^0}\stackrel{\rm w}{=}
-\underline{a}_{(\Psi)}\Psi(y)\delta\left(x^1-y^1\right),
\label{J_0} \\
&&\!\!\!\!\!\!\!\!\!\!\!\!\!\!\!\!\!\!
\left[\widehat{J}^1_{(\Psi)} (x),\Psi(y)\right]\Bigr|_{x^0=y^0}\stackrel{\rm w}{=}
-\overline{a}_{(\Psi)}\gamma^5 \Psi(y)\delta\left(x^1-y^1\right),
\label{J_1}\\
&&\!\!\!\!\!\!\!\!\!\!\!\!\!\!\!\!\!\!
\left[\widehat{J}^0_{(\Psi)}(x),\widehat{J}^1_{(\Psi)}(y)\right]\Bigr|_{x^0=y^0}
\stackrel{\rm w}{=} -ic_{(\Psi)}\partial_{x^1}\delta\left(x^1-y^1\right),
\label{J_S} 
\end{eqnarray}
and upon the above accepted definitions (\ref{bos-111})--(\ref{K_Z}), (\ref{K_Za}), and 
relations (\ref{K_ZaZx}), (\ref{Kab}),  we obtain:
\begin{eqnarray}
&&
\underline{a}_{(\Psi)}=1,\quad \overline{a}_{(\Psi)}=\frac{\beta^2}{4\pi}, 
\quad c_{(\Psi)}=\frac{\beta^2}{4\pi^2},\; \mbox{ and find,}
\label{J_3} \\
&&
\mbox{that: }\;\underline{a}_{(\Psi)}\,\overline{a}_{(\Psi)}=\pi c_{(\Psi)},\quad 
\underline{a}_{(\Psi)}-\overline{a}_{(\Psi)}=gc_{(\Psi)},
\label{J_33}
\end{eqnarray}
in agreement with \cite{s_w}--\cite{col}. On the other hand, in accordance with
\cite{man}, \cite{ks}, \cite{sok}, the algebra of the Heisenberg operator of the
conserved fermionic (vector) charge and the Thirring field, 
by virtue of (\ref{J_0}), (\ref{J_3}), coincides with the algebra of the conserved 
fermionic (vector) charge $O_{(\chi)}/\sqrt{\pi}$ and the free trial physical field 
$\chi(x)$ defined by Eqs. (\ref{nweyi19}), (\ref{K_O}), and/or 
Eqs. (\ref{E_1})--(\ref{E_6}), as: 
\begin{eqnarray}
&&\!\!\!\!\!\!\!\!\!\!\!\!\!\!\!\!\!\!
{\rm Q}_{(\Psi)}=
\frac{ {O}_{(\Psi)} }{\sqrt{\pi}}=\int\limits_{-\infty}^\infty dx^1
\widehat{J}^0_{(\Psi)}(x^1,x^0)\stackrel{\rm w}{=}\frac{\beta}{2\pi}\,O_{(\chi)}, 
\quad {\rm Q}_{(\chi)}=
\frac{ O_{(\chi)}}{\sqrt{\pi}}\equiv \frac{O}{\sqrt{\pi}}=
\int\limits_{-\infty}^\infty dx^1 \widehat{J}^0_{(\chi)}(x^1,x^0), 
\label{O_J_0} \\
&&\!\!\!\!\!\!\!\!\!\!\!\!\!\!\!\!\!\!
\mbox{whence: }\;
\left[{\rm Q}_{(\Psi)},\Psi(y)\right]=-\Psi(y), \quad 
\left[{\rm Q}_{(\chi)},\chi(y)\right]=-\chi(y),\;\mbox{ for }\; 
\underline{a}_{(\Psi)}=\underline{a}_{(\chi)}=1. 
\label{Q_J_Pchi} 
\end{eqnarray}
The pseudoscalar (pseudovector) charges $O_{5(\Psi)}$ and $O_{5(\chi)}\equiv O_5$ of Eqs. 
(\ref{nweyi19}), (\ref{K_O}) are related analogously. 
But because of Eq. (\ref{J_1}), $\overline{a}_{(\Psi)}\neq \overline{a}_{(\chi)}=1$, 
and their algebras with corresponding fields are different:
\begin{eqnarray}
&&\!\!\!\!\!\!\!\!\!\!\!\!\!\!\!\!\!\!
{\rm Q}_{5(\Psi)}=
\frac{ {O}_{5(\Psi)} }{\sqrt{\pi}}=\int\limits_{-\infty}^\infty dx^1
\widehat{J}^1_{(\Psi)}(x^1,x^0)\stackrel{\rm w}{=}\frac{\beta}{2\pi}\,O_{5(\chi)}, 
\quad {\rm Q}_{5(\chi)}=
\frac{ O_{5(\chi)}}{\sqrt{\pi}}=
\int\limits_{-\infty}^\infty dx^1 \widehat{J}^1_{(\chi)}(x^1,x^0),
\label{O5_J_1} \\
&&\!\!\!\!\!\!\!\!\!\!\!\!\!\!\!\!\!\!
\mbox{whence: }\;
\left[{\rm Q}_{5(\Psi)},\Psi(y)\right]=-\overline{a}_{(\Psi)}\gamma^5\Psi(y), \quad 
\left[{\rm Q}_{5(\chi)},\chi(y)\right]=-\gamma^5\chi(y).
\label{Q5_J_Pchi} 
\end{eqnarray}
Note, the use of values (\ref{J_0}), (\ref{J_1}) for calculation of the commutator in 
Eq. (\ref{45bn6l4}) also violates the equations of motion (\ref{45bn6l4}), 
(\ref{45bn6l5}), as well as the above-mentioned attempt to use the commutator (\ref{J_S}) 
in equation (\ref{bnemti5}), what may be compared with \cite{ruij}. 

Thus, we come to conclusions, that Thirring model \cite{thi}--\cite{d_f_z}, 
as well as the Federbush one \cite{ks}, is exactly solvable due to intrinsic hidden exact 
linearizability of its HEq, and that operator bosonization rules make sense only 
among the free fields operators (\ref{nweyi19}) with unambiguously defined procedure 
of normal ordering. For the Heisenberg currents these rules are applicable only in a weak 
sense (\ref{K_Za}). 

The natural manifestation of inequivalent representations (\ref{nblaie12}), (\ref{M_1}) 
and (\ref{nweyi29}) of free and Heisenberg  fermionic field and their currents 
are also the various values of Schwinger's terms (\ref{J_S}), (\ref{J_3}) 
and dynamical dimension (\ref{ZLa_Z}): 
\begin{eqnarray}
c_{(\chi)}=c_{(\lambda)}=\frac{1}{\pi}, \quad c_{(\Psi)}=\frac{1}{\pi+g},\quad 
{\rm d}_{(\chi)}=\frac 12, \quad 
{\rm d}_{(\Psi)}-\frac 12=\frac {\overline{\beta}^2}{4\pi}=
\frac {g^2}{4\pi}c_{(\Psi)}=\frac {g^2}{4\pi}\,\frac 1{(\pi+g)},
\label{Sc_t}
\end{eqnarray}
in agreement with \cite{klaib,d_f_z,ot,bip}. Similarly to the solution \cite{ks} of 
Federbush model, the
linear homogeneous HEq (\ref{nweyi13}) does not define the normalization (\ref{ZLa_Z}) of 
HF (\ref{nweyi29}), (\ref{Kab}), which, as well as for the free fields $\chi(x)$, 
$\lambda(x)$ is fixed \cite{fab-iva-09} only by the anticommutation relations 
(\ref{vbnmei2}) $\mapsto$ (\ref{Ps_d_Z}). 

We would like to emphasize again, that unlike \cite{mtu}--\cite{green} the bosonization 
procedure of Refs. \cite{col}--\cite{blot} is considered here as a particular case of 
dynamical mapping onto the Schr\"odinger physical field \cite{vklt}--\cite{ks} 
defined at $t=0$. From this view point the results of Refs. \cite{raja} and 
\cite{fab-iva} may be considered as 
DM of Thirring field onto the free massive scalar field $\phi_m(x)$ and onto the free 
massive Dirac field $\psi_M(x)$ respectively. The general form of solution 
(\ref{nweyi21}) should give a possibility to describe all phases of the theory under 
consideration. We show such example for finite temperature in the next section.

Now we wish to connect the Oksak solution with another known solutions of 
Thirring model \cite{man,mps_2}. To this end we use the formally unitary transformation  
of conformal shift for scalar fields from Ref. \cite{oksak}, generalized by the following 
way. By making use of the relations (\ref{nblaie16}), (\ref{nblaie19}), (\ref{UeU}), 
(\ref{w_2}) and keeping in mind the Eqs. (\ref{Kab}), we consider the family of solutions 
marked by arbitrary real parameter $\sigma$: 
\begin{eqnarray}
&&\!\!\!\!\!\!\!\!\!\!\!\!\!\!\!\!\!\!\!\!
{\rm K}_\sigma=\exp{\rm X}_\sigma,\quad 
{\rm X}_\sigma=i\sigma\frac{\overline{\xi}}4
\left(Q^{-\overline{\xi}}Q^{-\overline{\xi}}-Q^{\overline{\xi}}Q^{\overline{\xi}}\right)
=i\frac \sigma 4 OO_5, \quad \overline{\xi},\xi=\pm 
\label{K_X_s} \\
&&\!\!\!\!\!\!\!\!\!\!\!\!\!\!\!\!\!\!\!\!
\Psi(x,{\sigma})={\rm K}^{-1}_\sigma\Psi^{Ok}(x){\rm K}_\sigma,\quad 
\Psi_\xi(x,{\sigma})={\rm K}^{-1}_\sigma\Psi^{Ok}_\xi(x){\rm K}_\sigma=
{\cal N}_\varphi\left\{e^{R_\xi(x,\sigma)}\right\} v_{\xi}(\sigma),
\label{K_Psi_s} \\
&&\!\!\!\!\!\!\!\!\!\!\!\!\!\!\!\!\!\!\!\!
R_\xi(x,\sigma)=-i\left[\overline{\alpha}\varphi^{-\xi}\left(x^{-\xi}\right)
+\overline{\beta} \varphi^{\xi} \left(x^{\xi}\right)
+\frac{\xi}{4}(\overline{\alpha}+\sigma\overline{\beta}) Q^{\xi}
-\frac{\xi}{4}(\overline{\beta}+\sigma\overline{\alpha})Q^{-\xi}\right],
\label{R_Psi_s} \\
&&\!\!\!\!\!\!\!\!\!\!\!\!\!\!\!\!\!\!\!\!
v_\xi(\sigma)=
\left(\frac{\overline{\mu}}{2\pi}\right)^{1/2}
\left(\frac{\overline{\mu}}{\Lambda}\right)^{\overline{\beta}^2/{4\pi}}
\exp\left\{-a_0\frac{\pi}{8}\left[(\sigma^2+1)\cosh 2\eta+
2\sigma\sinh 2\eta\right]\right\}e^{i\varpi-i\xi\Theta/4}.
\label{v_xi_s} 
\end{eqnarray}
This solution obeys the same CAR (\ref{vbnmei3}), (\ref{vbnmei4}), (\ref{Ps_d_Z}), and 
the bosonization rule (\ref{K_Za}) with the same renormalization constant 
$Z_{(\Psi)}(a)$ (\ref{K_ZaZx}), (\ref{ZLa_Z}), for arbitrary $\sigma$, and the parameter 
$a_0$ may be again adsorbed to the parameter $\overline{\mu}$ by the rescaling 
substitution, which unlike the Oksak and free cases, now depends on Thirring coupling 
constant: 
\begin{eqnarray}
\overline{\mu}\longmapsto\overline{\overline{\mu}}\,
\exp\left\{a_0\frac \pi 4\left(\sigma^2+1+2\sigma\tanh 2\eta\right)\right\},  
\label{mu_a0eta}
\end{eqnarray}
By using Eqs. (\ref{K_5}), (\ref{K_O}), it is a simple matter to check that 
$\sigma=\pm 1$ gives the two types of Mandelstam solution \cite{man}, while  
$\sigma=-\coth 2\eta$ corresponds to normal form for solution of Morchio et al. 
\cite{mps_2}. This again demonstrates the advantages of normal ordered form of HF 
demanded by DM:
\begin{eqnarray}
&&\!\!\!\!\!\!\!\!\!\!\!\!\!\!\!\!\!\!\!\!
\Psi_\xi(x,1)=
{\cal N}_\varphi\left\{e^{R_\xi(x,1)}\right\} 
\left(\frac{\overline{\mu}}{2\pi}\right)^{1/2}
\left(\frac{\overline{\mu}}{\Lambda}\right)^{\overline{\beta}^2/{4\pi}}
\exp\left\{-a_0\frac{\pi^2}{\beta^2}\right\}e^{i\varpi-i\xi\Theta/4},
\label{Man_+1} \\
&&\!\!\!\!\!\!\!\!\!\!\!\!\!\!\!\!\!\!\!\!
R_\xi(x,1)=-i\left[\xi\frac\beta 2 \phi(x^1,x^0)-\frac{2\pi}\beta
\int\limits^{x^1}_{-\infty}dy^1\,\partial_0\phi(y^1,x^0)\right], \quad \sigma=1;
\label{Man_+1R} \\
&&\!\!\!\!\!\!\!\!\!\!\!\!\!\!\!\!\!\!\!\!
\Psi_\xi(x,-1)=
{\cal N}_\varphi\left\{e^{R_\xi(x,-1)}\right\} 
\left(\frac{\overline{\mu}}{2\pi}\right)^{1/2}
\left(\frac{\overline{\mu}}{\Lambda}\right)^{\overline{\beta}^2/{4\pi}}
\exp\left\{-a_0\frac{\beta^2}{16}\right\}e^{i\varpi-i\xi\Theta/4}, 
\label{Man_-1} \\
&&\!\!\!\!\!\!\!\!\!\!\!\!\!\!\!\!\!\!\!\!
R_\xi(x,-1)=
-i\left[\frac{2\pi}\beta\varphi(x^1,x^0)+
\xi\frac\beta 2\int\limits_{x^1}^{\infty}dy^1\,\partial_0\varphi(y^1,x^0)\right], 
\quad \sigma=-1;
\label{Man_-1R} \\
&&\!\!\!\!\!\!\!\!\!\!\!\!\!\!\!\!\!\!\!\!
\Psi_\xi(x,-\coth 2\eta)={\cal N}_\varphi\left\{e^{R_\xi(x,-\coth 2\eta)}\right\}
\left(\frac{\overline{\mu}}{2\pi}\right)^{1/2}
\left(\frac{\overline{\mu}}{\Lambda}\right)^{\overline{\beta}^2/{4\pi}}
\exp\left\{-a_0\frac\pi 8\frac{\cosh 2\eta}{\sinh^2 2\eta}\right\}
e^{i\varpi-i\xi\Theta/4}, 
\label{MPS} \\
&&\!\!\!\!\!\!\!\!\!\!\!\!\!\!\!\!\!\!\!\!
R_\xi(x,-\coth 2\eta)=-i\left[\overline{\alpha}\varphi^{-\xi}\left(x^{-\xi}\right)
+\overline{\beta}\varphi^{\xi}\left(x^{\xi}\right)
+\xi \frac{\pi}{2}\left(\frac{ Q^{\xi}}{\overline{\alpha}}+
\frac{Q^{-\xi}}{\overline{\beta}}\right)\right], \quad \sigma=-\coth 2\eta.
\label{MPS_R} 
\end{eqnarray} 
We would like to point out that $\sigma=1$ corresponds to DM (\ref{Man_+1}), (\ref{Man_+1R}) 
onto the pseudoscalar field $\phi(x)$, while $\sigma=-1$ gives another form of solution 
(\ref{Man_-1}), (\ref{Man_-1R}) with bosonization onto the scalar field $\varphi(x)$, and 
that, unlike (\ref{MPS}), (\ref{MPS_R}), the original solution of Morchio et al. 
\cite{mps_2} has $a_0=0$, and contains all Klein factors outside the normal form, thus 
its renormalization constant remains to be unknown. Here and below we use for brevity the 
mixed notations from identities (\ref{nweyi32}), (\ref{Kab}), and the following 
relations: 
\begin{eqnarray}
&&\!\!\!\!\!\!\!\!\!\!\!\!\!\!\!\!\!\!\!\!
\overline{\alpha}+\overline{\beta}=\frac{4\pi}{\beta}=2\sqrt{\pi}e^\eta,\quad 
\overline{\alpha}-\overline{\beta}=\beta=2\sqrt{\pi}e^{-\eta}, \quad 
\overline{\alpha}^2-\overline{\beta}^2=4\pi, 
\nonumber \\
&&\!\!\!\!\!\!\!\!\!\!\!\!\!\!\!\!\!\!\!\!
\frac {\overline{\alpha}\overline{\beta}}{4\pi}=
\frac 14 \left(\frac{4\pi}{\beta^2}-\frac{\beta^2}{4\pi}\right)=\frac{\sinh 2\eta}2, 
\quad\;
\frac 12+\frac{\overline{\beta}^2}{4\pi}=
\frac{ \overline{\alpha}^2+\overline{\beta}^2}{8\pi}=
\frac 14 \left(\frac{4\pi}{\beta^2}+\frac{\beta^2}{4\pi}\right)=
\frac{\cosh 2\eta}2={\rm d}_{(\Psi)}. 
\label{ID_ID} 
\end{eqnarray} 
Now let us turn to the VEV \cite{wai} of the strings of 
these fields (\ref{K_Psi_s}). Following \cite{blot} it is enough and convenient to 
consider only the product: 
\begin{eqnarray}
&&\!\!\!\!\!\!\!\!\!\!\!\!\!\!\!\!\!\!\!\!
\left\langle 0\left|\prod\limits_{i=1}^p \Psi_{\xi_i}^{(l_i)}(x_i,\sigma)
\right|0 \right\rangle =\mbox{? },\;\;\mbox{ where: }\; l=+1,\;\mbox{ for }\; \Psi, \;\; 
\mbox{ and }\;l=-1, \;\mbox{ for }\; \Psi^\dagger.
\label{Psi_ppp} 
\end{eqnarray}
By virtue of generalized formula (\ref{AB+-}), which for any 
${\rm R}^{(l_i)}_{\xi_i}(x_i,\sigma)\equiv l_iR_{\xi_i}(x_i,\sigma)\longmapsto 
{\cal R}_i={\cal R}_i^{(+)}+{\cal R}_i^{(-)}$, may be easy checked by induction: 
\begin{eqnarray}
&&\!\!\!\!\!\!\!\!\!\!\!\!\!\!\!\!\!\!\!\!
{\cal N}\Bigl\{\exp\left[{{\cal R}_1}\right]\Bigr\}\cdots
{\cal N}\Bigl\{\exp\left[{{\cal R}_p}\right]\Bigr\}\equiv 
\prod\limits_{i=1}^p {\cal N}\Bigl\{\exp\left[{{\cal R}_i}\right]\Bigr\}= 
\exp\left\{\sum\limits_{i<k}^p\left[{\cal R}_i^{(+)},{\cal R}_k^{(-)}\right]\right\}
{\cal N} \left\{\exp\left(\sum\limits_{j=1}^p {\cal R}_j\right)\right\},
\label{AAA_BBB} 
\end{eqnarray}
and obvious relations for any numbers $\gamma_i$, $i=1\div p$:
\begin{eqnarray}
&&\!\!\!\!\!\!\!\!\!\!\!\!\!\!\!\!\!\!\!\!
\sum\limits_{i<k}^p\gamma_i\gamma_k=\frac{1}{2}\sum\limits_{i\neq k}^p\gamma_i\gamma_k
=\frac{1}{2}\left(\sum\limits_{i=1}^p \gamma_i\right)^2
- \frac{1}{2}\sum\limits_{i=1}^p \gamma_i^2,
\label{LL_LL} 
\end{eqnarray}
with the help of formulas (\ref{x_xi_AB})--(\ref{A_D_-}) from Appendix B, 
one can obtain for the solutions (\ref{K_Psi_s}): 
\begin{eqnarray}
&&\!\!\!\!\!\!\!\!\!\!\!\!\!\!\!\!\!\!\!\!
\left\langle 0\left| \prod\limits_{i=1}^p \Psi_{\xi_i}^{(l_i)}(x_i,\sigma)
\right|0\right\rangle= 
\left\langle 0\left|{\cal N}_\varphi\left\{\exp\left(\sum\limits_{j=1}^p
{\cal R}_j\right)\right\} \right|0 \right\rangle 
\exp\left\{i\varpi\sum\limits_{i=1}^p l_i-i\frac{\Theta}{4}\sum\limits_{i=1}^p l_i \xi_i
\right\}
\nonumber \\
&&\!\!\!\!\!\!\!\!\!\!\!\!\!\!\!\!\!\!\!\!
\cdot 
\left(\frac{\overline{\mu}}{2\pi}\right)^{\frac{p}{2}}
\left(\frac{\overline{\mu}}{\Lambda}\right)^{\frac{p}{4\pi} \overline{\beta}^2}
\overline{\mu}^{\frac{1}{2\cdot8\pi}\left(\overline{\alpha}^2+
\overline{\beta}^2\right)\left[\left(\sum\limits_{i=1}^p l_i\right)^2+ 
\left(\sum\limits_{i=1}^p l_i\xi_i\right)^2-2p\right]}
\overline{\mu}^{\frac{2}{2\cdot8\pi}
\overline{\alpha}\overline{\beta}\left[\left(\sum\limits_{i=1}^p l_i\right)^2-
\left(\sum\limits_{i=1}^p l_i\xi_i\right)^2\right]}
\nonumber \\
&&\!\!\!\!\!\!\!\!\!\!\!\!\!\!\!\!\!\!\!\!
\cdot e^{-pa_0\frac{\pi}{8} \left[(\sigma^2+1)\cosh 2\eta+2\sigma\sinh 2 \eta\right]}
e^{-\frac{1}{4^3}(1+\sigma)^2(\overline{\alpha}+\overline{\beta})^2 a_0
\left[\left(\sum\limits_{i=1}^p l_i\right)^2-p\right]-\frac{1}{4^3}(1-\sigma)^2
(\overline{\alpha}-\overline{\beta})^2 a_0\left[\left(
\sum\limits_{i=1}^p l_i\xi_i\right)^2-p\right]}
\nonumber \\
&&\!\!\!\!\!\!\!\!\!\!\!\!\!\!\!\!\!\!\!\!
\cdot \prod\limits_{i<k}^p \biggl[e^{i 2\pi^2 (\xi_i-\xi_k)}
\left(i\{x_i^{- \xi_i}-x_k^{-\xi_i}\}+0\right)^{
[(\overline{\alpha}^2+\overline{\alpha}\overline{\beta})+
(\overline{\alpha}^2-\overline{\alpha}\overline{\beta})\xi_i\xi_k]} 
\left(i\{x_i^{\xi_i}-x_k^{\xi_i}\}+0\right)^{
[(\overline{\beta}^2+\overline{\alpha}\overline{\beta})+
(\overline{\beta}^2-\overline{\alpha}\overline{\beta})\xi_i\xi_k]
}\biggl]^{\frac{1}{8\pi} l_i l_k},
\nonumber 
\end{eqnarray}
\newpage \noindent 
or, after some simplifications\footnote{Here we put $\xi_i,\xi_k=\pm 1$, for 
$\xi_i+\xi_k$, and $\xi_i-\xi_k$, and use: $\delta_{\xi_i,\xi_k}=(1+\xi_i\xi_k)/2$, 
$\delta_{\xi_i,1}=(1+\xi_i)/2$, $\delta_{\xi_i,-1}=(1-\xi_i)/2$.}: 
\begin{eqnarray}
&&\!\!\!\!\!\!\!\!\!\!\!\!\!\!\!\!\!\!\!\!
\left(\Lambda^{\overline{\beta}^2/4\pi}\sqrt{2\pi}\right)^p
\left\langle 0\left|\prod\limits_{i=1}^p\Psi_{\xi_i}^{(l_i)}(x_i,\sigma)
\right|0\right\rangle= 
\left\langle 0\left|{\cal N}_\varphi\left\{\exp\left(\sum\limits_{j=1}^p
{\cal R}_j\right)\right\}\right|0\right\rangle 
\exp\left\{i\varpi\sum\limits_{i=1}^p l_i-i\frac{\Theta}{4}\sum\limits_{i=1}^p l_i \xi_i
\right\}
\nonumber \\
&&\!\!\!\!\!\!\!\!\!\!\!\!\!\!\!\!\!\!\!\!
\cdot 
\left(\overline{\mu}\,\exp\left\{-a_0\frac{\pi}{4}(1+\sigma)^2\right\}\right)^
{\left(\sum\limits_{i=1}^p l_i\right)^2(\pi/\beta^2)}
\left(\overline{\mu}\,\exp\left\{-a_0\frac{\pi}{4}(1-\sigma)^2\right\}\right)^
{\left(\sum\limits_{i=1}^p l_i\xi_i\right)^2(\beta^2/16\pi)}
\nonumber \\
&&\!\!\!\!\!\!\!\!\!\!\!\!\!\!\!\!\!\!\!\!
\cdot\prod\limits_{i<k}^p \left\{e^{i \pi(\xi_i-\xi_k)}
\left[\frac{ i(x_i^{-}-x_k^{-})+0}{i(x_i^{+}-x_k^{+})+0}\right]^{\xi_i+\xi_k}
\left[i0\, \varepsilon (x_i^{0}-x_k^{0})-(x_i-x_k)^2 \right]^
{(4\pi/\beta^2)+\xi_i\xi_k(\beta^2/4\pi)}\right\}^{l_i l_k/4}.
\label{Ps_p_3}
\end{eqnarray}
The first line and second line  
of this expression provide at least five independent sources of the superselection rules 
\cite{wai,blot}, which usually are associated only with conservation of scalar field's 
(vector current's) charge $O$ (\ref{O_J_0}), and pseudoscalar field's (pseudovector 
current's) charge $O_5$ (\ref{O5_J_1}), respectively:
\begin{eqnarray}
&&\!\!\!\!\!\!\!\!\!\!\!\!\!\!\!\!\!\!\!\!
\sum\limits_{i=1}^p l_i=0, 
\label{ssr_1} \\
&&\!\!\!\!\!\!\!\!\!\!\!\!\!\!\!\!\!\!\!\!
\sum\limits_{i=1}^p l_i \xi_i=0. 
\label{ssr_2}
\end{eqnarray}   
The first one defined by Oksak and Morchio et al., due to above mentioned charge 
conservation, originates from the VEV of normal exponential in r.h.s. of the first
line, taken instead $|0\rangle$ for the vacuum state $|\widehat{\upsilon}\rangle$ 
averaged with respect to 
the field-translation gauge group (\ref{gauge}), leading to \cite{mps_2,blot,oksak}: 
\begin{eqnarray}
&&\!\!\!\!\!\!\!\!\!\!\!\!\!\!\!\!\!\!\!\!
\left\langle \widehat{\upsilon}\left|{\cal N}_\varphi\left\{\exp\left(\sum\limits_{j=1}^p
{\cal R}_j\right)\right\}\right|\widehat{\upsilon}\right\rangle\Longrightarrow 
\delta_{\sum\limits_{i=1}^p l_i,0}\,\delta_{\sum\limits_{i=1}^p l_i\xi_i,0}.
\label{ssr_3} 
\end{eqnarray}
The VEV of this normal form for the usual non-degenerate vacuum state $|0\rangle$ is equal 
to 1 identically \cite{nak,fab-iva,fab-iva-09}. Nevertheless, these rules arise from the 
second line 
at the limit $\overline{\mu}\to 0$ as the natural conditions of nonzero result 
\cite{fab-iva}. 
We can suggest now three additional sources of these rules: the third one is the 
$\varpi$ - and $\Theta$ - independence condition for the VEV (\ref{Psi_ppp}), 
(\ref{Ps_p_3}),   
the fourth one follows from its above mentioned independence on the parameter $a_0$, 
and the fifth one follows from its independence on $\sigma$, if the transformation 
(\ref{K_X_s}) leaves the vacuum invariant. Obviously independence on the $a_0$ 
automatically means the independence on $\sigma$ and vice versa. 

The independence on the initial random values of overall and relative phases has 
purely fermionic nature and does not reduce to the (pseudo) scalar field-translation 
gauge symmetry (\ref{gauge}), which can only shift their arbitrary initial values. 
The fate of $a_0$ - 
independence of the VEV is more delicate, because for Oksak solution, $\sigma=0$, it 
may be again eliminated from the second line of (\ref{Ps_p_3}) by redefinition 
(\ref{mu_a0}) of the $\overline{\mu}$. But that is not the case for arbitrary $\sigma$. 
It is worth to note the superselection rules (\ref{ssr_1}), (\ref{ssr_2}) leave necessary 
for VEV (\ref{Psi_ppp}) only the ultraviolet renormalization in the l.h.s of the first 
line of Eq. (\ref{Ps_p_3}). Whence, only the third line of (\ref{Ps_p_3}) survives 
\cite{blot}, which is always independent of parameters $\overline{\mu}$, $\sigma$, 
$a_0$, $\varpi$, $\Theta $: 
\begin{eqnarray}
&&\!\!\!\!\!\!\!\!\!\!\!\!\!\!\!\!\!\!\!\!
\left(\!\Lambda^{\overline{\beta}^2/4\pi}\!\right)^p\!\!
\left\langle\! 0\left|\prod\limits_{i=1}^p\Psi_{\xi_i}^{(l_i)}(x_i,\sigma)
\right|0\!\right\rangle=
\left\langle\!0\left|\prod\limits_{i=1}^p\chi_{\xi_i}^{(l_i)}\left(x_i\right)
\right|0\!\right\rangle \!
\prod\limits_{i<k}^p \!
\Biggl[\! i0\,\varepsilon (x_i^{0}-x_k^{0})-(x_i-x_k)^2 
\Biggr
]^
{ \displaystyle \frac g4\!\left[ \frac 1{\pi}- \frac {\xi_i\xi_k}{\pi+g}\right]
\!l_i l_k}\!\!, \;\,
\label{Ps_p_5} \\
&&\!\!\!\!\!\!\!
\mbox{where: }\;\left(2\pi\right)^{p/2}
\left\langle 0\left|\prod\limits_{i=1}^p\chi_{\xi_i}^{(l_i)}\left(x_i\right)
\right|0\right\rangle \Longrightarrow 
\delta_{\sum\limits_{i=1}^p l_i,0}\,\delta_{\sum\limits_{i=1}^p l_i\xi_i,0}
\nonumber  \\
&&\!\!\!\!\!\!\!
\cdot \prod\limits_{i<k}^p \left\{e^{i\pi(\xi_i-\xi_k)}
\left[i(x_i^{-}-x_k^{-})+0\right]^{(1+\xi_i)(1+\xi_k)}
\left[i(x_i^{+}-x_k^{+})+0\right]^{(1-\xi_i)(1-\xi_k)}\right\}^{l_i l_k/4}.
\label{chi_p_6} 
\end{eqnarray}
This gives the well known expressions for two-point functions \cite{klaib,bip} with the 
dynamical dimension (\ref{Sc_t}).

We see that, unlike Schwinger model \cite{blot,shif,rub}, as long as we deal with 
solutions of ``phase decoupled'' HEqs (\ref{45bn6l5}) or (\ref{nweyi13}), that preserve 
the $\varpi$ 
and $\Theta$ arbitrariness, both the superselection rules (\ref{ssr_1}), (\ref{ssr_2}) 
with the conservation of both {\bf currents} should be fulfilled independently of chosen 
phase of the theory, including the phase with spontaneously breaking of chiral symmetry 
\cite{fujita,fujita_2}. 
From this view point the breaking of the rule (\ref{ssr_2}) may be achieved 
formally only or by introducing the mass term into HEq (\ref{45bn6l4}) ``by hand'' 
\cite{fab-iva}, or otherwise, by excluding $a_0$ via taking the Mandelstam's solution 
with $\sigma=1$ supplemented with fixing of the values $\overline{\mu}\mapsto M$ and 
$\Theta$ \cite{fab-iva,fab-iva-07}. 
However, as we will see below, the latter way is impossible for the finite temperature 
case. Moreover, if one of the gauge symmetries remains unbroken: 
$O|0\rangle\Rightarrow 0$, being connected by transformation (\ref{K_X_s}) all 
the above solutions refer to the same vacuum state $|0\rangle$. 

Keeping in mind the correspondence (\ref{Lamu}), for $\varepsilon^0\to 0$, 
$\varepsilon^1\to 1/\Lambda$, the renormalized operator of scalar ``condensate'' read: 
\begin{eqnarray}
&&\!\!\!\!\!\!\!\!\!\!\!\!\!\!\!\!\!\!\!\!
\overline{\Psi}(x+\varepsilon)\Psi (x)=\frac{\overline{\mu}}{\pi}
\left(\frac{\overline{\mu}^2}{\Lambda^2}\right)^{\overline{\beta}^2/4\pi}
\left(-\overline{\mu}^2\varepsilon^2+i0\varepsilon^0\right)^
{-\overline{\alpha}\overline{\beta}/4\pi}
\exp\left\{-a_0\frac{\pi}{4}(1-\sigma)^2\frac{\beta^2}{4\pi}\right\}
\nonumber \\ 
&&\!\!\!\!\!\!\!\!\!\!\!\!\!\!\!\!\!\!\!\!
\cdot{\cal N}_\phi\left\{\sin\left(\beta\left[\phi(x)+(1-\sigma)\frac{O}{4}\right]+
\frac{\Theta}2\right)\right\}, \quad \left(\overline{\Psi}(x)\Psi (x)\right)_{ren}=
\lim_{\varepsilon\to 0}\left(-\Lambda^2\varepsilon^2\right)^
{\overline{\alpha}\overline{\beta}/4\pi}\overline{\Psi}(x+\varepsilon)\Psi (x),
\label{Psi_Psi_ep} \\ 
&&\!\!\!\!\!\!\!\!\!\!\!\!\!\!\!\!\!\!\!\!
\left(\overline{\Psi}(x)\Psi (x)\right)_{ren}
\Longrightarrow \frac{\Lambda}{\pi}
\left(\frac{\overline{\mu}}{\Lambda}\,\exp\left\{-a_0\frac{\pi}{4}(1-\sigma)^2\right\}
\right)^{\beta^2/4\pi}
{\cal N}_\phi\left\{\sin\left(\beta\left[\phi(x)+(1-\sigma)\frac{O}{4}\right]+
\frac{\Theta}2\right)\right\},  
\label{Psi_Psi} \\ 
&&\!\!\!\!\!\!\!\!\!\!\!\!\!\!\!\!\!\!\!\!
\mbox{where: }\;\sum\limits_{i=1}^2 l_i =0,\quad 
\left(\sum\limits_{i=1}^2 l_i \xi_i\right)^2=4,  
\quad \left(-\overline{\mu}^2\varepsilon^2+i0\varepsilon^0\right)^
{-\overline{\alpha}\overline{\beta}/4\pi}
\Longrightarrow \left(\overline{\mu}\varepsilon^1\right)^
{-\overline{\alpha}\overline{\beta}/2\pi}\longmapsto 
\left(\frac{\overline{\mu}}{\Lambda}\right)^{-\overline{\alpha}\overline{\beta}/2\pi}. 
\label{l_2_mu_L}
\end{eqnarray}
This operator simplifies for Mandelstam case $\sigma=1$, and by virtue of 
(\ref{l_2_mu_L}) its VEV for the vacuum state (\ref{ssr_3}), of course is zero 
\cite{mps_2,fujita_2}, contrary to \cite{fab-iva-09}. For the non degenerate vacuum state 
$|0\rangle$ it reads: 
\begin{eqnarray}
&&\!\!\!\!\!\!\!\!\!\!\!\!\!\!\!\!\!\!\!\!
\left\langle 0\left|\left(\overline{\Psi}(x)\Psi (x)\right)_{ren}\right|0\right\rangle
\Longrightarrow \frac{\Lambda}{\pi}
\left(\frac{\overline{\mu}}{\Lambda}\,\exp\left\{-a_0\frac{\pi}{4}(1-\sigma)^2\right\}
\right)^{\beta^2/4\pi}\sin \frac{\Theta}2.  
\label{Psi_Psi_0}
\end{eqnarray}
Analogously for pseudoscalar case the same relations (\ref{l_2_mu_L}) take place, leading 
to:
\begin{eqnarray}
&&\!\!\!\!\!\!\!\!\!\!\!\!\!\!\!\!\!\!\!\!
\overline{\Psi}(x+\varepsilon)\gamma^5\Psi (x)=i\, \frac{\overline{\mu}}{\pi}
\left(\frac{\overline{\mu}^2}{\Lambda^2}\right)^{\overline{\beta}^2/4\pi}
\left(-\overline{\mu}^2\varepsilon^2+i0\varepsilon^0\right)^
{-\overline{\alpha}\overline{\beta}/4\pi}
\exp\left\{-a_0\frac{\pi}{4}(1-\sigma)^2\frac{\beta^2}{4\pi}\right\}
\nonumber \\
&&\!\!\!\!\!\!\!\!\!\!\!\!\!\!\!\!\!\!\!\!
\cdot{\cal N}_\phi\left\{\cos\left(\beta\left[\phi(x)+(1-\sigma)\frac{O}{4}\right]+
\frac{\Theta}2\right)\right\},\quad 
\left(\overline{\Psi}(x)\gamma^5\Psi (x)\right)_{ren}\!\!=
\lim_{\varepsilon\to 0}\left(-\Lambda^2\varepsilon^2\right)^
{\overline{\alpha}\overline{\beta}/4\pi}\overline{\Psi}(x+\varepsilon)\gamma^5\Psi (x),
\quad 
\label{Psi_5_Psi_ep} \\ 
&&\!\!\!\!\!\!\!\!\!\!\!\!\!\!\!\!\!\!\!\!
\left(\overline{\Psi}(x)\gamma^5\Psi (x)\right)_{ren}
\Longrightarrow i\, \frac{\Lambda}{\pi}
\left(\frac{\overline{\mu}}{\Lambda}\,\exp\left\{-a_0\frac{\pi}{4}(1-\sigma)^2\right\}
\right)^{\beta^2/4\pi}
{\cal N}_\phi\left\{\cos\left(\beta\left[\phi(x)+(1-\sigma)\frac{O}{4}\right]+
\frac{\Theta}2\right)\right\},  
\label{Psi_5_Psi} \\
&&\!\!\!\!\!\!\!\!\!\!\!\!\!\!\!\!\!\!\!\!
\left\langle 0\left|\left(\overline{\Psi}(x)\gamma^5\Psi (x)\right)_{ren}
\right|0\right\rangle
\Longrightarrow i\,\frac{\Lambda}{\pi}
\left(\frac{\overline{\mu}}{\Lambda}\,\exp\left\{-a_0\frac{\pi}{4}(1-\sigma)^2\right\}
\right)^{\beta^2/4\pi}\cos\frac{\Theta}2.  
\label{Psi_5_Psi_0}
\end{eqnarray}
Besides $\overline{\mu}$, $\Lambda$, and $\Theta$, for $\sigma\neq 1$, 
these matrix elements (\ref{Psi_Psi_0}), (\ref{Psi_5_Psi_0}) depend on additional 
dimensionless non-physical volume cut-off regularization parameter $a_0$, and their final 
value depend on the order of limits $\overline{\mu}\to 0$ and $\Lambda\to\infty$ and on 
the sign of $\eta$ (\ref{Kab}), that is the sign of $g$. For the free case: $g=0=\eta$, 
$\beta^2=4\pi$, the $\Lambda$ - dependence disappears as it should be for the free field 
(\ref{nblaie12}), (\ref{u_xi}). But for both the HF and free fields the ratio of these 
condensates (\ref{Psi_Psi_0}), (\ref{Psi_5_Psi_0}) remains still equal to 
$(-i)\tan(\Theta/2)$. 

Why the discussed above breaking of the rule (\ref{ssr_2}), remaining meaningless  
for zero temperature case will be impossible all the more at finite temperature? 
The reason is the existence of another important formally unitary 
transformation of the solutions (\ref{K_Psi_s}), which introduce the two-parametric 
extension of Oksak solution (\ref{nweyi29}), (\ref{Psi_varho}) obeys again the same 
CAR (\ref{vbnmei3}), (\ref{vbnmei4}), (\ref{Ps_d_Z}), and the bosonization rule 
(\ref{K_Za}) with the same renormalization constant $Z_{(\Psi)}(a)$ (\ref{ZLa_Z}) for 
arbitrary $\sigma, \rho$, and for $\overline{\xi},\xi=\pm $:
\begin{eqnarray}
&&\!\!\!\!\!\!\!\!\!\!\!\!\!\!\!\!\!\!\!\!
{\cal L}_{\rho}=\exp\left\{-\frac i2 \rho\,Q^{\overline{\xi}}Q^{-\overline{\xi}}\right\}=
\exp\left\{-\,\frac i8 \rho\left(O^2-O^2_5\right)\right\}, \quad 
\Psi(x,\sigma,\rho)={\cal L}_{\rho}^{-1}\Psi(x,\sigma){\cal L}_{\rho}, 
\label{cL_rho} \\
&&\!\!\!\!\!\!\!\!\!\!\!\!\!\!\!\!\!\!\!\!
\Psi_\xi (x,\sigma,\rho)=
{\cal L}_{\rho}^{-1}\Psi_\xi (x,\sigma){\cal L}_{\rho}=
{\rm K}^{-1}_\sigma\Psi_\xi (x,0,\rho){\rm K}_\sigma=
{\cal N}_\varphi\left\{e^{R_\xi(x,\sigma,\rho)}\right\} v_{\xi}(\sigma,\rho),
\;\;\mbox{ where:}
\label{psi_rho_sigm} \\
&&\!\!\!\!\!\!\!\!\!\!\!\!\!\!\!\!\!\!\!\!
R_\xi(x,\sigma,\rho)=-
 i2\sqrt{\pi}\left[\varrho^{-\xi}(x)+\frac{\sigma^\xi_0}{4}{\cal W}^{-\xi}
+\frac{\sigma^\xi_1}{4}{\cal W}^{\xi}\right]=
-i\left[\overline{\alpha}\varphi^{-\xi}(x^{-\xi})+
\overline{\beta}\varphi^{\xi}(x^{\xi})+\frac{\Sigma^\xi_0}{4}Q^{-\xi}
+\frac{\Sigma^\xi_1}{4}Q^{\xi}\right]\!,
\label{R_rh_Sg_Q} \\
&&\!\!\!\!\!\!\!\!\!\!\!\!\!\!\!\!\!\!\!\!
v_\xi(\sigma,\rho)=
\left(\frac{\overline{\mu}}{2\pi}\right)^{1/2}
\left(\frac{\overline{\mu}}{\Lambda}\right)^{\overline{\beta}^2/{4\pi}}
e^{i\varpi-i\xi\Theta/4}
\exp\left\{-\frac{a_0}{32}\left[\left(\Sigma^\xi_0\right)^2+
\left(\Sigma^\xi_1\right)^2\right]\right\}, \;\mbox { or:}
\label{v_xi_r_S} \\
&&\!\!\!\!\!\!\!\!\!\!\!\!\!\!\!\!\!\!\!\!
v_\xi(\sigma,\rho)=
\left(\frac{\overline{\mu}}{2\pi}\right)^{1/2}
\left(\frac{\overline{\mu}}{\Lambda}\right)^{\overline{\beta}^2/{4\pi}}
e^{i\varpi-i\xi\Theta/4}
\exp\left\{-a_0\frac{\pi}{8}\left(
\left[(\sigma^\xi_0)^2+(\sigma^\xi_1)^2\right]\cosh 2\eta-
2\sigma^\xi_0\sigma^\xi_1\sinh 2\eta\right)\right\},
\label{v_xi_r_s} \\
&&\!\!\!\!\!\!\!\!\!\!\!\!\!\!\!\!\!\!\!\!
\mbox{with: }\;\sigma^\xi_0=\sigma^\xi_0(\sigma)=- \xi\sigma, \quad 
\sigma^\xi_1=\sigma^\xi_1(\rho)=\xi 1+\rho,\;\;\, \mbox { and: }\;
\Sigma^\xi_0=\overline{\alpha}\sigma^\xi_0-\overline{\beta}\sigma^\xi_1, \quad 
\Sigma^\xi_1=\overline{\alpha}\sigma^\xi_1-\overline{\beta}\sigma^\xi_0. 
\label{sgm_Sgm_01} 
\end{eqnarray}
Remember that for divergent value of $a_0$, as for usual box, the appeared $\xi$ - 
dependence of last exponential of c - number spinor (\ref{v_xi_r_s}) leads in general 
to non-physical $\xi$ - dependent and thus non-renormalizable divergences for every 
components of the field $\Psi_\xi (x,\sigma,\rho)$. 
Thus appeared $\xi$ - dependence of the last exponential of c - number spinor 
(\ref{v_xi_r_s}) eliminates for arbitrary $\rho$ only for the case\footnote{For $\rho=0$  
in (\ref{K_Psi_s}) such of dependence was absent for arbitrary $\sigma$.} (\ref{MPS}) of 
Morchio et al. with $\sigma=-\coth 2\eta$, leading to: 
\begin{eqnarray}
&&\!\!\!\!\!\!\!\!\!\!\!\!\!\!\!\!\!\!\!\!
R_\xi(x,-\coth 2\eta,\rho)\Longrightarrow 
-i\left[\overline{\alpha}\varphi^{-\xi}\left(x^{-\xi}\right)
+\overline{\beta}\varphi^{\xi}\left(x^{\xi}\right)+
\xi\frac{\pi}{2}\left(\frac{Q^{\xi}}{\overline{\alpha}}+
\frac{Q^{-\xi}}{\overline{\beta}}\right)+\rho\frac{\sqrt{\pi}}2{\cal W}^{\xi}\right], 
\label{Mor_R_r} \\
&&\!\!\!\!\!\!\!\!\!\!\!\!\!\!\!\!\!\!\!\!
v_\xi(-\coth 2\eta,\rho)\Longrightarrow 
\left(\frac{\overline{\mu}}{2\pi}\right)^{1/2}
\left(\frac{\overline{\mu}}{\Lambda}\right)^{\overline{\beta}^2/{4\pi}}
e^{i\varpi-i\xi\Theta/4}
\exp\left\{-a_0\frac{\pi}4\left(\frac 12+\frac{\overline{\beta}^2}{4\pi}\right)
\left[\frac{1}{\sinh^2 2\eta}+\rho^2\right]\right\}. 
\label{v_Mor_r} 
\end{eqnarray}
For further reference we write down here also the transformations (\ref{K_Psi_s}), 
(\ref{psi_rho_sigm}) for the free case: 
\begin{eqnarray}
&&\!\!\!\!\!\!\!\!\!\!\!\!\!\!\!\!\!\!\!\!
\chi(x,{\sigma})={\rm K}^{-1}_\sigma\chi(x){\rm K}_\sigma,\quad 
\chi_\xi(x^{-\xi},\sigma)={\rm K}^{-1}_\sigma\chi_\xi(x^{-\xi}){\rm K}_\sigma=
{\cal N}_\varphi\left\{e^{B_\xi(x^{-\xi},\sigma)}\right\}u_{\xi}(\sigma),
\label{chi_sigm} \\
&&\!\!\!\!\!\!\!\!\!\!\!\!\!\!\!\!\!\!\!\!
B_\xi(x^{-\xi},\sigma)
=-i\left[2\sqrt{\pi}\varphi^{-\xi}\left(x^{-\xi}\right)
-\xi\sigma\frac{\sqrt{\pi}}{2} Q^{-\xi}+\xi\frac{\sqrt{\pi}}{2}Q^{\xi}\right],
\label{R_sigm} \\
&&\!\!\!\!\!\!\!\!\!\!\!\!\!\!\!\!\!\!\!\!
u_\xi(\sigma)=
\left(\frac{\overline{\mu}}{2\pi}\right)^{1/2}e^{i\varpi-i\xi\Theta/4}
\exp\left\{-a_0\frac{\pi}{8}(\sigma^2+1)\right\}, 
\label{u_sigm} \\
&&\!\!\!\!\!\!\!\!\!\!\!\!\!\!\!\!\!\!
\chi_\xi(x^{-\xi},\sigma,\rho)=
{\cal L}_{\rho}^{-1}\chi_\xi(x^{-\xi},\sigma){\cal L}_{\rho}=
{\rm K}^{-1}_\sigma\chi_\xi(x^{-\xi},0,\rho){\rm K}_\sigma=
{\cal N}_\varphi\left\{e^{B_\xi(x^{-\xi},\sigma,\rho)}\right\}u_{\xi}(\sigma,\rho),
\label{chi_r_sig} \\
&&\!\!\!\!\!\!\!\!\!\!\!\!\!\!\!\!\!\!
B_\xi(x^{-\xi},\sigma,\rho)=- i2\sqrt{\pi}
\left[\varphi^{-\xi}(x^{-\xi})+\frac{\sigma^\xi_0}{4}Q^{-\xi}
+\frac{\sigma^\xi_1}{4}Q^{\xi}\right], 
\label{chi_rho_sigm} \\
&&\!\!\!\!\!\!\!\!\!\!\!\!\!\!\!\!\!\!
u_\xi(\sigma,\rho)=
\left(\frac{\overline{\mu}}{2\pi}\right)^{1/2}e^{i\varpi-i\xi\Theta/4}
\exp\left\{-a_0\frac{\pi}{8}\left[(\sigma^\xi_0)^2+(\sigma^\xi_1)^2\right]\right\}, 
\label{u_rho_sigm}
\end{eqnarray}
which obeys the CAR (\ref{vbnmei2})--(\ref{vbnmei4}) and operator bosonization rules 
(\ref{nweyi19}) with $Z_{(\chi)}(a)=1$ for arbitrary $\sigma, \rho$. 

The corresponding VEV of the string of the fields 
(\ref{psi_rho_sigm})--(\ref{sgm_Sgm_01}) takes the following form: 
\begin{eqnarray}
&&\!\!\!\!\!\!\!\!\!\!\!\!\!\!\!\!\!\!
\left(\Lambda^{\overline{\beta}^2/4\pi}\sqrt{2\pi}\right)^p
\left\langle 0\left|\prod\limits_{i=1}^p
\Psi_{\xi_i}^{(l_i)}(x_i,\sigma,\rho)\right|0 \right\rangle=
\left\langle 0\left|{\cal N}_\varphi\left\{\exp\left(\sum\limits_{j=1}^p
{\cal R}_j\right)\right\}\right|0\right\rangle 
\exp\left\{i\varpi\sum\limits_{i=1}^p l_i - i \frac{\Theta}{4}
\sum \limits_{i=1}^p l_i \xi_i\right\} 
\nonumber \\
&&\!\!\!\!\!\!\!\!\!\!\!\!\!\!\!\!\!\!
\cdot \left(\overline{\mu}\exp\left\{-a_0\frac{\pi}{4}
\left[(1+\sigma)^2+\left(\beta^2/4\pi\right)^2\rho^2\right]
\right\}\right)^{\left(\sum\limits_{i=1}^p l_i\right)^2(\pi/\beta^2)}
\nonumber \\
&&\!\!\!\!\!\!\!\!\!\!\!\!\!\!\!\!\!\!
\cdot \left(\overline{\mu}\exp\left\{-a_0\frac{\pi}{4}
\left[(1-\sigma)^2+\left(4\pi/\beta^2\right)^2\rho^2\right]
\right\}\right)^{\left(\sum\limits_{i=1}^p l_i\xi_i\right)^2 (\beta^2/16\pi)}
\nonumber \\
&&\!\!\!\!\!\!\!\!\!\!\!\!\!\!\!\!\!\!
\cdot \exp\left\{-a_0\frac{\pi}{4}\rho\left[\cosh 2\eta+\sigma\sinh 2\eta \right]
\left(\sum\limits_{i=1}^p l_i\xi_i\right)\left(\sum\limits_{j=1}^p l_j\right)\right\}
\nonumber \\
&&\!\!\!\!\!\!\!\!\!\!\!\!\!\!\!\!\!\!
\cdot
\prod\limits_{i<k}^p \left\{e^{i \pi(\xi_i-\xi_k)}
\left[\frac{i(x_i^{-}-x_k^{-})+0}{i(x_i^{+}-x_k^{+})+0}\right]^{\xi_i+\xi_k}
\left[i0\, \varepsilon (x_i^{0}-x_k^{0})-(x_i-x_k)^2 \right]^
{(4\pi/\beta^2)+\xi_i\xi_k(\beta^2/4\pi)}\right\}^{l_i l_k/4}.
\label{VV_rho}
\end{eqnarray}
Here for $\rho\neq 0$ it is impossible to remove the $a_0$ - dependence from the third 
line already for any of above mentioned solutions of Mandelstam with $\sigma=\pm 1$, or 
Morchio et al. with $\sigma=-\coth 2\eta$. It may be only adsorbed into the parameter 
$\overline{\mu}$ if the first role (\ref{ssr_1}) fulfills. 
Now only for the both fulfilled superselection rules (\ref{ssr_1}), (\ref{ssr_2}) 
this VEV reduces again to above expression in the last line of Eq. (\ref{Ps_p_3}) 
or Eqs. (\ref{Ps_p_5}), (\ref{chi_p_6}), which is exactly the last line of Eq. 
(\ref{VV_rho}), and thus, does not depend on any of the regularization and transformation 
parameters: $\overline{\mu}$, $a_0$, and $\sigma$, $\rho$, $\varpi$, $\Theta$, and on the 
any choice of volume cut-off regularization function. Thus, the discarding of 
superselection rule (\ref{ssr_2}) inevitably spoils the $\sigma$, $\rho$, and $\Theta $ 
-- invariance of this $n$- point fermionic Wightman function and its independence on 
the parameters $\overline{\mu}$ and $a_0$. So, the latters should be fixed by some 
additional conditions \cite{fab-iva}, what, however, seems impossible, at least for 
regularization dependent value of $a_0$ (see Appendix C).

\newpage 

\section{Thirring model for nonzero temperature}

\subsection{Thermodynamics of ideal 1D gases}

From the standard courses \cite{isih} it may be easily shown, the equilibrium 
thermodynamics of the free massless bosons in the 1 -dimension box of length $L$ 
coincides with that of the free massless spin $1/2$ fermions at the same temperature 
${\rm k}_B T=1/\varsigma$ only for both zero chemical potentials $\mu_{(B)}=\mu_{(F)}=0$, 
giving a simplest example of thermal bosonization \cite{gom_ste} for pressure $P$, 
densities of internal energy ${\cal U}$ and entropy $S$: 
\begin{eqnarray}
&&\!\!\!\!\!\!\!\!\!\!\!\!\!\!\!\!\!\!\!\!
P_{(B),(F)}=\frac{{\cal U}_{(B),(F)}}L=\frac{\pi^2}{3\varsigma^2 hc}, \quad 
\frac{S_{(B),(F)}}{{\rm k}_B L}=
\left(\frac{\partial P_{(B),(F)}}{\partial(1/\varsigma)}\right)_{\mu}=
\frac{2\pi^2}{3\varsigma hc},\;\;\mbox{ however, for given}
\label{PBFS} \\
&&\!\!\!\!\!\!\!\!\!\!\!\!\!\!\!\!\!\!\!\!
\mbox{particles densities: }\; 
\overline{n}_{(B)}=\frac{N_{(B)}}L, \quad \;
\overline{n}^\pm_{(F)}=\frac{N^\pm_{(F)}}L,\; \mbox{ with }\;  
h=2\pi\hbar,\;\; c\;\mbox{ - speed of light:} 
\label{nn} \\
&&\!\!\!\!\!\!\!\!\!\!\!\!\!\!\!\!\!\!\!\!
\mu_{(B)}(T,\overline{n}_{(B)})=
\frac 1\varsigma \ln\left(1-e^{-\overline{n}_{(B)}\varsigma hc/2}\right), \quad
\mu^\pm_{(F)}(T,\overline{n}^\pm_{(F)})=
\pm\,\frac 1\varsigma \ln\left(e^{\overline{n}^\pm_{(F)}\varsigma hc/2}-1\right).
\label{MuB_MuF}
\end{eqnarray}
This qualitative ``equilibrium'' corpuscular picture means, the both systems for the 
same $\varsigma$ and $L$ have the same $P,\,{\cal U}\,,S$, and also another thermodynamic 
potentials. The condition $\mu_{(B)}=0$ for arbitrary temperature implies an infinite 
boson density, $\overline{n}_{(B)}\mapsto\infty$, corresponding to specific case of 
thermodynamic limit: $N_{(B)}\to\infty$, $L\to\infty$ for the ``bosonic picture''. 
The ``equilibrium'' fermion pressure (\ref{PBFS}) of the ``fermionic picture'' actually 
is a sum of partial ones of $N^{+}_{(F)}$ fermions ${\rm b}(p^1)$ and $N^{-}_{(F)}$ 
antifermions ${\rm f}(p^1)$, defined in (\ref{L_R_chi}) and Appendix E, with opposite 
values of chemical potentials $\mu^{\pm}_{(F)}=\pm\mu_{(F)}$ \cite{blin,alv-gom}: 
\begin{eqnarray}
&&\!\!\!\!\!\!\!\!\!\!\!\!\!\!\!\!\!\!\!\!  
P_{(F)}(T,\mu_{(F)})=\frac{{\cal U}_{(F)}}L=
P^{+}_{(F)}(T,\mu^+_{(F)})+P^{-}_{(F)}(T,\mu^-_{(F)})=
\frac{\pi^2}{3\varsigma^2 hc}+\frac{\mu^2_{(F)}}{hc}, 
\label{P_xi_P} \\
&&\!\!\!\!\!\!\!\!\!\!\!\!\!\!\!\!\!\!\!\!
\mbox{with charge density: }\;
\frac {{\rm Q}_{(F)}}{L}=\overline{n}^+_{(F)}-\overline{n}^-_{(F)}=
\left(\frac{\partial P_{(F)}}{\partial\mu_{(F)}}\right)_{\varsigma} 
=\frac{2\mu_{(F)}}{hc},\;\mbox{ where: }\;
{\rm Q}_{(F)}=\left\langle\!\!\!\left\langle \frac{O_{(\chi)} }{\sqrt{\pi}}
\right\rangle\!\!\!\right\rangle, 
\label{N_L_N_R} 
\end{eqnarray}
is averaged total charge. For any values of $\mu_{(F)},\;\mu_{(B)}$ this recasts the 
``equilibrium'' Gibbs potentials as: 
\begin{eqnarray}
&&\!\!\!\!\!\!\!\!\!\!\!\!\!\!\!\!\!\!\!\!
{\cal G}_{(F)}\equiv 
{\cal U}_{(F)}+P_{(F)}L-TS_{(F)}=N^{+}_{(F)}\mu^+_{(F)}+N^{-}_{(F)}\mu^-_{(F)}=
\left(N^{+}_{(F)}-N^{-}_{(F)}\right)\mu_{(F)}=
{\rm Q}_{(F)}\mu_{(F)}=\frac{2L \mu^2_{(F)}}{hc}, 
\label{Nmu_Nmu} \\
&&\!\!\!\!\!\!\!\!\!\!\!\!\!\!\!\!\!\!\!\!
{\cal G}_{(B)}=N_{(B)}\mu_{(B)}. \;\;\mbox { Therefore: }\;
{\cal G}_{(F)}\Longrightarrow{\cal G}_{(B)}\longrightarrow 0,
\label{muF_muB}
\end{eqnarray}
only for $\mu_{(F)}=0$, with 
$\overline{n}^+_{(F)}=\overline{n}^-_{(F)}=\overline{n}^0_{(F)}= 
2\ln 2/(\varsigma hc)$, i.e. for sector with zero total charge ${\rm Q}_{(F)}=0$. 
Similarly to radiation \cite{isih} (see Appendix E for details) the equilibrium pressure 
of massless particles on the wall originates from their adsorption and emission, thus 
with $\mu_{(B)}=\mu_{(F)}=0$. For example, the left wall adsorbs the left moving 
$N^+_{\rm L}$ fermions and $N^-_{\rm L}$ antifermions, with total charge 
${\rm Q}_{\rm L}$, and emits the right moving $N^+_{\rm R}$ fermions and 
$N^-_{\rm R}$ antifermions, with total charge ${\rm Q}_{\rm R}$ \cite{rub}. Then, the  
equilibrium for this wall means ${\rm Q}_{\rm R}-{\rm Q}_{\rm L}={\rm Q}_{5(F)}=0$. 
Since for $\mu_{(F)}=0$ the total charge (\ref{N_L_N_R}) also vanishes 
${\rm Q}_{\rm R}+{\rm Q}_{\rm L}={\rm Q}_{(F)}=0$, from Eqs. (\ref{E_4_1}), 
(\ref{E_6_1}), we may conclude, that for such of equilibrium state: 
$N^+_{\rm R,L}=N^-_{\rm R,L}$, providing an exact right and left fermions - antifermions 
pairing into the right and left moving bosons respectively. 
So, this  qualitative ``equilibrium'' picture admits virtual nonzero fermionic 
density at finite temperature $T>0$, which vanishes only with $T=0$, corresponding to 
``clean'' fermionic bosonization with zero Fermi energy 
$\mu^+_{(F)}(0,\overline{n}^+_{(F)})=\overline{n}^+_{(F)}hc/2$. 

We would like to point out, that for nonzero temperature the previous purely abstract 
infrared regularization parameter $L$ acquires a physical meaning as a macroscopic 
thermodynamic parameter (\ref{PBFS}), (\ref{nn}) \cite{isih} of the real or effective ``box size'' 
for  the thermodynamic system under consideration. So, the corresponding dependence 
requires additional care, because any function like (\ref{V_L_vs}) of the appeared new 
dimensionless variable $L/\varsigma$ has different limits at $L\to\infty$ or at 
$\varsigma\to\infty$. 
The volume cut-off regularization function $\Delta(y^1/L)$, (\ref{K_O}), (\ref{T_K_O}), 
divided by Const$\cdot L$ (with corresponding Const $=2$, or $\sqrt{\pi}$, or etc., 
depends on the functions listed at the Table in Appendix C), acquires a physical meaning 
of probability density to find the particle at the point $y^1$ in external field of the 
``walls'' of this ``box''. Of course, the physics should not depends on the choice of 
this regularization function, what is analysed in Appendix C. But it seems, that in any 
case the {\bf box should have a size} $L$, -- as a physical parameter to provide a 
possibility of some kind of thermodynamic limit for (\ref{K_O}), (\ref{D_a0}), 
(\ref{Q_pm}), and (\ref{PBFS}), (\ref{nn}), as well as for (\ref{QQ_T_p}), 
(\ref{QQ_Tt_p}), (\ref{DDddDD}), (\ref{a_1_0}), (\ref{a_2_}), (\ref{V_L_vs}) below. 
From this view point the charge regularization of Refs. \cite{mps_1,blot} 
(see Appendix C) belongs to the different type since their charge 
definition has nothing to do with any kind of thermodynamic limit. 

\subsection{On fermionic tilde conjugation rules}

Following to Ojima \cite{ojima} let us start with simplest fermionic oscillator 
(for one fixed mode $k^1$), which has only two normalized states  $|0\rangle$ and  
$|1\rangle$, with energy $0$ and $\omega$, annihilated/created by fermionic operators 
$b,b^\dagger$: $b|0\rangle=0$, and $|1\rangle=b^\dagger|0\rangle$, $\{b,b^\dagger\}=1$, 
$\{b,b\}=0$. The thermal vacuum appears as a normalized sum of tensor products of two 
independent copies of these states: 
$|0\widetilde{0}\rangle=|0\rangle\otimes|\widetilde{0}\rangle$, 
$|1\widetilde{1}\rangle=|1\rangle\otimes|\widetilde{1}\rangle$, weighted with 
corresponding Gibbs and relative phase exponential factors \cite{ojima}, so that for 
$\{b,\widetilde{b}^\#\}=0$, $(\widetilde{b}^\#=\widetilde{b},\widetilde{b}^\dagger)$, 
omitting for brevity below as above the evident index $k^1$, which label all the 
states and another operators used here, it reads: 
\begin{eqnarray}
&&\!\!\!\!\!\!\!\!\!\!\!\!\!\!\!\!\!\!\!\!
|0(\varsigma)\rangle_{(F)}=
\frac{|0\widetilde{0}\rangle+e^{i\Phi}e^{-\varsigma\omega/2}|1\widetilde{1}\rangle}
{\left[\langle 0\widetilde{0}|0\widetilde{0}\rangle+e^{-\varsigma\omega}
\langle 1\widetilde{1}|1\widetilde{1}\rangle\right]^{1/2}}\equiv 
\cos\vartheta(k^1)\left(1+e^{i\Phi}\tan\vartheta(k^1)b^\dagger\widetilde{b}^\dagger\right)
|0\widetilde{0}\rangle={\rm V}^{-1}_{\vartheta(k^1)(F)}|0\widetilde{0}\rangle,
\label{OJ_} \\
&&\!\!\!\!\!\!\!\!\!\!\!\!\!\!\!\!\!\!\!\!
\mbox{where, for: }\; \vartheta(k^1)\equiv \vartheta(k^1,\varsigma), \quad 
\tan^2\vartheta(k^1,\varsigma)=e^{-\varsigma\omega}, \quad \omega=\omega(k^1):  
\label{OJ_01} \\
&&\!\!\!\!\!\!\!\!\!\!\!\!\!\!\!\!\!\!\!\!
{\rm V}^{-1}_{\vartheta(k^1)(F)}=\exp\left\{e^{i\Phi}\tan\vartheta(k^1)G_+\right\}
\exp\left\{-\ln\left(\cos^2\vartheta(k^1)\right)G_3\right\}
\exp\left\{-e^{-i\Phi}\tan\vartheta(k^1)G_-\right\},
\label{OJ_1} \\
&&\!\!\!\!\!\!\!\!\!\!\!\!\!\!\!\!\!\!\!\!
G_+=b^\dagger \widetilde{b}^\dagger, \quad  
G_-=\widetilde{b}b=\left(G_+\right)^\dagger,\quad 
G_3=\frac 12\left(b^\dagger b-\widetilde{b}\widetilde{b}^\dagger\right)
=\frac 12\left(b^\dagger b+\widetilde{b}^\dagger \widetilde{b}-1\right),\;\mbox{ with:}
\label{OJ_2} \\
&&\!\!\!\!\!\!\!\!\!\!\!\!\!\!\!\!\!\!\!\!
\left[G_+,G_-\right]=2G_3,\quad \left[G_3,G_\pm\right]=\pm G_\pm,\quad 
G_\pm=G_1\pm i G_2,\quad \vec{G}^2=G^2_3-G_3+G_+G_- \Rightarrow j(j+1)\widehat{I},
\label{OJ_02} \\
&&\!\!\!\!\!\!\!\!\!\!\!\!\!\!\!\!\!\!\!\!
\mbox{thus: }\;
{\rm V}^{-1}_{\vartheta(k^1)(F)}=
\exp\left\{\vartheta(k^1)\left[e^{i\Phi}G_+-e^{-i\Phi}G_-\right]\right\}=
{\rm V}_{-\vartheta(k^1)(F)}={\rm V}^{\dagger}_{\vartheta(k^1)(F)}, 
\label{OJ_3}  
\end{eqnarray}
-- is a standard form of operator of the coherent state for group $SU(2)$ 
\cite{perelom}, where the relations (\ref{kir_1}), (\ref{kir_2}) are used. 
This observation allows to identify the algebra (\ref{OJ_02}) as ``quasispin'' 
algebra \cite{lipkin}, with  the ``cold'' vacuum $|0\widetilde{0}\rangle$ as its 
lowest state for representation with ``quasispin'' $j=1/2$, and the state 
$|1\widetilde{1}\rangle$ as the highest one: 
\begin{eqnarray}
&&\!\!\!\!\!\!\!\!\!\!\!\!\!\!\!\!\!\!\!\!
|0\widetilde{0}\rangle\Rightarrow \left|\frac 12,- \frac 12\right\rangle, \quad 
|1\widetilde{1}\rangle\Rightarrow \left|\frac 12, \frac 12\right\rangle,
\label{OJ_4} \\
&&\!\!\!\!\!\!\!\!\!\!\!\!\!\!\!\!\!\!\!\!
G_3\left|\frac 12, \pm \frac 12 \right\rangle=
\pm \frac 12\left|\frac 12,\pm \frac 12 \right\rangle, \quad 
G_\pm\left|\frac 12, \pm \frac 12\right\rangle=0. 
\label{OJ_04} 
\end{eqnarray}
The unique arisen arbitrary relative phase $\Phi$ reflects now the fact: the 
quantum state is not the vector, rather the ray. Thus, the thermal vacuum 
(\ref{OJ_}), as a coherent state \cite{perelom}, is annihilated by operator:
\begin{eqnarray}
&&\!\!\!\!\!\!\!\!\!\!\!\!\!\!\!\!\!\!\!\!
G_-(\varsigma)={\rm V}^{-1}_{\vartheta(k^1)(F)}G_-{\rm V}_{\vartheta(k^1)(F)}=
\cos^2\vartheta G_- +e^{i\Phi}\sin 2\vartheta G_3-e^{2i\Phi}\sin^2\vartheta G_+ =\,
\stackunder{\sim}{b}(\varsigma)b(\varsigma),
\label{cG_G}\\
&&\!\!\!\!\!\!\!\!\!\!\!\!\!\!\!\!\!\!\!\!
\mbox{as well as by operators: } \;\left\{ 
\begin{array}{c}
b(\varsigma)={\rm V}^{-1}_{\vartheta(k^1)(F)}\, b\, {\rm V}_{\vartheta(k^1)(F)}=
\cos\vartheta(k^1)\,b- e^{i\Phi}\sin\vartheta(k^1)\,\widetilde{b}^\dagger , \\
\stackunder{\sim}{b}(\varsigma)=
{\rm V}^{-1}_{\vartheta(k^1)(F)}\,\widetilde{b}\,{\rm V}_{\vartheta(k^1)(F)}=
\cos\vartheta(k^1)\,\widetilde{b} + e^{i\Phi}\sin\vartheta(k^1)\,b^\dagger .
\end{array}
\right.
\label{OJ_5} 
\end{eqnarray}
Up to now $\widetilde{b}^\#$ is only notation, which does not define any operation. 
To fix it as an operation: 
$\stackunder{\sim}{b}(\varsigma)\mapsto\widetilde{b}(\varsigma)$, one should 
choose the value of $\Phi$. The popular choice $\Phi=0$ leads to complicated tilde 
conjugation rules for the fermionic case, different from the bosonic one \cite{mtu}. 
The Ojima choice $\Phi=-\pi/2$ gives fermionic rules the same as for bosonic 
case \cite{ojima}. 
We see now, the choice $\Phi=\pi/2$ is also good and, as well as the original 
Ojima's one, satisfies the properties of antilinear homomorphism and the condition 
$\widetilde{\widetilde{b}}(\varsigma)=b(\varsigma)$. 
It seems convenient for the purposes of bosonization, the tilde operation has 
the same properties for both Fermi and Bose cases. As a byproduct, we observe a useful 
interpretation of the thermal vacuum, defined by Bogoliubov transformation (\ref{OJ_}), as 
a coherent state, obtained by coherent $SU(2)$ rotation of vacuum states of all Fermi 
oscillators $|0_{k^1}\widetilde{0}_{k^1}\rangle$ for different $k^1$ as a lowest 
quasispin states, around one and the same unit vector $\vec{\rm u}=(\sin\Phi,\cos\Phi,0)$, 
on the different angles $-2\vartheta(k^1,\varsigma)$: 
${\rm V}^{-1}_{\vartheta(k^1)(F)}=
\exp\left[i2\vartheta(k^1,\varsigma)\left(\vec{\rm u}\cdot\vec{G}\right)\right]$ 
\cite{perelom}. 

Analogous picture may be obtained from \cite{ojima} for thermal Bogoliubov transformation 
${\rm V}_{\vartheta(k^1)(B)}$ of simplest bosonic oscillator for one fixed mode $k^1$, 
leading to connection between the bosonic thermal vacuum and coherent state 
\cite{perelom} for the discrete series representation of group $SU(1,1)$, what is similar 
to the ``small'' case discussed in Appendix A (we again omit for brevity the label 
$k^1$ for the states and another operators): 
\begin{eqnarray}
&&\!\!\!\!\!\!\!\!\!\!\!\!\!\!\!\!\!\!\!\!
|0(\varsigma)\rangle_{(B)}=
\frac{\sum\limits^\infty_{n=0}e^{i\Phi_n}e^{-n\varsigma\omega/2}|n\widetilde{n}\rangle}
{\left[\sum\limits^\infty_{n=0}e^{-n\varsigma\omega}
\langle n\widetilde{n}|n\widetilde{n}\rangle\right]^{1/2}}\;
\stackunder{\Phi_n\mapsto n\Phi}{\Longrightarrow}\;\frac{1}{\cosh\vartheta(k^1)}
 \exp\left(e^{i\Phi}\tanh\vartheta(k^1)\,a^\dagger \widetilde{a}^\dagger\right)
|0\widetilde{0}\rangle={\rm V}^{-1}_{\vartheta(k^1)(B)}|0\widetilde{0}\rangle,
\label{OJ_B} \\
&&\!\!\!\!\!\!\!\!\!\!\!\!\!\!\!\!\!\!\!\!
[a,a^\dagger]=[\widetilde{a},\widetilde{a}^\dagger]=1,\quad
[a,\widetilde{a}^\#]=0,\quad 
|n\rangle=\frac{(a^\dagger)^n}{\sqrt{n!}}|0\rangle,\quad 
|\widetilde{n}\rangle=\frac{(\widetilde{a}^\dagger)^n}{\sqrt{n!}}
|\widetilde{0}\rangle, \quad \tanh^2\vartheta(k^1,\varsigma)=e^{-\varsigma\omega(k^1)},  
\label{OJ_01B} \\
&&\!\!\!\!\!\!\!\!\!\!\!\!\!\!\!\!\!\!\!\!
Y_+=a^\dagger \widetilde{a}^\dagger, \quad  
Y_-=\widetilde{a}a=\left(Y_+\right)^\dagger,\quad 
Y_0=\frac 12\left(a^\dagger a+\widetilde{a}\widetilde{a}^\dagger\right)
=\frac 12\left(1+a^\dagger a+\widetilde{a}^\dagger \widetilde{a}\right),
\label{OJ_2B} \\
&&\!\!\!\!\!\!\!\!\!\!\!\!\!\!\!\!\!\!\!\!
\left[Y_-,Y_+\right]=2Y_0,\quad \left[Y_0,Y_\pm\right]=\pm Y_\pm, \quad 
C_2= Y^2_0-Y_0-Y_+Y_-=
\frac 14\left[-1+\left(a^\dagger a-\widetilde{a}^\dagger\widetilde{a}\right)^2\right]
\Rightarrow \kappa(\kappa-1)\widehat{I},
\label{OJ_02B} \\
&&\!\!\!\!\!\!\!\!\!\!\!\!\!\!\!\!\!\!\!\!
C_2|\kappa,\nu\rangle=\kappa|\kappa,\nu\rangle, \quad 
Y_0|\kappa,\nu\rangle=\nu|\kappa,\nu\rangle, \quad \nu=\kappa+m,\quad 
\nu\Rightarrow\kappa=\frac 12, \quad 
|0\widetilde{0}\rangle \Rightarrow\left|\frac 12,\frac 12\right\rangle,
\label{OJ_vacB} \\
&&\!\!\!\!\!\!\!\!\!\!\!\!\!\!\!\!\!\!\!\!
{\rm V}^{-1}_{\vartheta(k^1)(B)}=\exp\left\{e^{i\Phi}\tanh\vartheta(k^1)\,Y_+\right\}
\exp\left\{-\ln(\cosh^2\vartheta(k^1))\,Y_0\right\}
\exp\left\{-e^{-i\Phi}\tanh\vartheta(k^1)\,Y_-\right\},
\label{OJ_1B} \\
&&\!\!\!\!\!\!\!\!\!\!\!\!\!\!\!\!\!\!\!\!
\mbox{thus: }\;
{\rm V}^{-1}_{\vartheta(k^1)(B)}=
\exp\left\{\vartheta(k^1)\left[e^{i\Phi}Y_+ -e^{-i\Phi}Y_-\right]\right\}=
{\rm V}_{-\vartheta(k^1)(B)}=
{\rm V}^\dagger_{\vartheta(k^1)(B)}, \;\mbox{ and so on.}
\label{OJ_3B}  
\end{eqnarray}
However for this case the numerator in (\ref{OJ_B}) contains a countable number of terms 
with countable number of arbitrary phases $\Phi_n$ \cite{ojima}. By means of the 
relations (\ref{kir_1}), (\ref{kir_2}), the coherent state (\ref{OJ_1B}), (\ref{OJ_3B}) 
would be obtained now only for countable number of coherent choices, 
$\Phi_n\mapsto n\Phi$, $n=0,1,2,\ldots$, already for every one simplest bosonic 
oscillator for given $k^1$ only. We did not find a reason to prefer this choice instead 
of the usual one $\Phi_n=0$ \cite{ojima}, again fixes the tilde-operation as 
antilinear homomorphism with the condition 
$\widetilde{\widetilde{a}}(\varsigma)=a(\varsigma)$ \cite{mtu}. Note the inequivalent 
vacuum again appears here as a coherent state, as well as in (\ref{VEV_0}) 
\cite{i_z} or as for the case of the c- number field's shift in Refs. \cite{mtu,i_z}. 

Analogously to (\ref{cK_pm})--(\ref{U_cKet}), the total thermal transformation is given 
by infinite product of operators (\ref{OJ_3B}) for 
$a,\widetilde{a}=a(k^1),\widetilde{a}(k^1)$, and the transformed vacuum state is an 
infinite product of one mode states similar to (\ref{U_cKet}):
\begin{eqnarray}
&&\!\!\!\!\!\!\!\!\!\!\!\!\!\!\!\!\!\!\!\!
\mbox{with: }\; a(k^1)\Rightarrow \frac{c(k^1)}{\sqrt{2k^0L}},\;\mbox{ as: }\;
{\cal V}^{-1}_{\vartheta(B)}=\prod\limits^\infty_{k^1=-\infty}
{\rm V}^{-1}_{\vartheta(k^1)(B)}=\exp\left\{-{\cal X}_\vartheta\right\}=
{\cal V}^\dagger_{\vartheta(B)}, \quad 
{\cal V}_{\vartheta(B)} = e^{{\cal X}_\vartheta}, 
\label{V_X} \\
&&\!\!\!\!\!\!\!\!\!\!\!\!\!\!\!\!\!\!\!\!
{\cal X}_\vartheta = \frac{1}{2\pi}\int\limits_{-\infty}^{+ \infty}
\frac{d q^1}{2 q^0}\vartheta\left(q^1,\varsigma\right)
\biggl(\widetilde{c}\left(q^1\right)c\left(q^1\right) - 
c^\dagger\left(q^1\right)\widetilde{c}^\dagger\left(q^1\right)\biggl)
=\widetilde{{\cal X}}_\vartheta=-{\cal X}^\dagger_\vartheta, \quad \;
|0(\varsigma)\rangle={\cal V}_{\vartheta(B)}^{-1} |0\widetilde{0}\rangle,
\label{X_X} \\
&&\!\!\!\!\!\!\!\!\!\!\!\!\!\!\!\!\!\!\!\!
{\cal V}^{-1}_{\vartheta(B)}=
\exp\left\{\frac{1}{2\pi}\int\limits_{-\infty}^{+\infty}
d k^1\tanh\vartheta(k^1,\varsigma)
\frac{c^\dagger\left(k^1\right)\widetilde{c}^\dagger\left(k^1\right)}{2k^0}\right\}
\nonumber \\
&&\!\!\!\!\!\!\!\!\!\!\!\!\!\!\!\!\!\!\!\!
\cdot \exp\left\{-\, \frac{1}{2\pi}\int\limits_{-\infty}^{+\infty}
\frac{d k^1}{2k^0}\ln\left(\cosh^2\vartheta(k^1,\varsigma)\right)
\frac 12\left(c^\dagger\left(k^1\right)c\left(k^1\right)+
\widetilde{c}\left(k^1\right)\widetilde{c}^\dagger\left(k^1\right)\right)\right\}
\nonumber \\
&&\!\!\!\!\!\!\!\!\!\!\!\!\!\!\!\!\!\!\!\!
\cdot \exp\left\{-\,\frac{1}{2\pi} \int\limits_{-\infty}^{+\infty}
d k^1\tanh\vartheta(k^1,\varsigma)
\frac{\widetilde{c}\left(k^1\right)c\left(k^1\right)}{2k^0} \right\}, 
\label{V_V_V} \\
&&\!\!\!\!\!\!\!\!\!\!\!\!\!\!\!\!\!\!\!\!
\begin{array}{c}
c \left(k^1; [\pm]\varsigma\right)={\cal V}_{\vartheta(B)}^{\mp 1}c\left(k^1\right)
{\cal V}^{\pm 1}_{\vartheta(B)}=
\cosh\vartheta c\left(k^1\right)[\mp]\sinh\vartheta\widetilde{c}^\dagger\left(k^1\right),
\\
\widetilde{c}\left(k^1;[\pm]\varsigma\right)={\cal V}_{\vartheta(B)}^{\mp 1}
\widetilde{c}\left(k^1\right){\cal V}^{\pm 1}_{\vartheta(B)}=
\cosh\vartheta\widetilde{c}\left(k^1\right)[\mp]\sinh\vartheta c^\dagger\left(k^1\right),
\end{array}
\qquad
\begin{array}{c}
c \left(k^1;[+]\varsigma\right)|0(\varsigma)\rangle =0, \\
\widetilde{c}\left(k^1;[+]\varsigma\right)|0(\varsigma)\rangle =0,
\end{array}
\label{c_tc_V} \\
&&\!\!\!\!\!\!\!\!\!\!\!\!\!\!\!\!\!\!\!\!
\begin{array}{c}
\left[c \left(k^1;[\pm]\varsigma\right), c^\dagger\left(q^1;[\pm]\varsigma\right)\right]=
(2\pi)\left(2 k^0\right)\delta\left(k^1-q^1\right),  \\
\left[\widetilde{c}\left(k^1;[\pm]\varsigma\right), 
\widetilde{c}^\dagger\left(q^1;[\pm]\varsigma\right)\right] =
(2\pi)\left(2 k^0\right)\delta\left(k^1-q^1\right),
\end{array} \qquad \tanh^2\vartheta(k^1,\varsigma)=e^{-\varsigma k^0}, \quad k^0=|k^1|.
\label{CCRcn_tcv} 
\end{eqnarray}
Thus, for ``hot'' pseudoscalar field with respect to the ``hot'' vacuum 
$|0(\varsigma)\rangle$ (\ref{X_X}), (\ref{c_tc_V}) one finds \cite{mtu,ojima}:
\begin{eqnarray}
\phi (x;[+]\varsigma)={\cal V}_{\vartheta(B)}^{-1}\phi (x){\cal V}_{\vartheta(B)}=
\frac{1}{2\pi}\int\limits_{-\infty}^\infty \frac{d k^1}{2 k^0}
\left[c\left(k^1;[+]\varsigma\right)e^{-i(kx)}+c^\dagger\left(k^1;[+]\varsigma\right)
e^{+i(kx)}\right].
\label{phi_t_c+} 
\end{eqnarray}
The meaning of the label $[\pm]$ of the field is explained below.
\subsection{Temperature equation of motion. ``Hot'' and ``cold'' thermofields}

So in the framework of thermofield dynamics \cite{mtu}, at finite temperature it is 
necessary to double the number of degrees of freedom by providing all the fields $\Psi$ 
with their tilde partners $\widetilde{\Psi}$. According to
\cite{mtu}, the resulting theory will be determined by the Hamiltonian 
$\widehat{H}[\Psi,\widetilde{\Psi}]=H[\Psi]-\widetilde{H}[\widetilde{\Psi}]$,
where $\widetilde{H}[\widetilde{\Psi}] =
H^{*}[\widetilde{\Psi}^{*}]$, with $H[\Psi]=H_{0[\Psi]}(x^0)+H_{I[\Psi]}(x^0)$ 
given by (\ref{K_0})-(\ref{K_3}), whence for Thirring model:  
$\widetilde{H}_{I[\widetilde{\Psi}]}=H_{I[\widetilde{\Psi}]}$,
and $\widetilde{H}_{0[\widetilde{\Psi}]}=-H_{0[\widetilde{\Psi}]}$. 
Though the substitution like (\ref{K_2}), (\ref{OJ_5}), for the free massless Dirac 
thermofields, $\chi(x)\mapsto\chi(x,\varsigma)$, also does not change the form \cite{mtu} 
of the free operator: 
$\widehat{H}_{0}[\chi,\widetilde{\chi}]=H_{0}[\chi]-\widetilde{H}_{0}[\widetilde{\chi}]$, 
these free fields, as explained above, generally speaking, are not now the physical 
fields of this QFT model \cite{alv-gom, blot}, and each of the terms in 
$\widehat{H}[\Psi,\widetilde{\Psi}]$ must be equivalent in a weak sense to the free 
Hamiltonian of massless (pseudo) scalar fields $(\phi(x))$, $\varphi(x)$, at least 
at zero temperature, $T=0$ \cite{blot,mtu}. 

For any functional ${\cal F}\left[\Psi\right]$ of HF $\Psi$ in the representation of 
given physical fields $\psi(x)$, i.e. for dynamical mapping $\Psi(x)=\Upsilon[\psi(x)]$  
at zero temperature, being interested in the matrix elements for the thermal 
vacuum of the type: 
\begin{eqnarray}
&&\!\!\!\!\!\!\!\!\!\!\!\!\!\!\!\!\!\!\!\!
\langle 0(\varsigma)|{\cal F}\left[\Psi(x)\right]|0(\varsigma)\rangle=
\langle 0\widetilde{0}|{\cal V}_{\vartheta}{\cal F}\left[\Psi(x)\right] 
{\cal V}^{-1}_{\vartheta}|0\widetilde{0}\rangle=
\langle 0\widetilde{0}|{\cal F}\left[{\cal V}_{\vartheta}\Psi(x)
{\cal V}^{-1}_{\vartheta}\right]|0\widetilde{0}\rangle, 
\label{VFV_1} \\
&&\!\!\!\!\!\!\!\!\!\!\!\!\!\!\!\!\!\!\!\!
\mbox{we come to formal mapping: }\; 
{\cal V}_{\vartheta}\Psi(x){\cal V}^{-1}_{\vartheta}=\Psi(x,[-]\varsigma)=
\Upsilon\left[{\cal V}_{\vartheta}\psi(x){\cal V}^{-1}_{\vartheta}\right]=
\Upsilon\left[\psi(x,[-]\varsigma)\right],
\label{VPV_1} \\
&&\!\!\!\!\!\!\!\!\!\!\!\!\!\!\!\!\!\!\!\!
\mbox{onto the ``cold'' physical thermofield: }\; 
\psi(x,[-]\varsigma)={\cal V}_{\vartheta}\psi(x){\cal V}^{-1}_{\vartheta},
\label{cold_psi}
\end{eqnarray}
essentially with the same coefficient functions, as for the initial DM 
$\Psi(x)=\Upsilon[\psi(x)]$, that, contrary to \cite{mtu, ojima}, thus transferring 
so all the temperature dependence from the ``hot'' vacuum state 
(\ref{K_0--}), (\ref{OJ_}), (\ref{OJ_B}), (\ref{X_X}), (\ref{c_tc_V}), onto these 
``cold'' physical thermofields. 
However, to compute the matrix element (\ref{VFV_1}) it is necessary to substitute 
into the r.h.s. of (\ref{VFV_1}), (\ref{VPV_1}) 
the ``cold'' physical thermofields (\ref{cold_psi}) again in terms of the initial 
physical fields $\psi(x)$ via obtained from (\ref{cold_psi}) their linear combinations 
(\ref{c_tc_V})$[-]$, analogous (but not the same!) to Eqs. (\ref{K_2}), (\ref{OJ_5}), 
(\ref{c_tc_V})$[+]$, and reorder again the so obtained operator with respect to the 
initial physical fields $\psi(x)$. 
The same operations also convert the formal mapping (\ref{VPV_1}) into temperature 
dependent DM with respect to the ``cold'' vacuum $|0\widetilde{0}\rangle$, and precisely 
in such of sense we call further the r.h.s. of (\ref{VPV_1}) again as a new DM 
$\widehat{\Upsilon}$, or e.g. $c(k^1)\mapsto {\rm b}(k^1), {\rm f}(k^1)$: 
\begin{eqnarray}
\Psi(x,[-]\varsigma)=\Upsilon\left[\psi(x,[-]\varsigma)\right]=
\Upsilon\left[{\cal V}_{\vartheta}\psi(x){\cal V}^{-1}_{\vartheta}\right]
\Longrightarrow \widehat{\Upsilon}\left[[-]\varsigma; \psi(x)\right]=
\widehat{\vec{\Upsilon}}\left[[-]\varsigma; c(k^1),\widetilde{c}(k^1)\right]. 
\label{cold_Psi_-}
\end{eqnarray}
On the contrary, the standard computation way \cite{mtu, ojima} implies the substitution 
of the inverse to (\ref{K_2}), (\ref{OJ_5}), (\ref{c_tc_V})$[+]$ linear expressions of 
physical fields 
$\psi(x)={\cal V}_{\vartheta}\psi(x,[+]\varsigma){\cal V}^{-1}_{\vartheta}$ in terms of 
the ``hot'' physical thermofields 
$\psi(x,[+]\varsigma)={\cal V}^{-1}_{\vartheta}\psi(x){\cal V}_{\vartheta}$, given by 
the (\ref{K_2}), (\ref{OJ_5}), (\ref{c_tc_V})$[+]$, 
into the l.h.s. of (\ref{VFV_1}) and reordering the so obtained operator 
with respect to this ``hot'' physical thermofield i.e. with respect to the thermal 
``hot'' -- vacuum 
(\ref{K_0--}), (\ref{OJ_}), (\ref{OJ_B}), (\ref{X_X}), (\ref{c_tc_V}). Of course, the 
same operations give the new DM $\widehat{\Upsilon}$ for the initial HF with respect to 
this ``hot'' thermal vacuum (\ref{c_tc_V}) \cite{mtu}: 
\begin{eqnarray}
\Psi(x)=\Upsilon[\psi(x)]=
\Upsilon[{\cal V}_{\vartheta}\psi(x,[+]\varsigma){\cal V}^{-1}_{\vartheta}]
\Longrightarrow\widehat{\Upsilon}\left[[+]\varsigma;\psi(x,[+]\varsigma)\right]=
\widehat{\vec{\Upsilon}}\left[[+]\varsigma; c(k^1,[+]\varsigma),
\widetilde{c}(k^1,[+]\varsigma)\right]. 
\label{hot_Psi_+}
\end{eqnarray}
We would like to point out, that this field does not equal to 
$\Psi(x,[+]\varsigma)={\cal V}^{-1}_{\vartheta}\Psi(x){\cal V}_{\vartheta}$, which 
will appear below as a byproduct of our further consideration\footnote{reading as: 
$\Psi(x,[+]\varsigma)=\Upsilon\left[\psi(x,[+]\varsigma)\right]
=\Upsilon\left[{\cal V}^{-1}_{\vartheta}\psi(x){\cal V}_{\vartheta}\right]
\Longrightarrow \widehat{\Upsilon}\left[[+]\varsigma; \psi(x)\right]=
\widehat{\vec{\Upsilon}}\left[[+]\varsigma; c(k^1),\widetilde{c}(k^1)\right]$, i.e. 
again with respect to the ``cold'' vacuum $|0\widetilde{0}\rangle$.}
similar to (\ref{cold_Psi_-}). So, to avoid some ambiguities \cite{abr,abr2,abr3} one 
should carefully distinguish the use of ``hot'' and ``cold'' physical thermofields 
$\psi(x,[\pm]\varsigma)$ with respect to corresponding vacua. 

The kinematic independence of fermionic tilde-conjugate fields $\widetilde{\Psi}$ for 
$T=0$ means: 
\begin{eqnarray}
\left\{\Psi_\xi(x),\widetilde{\Psi}^{\#}_{\xi'}(y)\right\}\Bigr|_{x^0= y^0}=0,\qquad 
\left\{\Psi_\xi(x),\widetilde{\Psi}^{\#}_{\xi'}(y)\right\}\Bigr|_{(x-y)^2<0}=0,
\label{T_vbnmei4}
\end{eqnarray}
and corresponds to above independence of their Hamiltonians and their HEqs for $T=0$. 
This allows to consider a solution only for the one of them. Since the thermal 
transformations ${\cal V}_{\vartheta(F)}$, ${\cal V}_{\vartheta(B)}$ are not depend on  
coordinates and time, they can be applied directly to zero temperature HEq of Thirring 
model (\ref{45bn6l4}), (\ref{45bn6l5}), resulting to the same HEqs for the new HF 
(\ref{VPV_1}), with the same kinematic independence condition (\ref{T_vbnmei4}) for 
finite temperature, where we omit for brevity the label $[\pm]$ of 
$\Psi(x;[\pm]\varsigma)$, where it is not important: 
\begin{eqnarray}
&&\!\!\!\!\!\!\!\!\!\!\!\!\!\!\!\!\!\!\!\!
i\partial_0\Psi(x,\varsigma)=\left[\Psi(x,\varsigma),  
\widehat{H}[\Psi,\widetilde{\Psi}]\,\right]=
\left[E(P^1)+g\gamma^0\gamma_\nu J_{(\Psi)}^\nu(x,\varsigma)
\right]\Psi(x,\varsigma),
\label{T_45bn6l4} \\
&&\!\!\!\!\!\!\!\!\!\!\!\!\!\!\!\!\!\!\!\!
\mbox{or: }\;
2\partial_\xi \Psi_\xi(x,\varsigma)=-igJ^{-\xi}_{(\Psi)}(x,\varsigma)
\Psi_\xi(x,\varsigma), \quad  \xi=\pm,
\label{T_45bn6l5} \\
&&\!\!\!\!\!\!\!\!\!\!\!\!\!\!\!\!\!\!\!\!
\mbox{so: }\;
2\partial_\xi\widetilde{\Psi}_\xi(x,\varsigma)=+ig
\widetilde{J}^{-\xi}_{(\widetilde{\Psi})}(x,\varsigma)
\widetilde{\Psi}_\xi(x,\varsigma), \quad  \xi=\pm,
\label{tT_45bn6l5} \\
&&\!\!\!\!\!\!\!\!\!\!\!\!\!\!\!\!\!\!\!\!
\left\{\Psi_\xi(x,\varsigma),
\widetilde{\Psi}^{\#}_{\xi'}(y,\varsigma)\right\}\Bigr|_{x^0= y^0}=0,\qquad 
\left\{\Psi_\xi(x,\varsigma)),
\widetilde{\Psi}^{\#}_{\xi'}(y,\varsigma))\right\}\Bigr|_{(x-y)^2<0}=0,
\label{tT_vbnmei4}
\end{eqnarray}
-- for each $\xi$-component of the fields $\Psi_\xi(x,\varsigma)$, 
$\widetilde{\Psi}_\xi(x,\varsigma)$, that are also formally related to the corresponding 
current components as:
\begin{eqnarray}
J_{(\Psi)}^\xi (x,\varsigma)=J_{(\Psi)}^0 (x,\varsigma)+\xi
J_{(\Psi)}^1(x;  \varsigma)\longmapsto
2\Psi_\xi^\dagger(x,\varsigma)\Psi_\xi (x,\varsigma), \quad \xi=\pm,
\label{T_vbnmei8}
\end{eqnarray}
and the same for tilde-conjugate currents and fields. 
Thus, to integrate these HEqs we can sequentially repeat all the previous zero 
temperature steps of the previous section, \cite{ks_tt,ks_ttt,ks_tttt}. Applying the same 
arguments based on the currents conservation: 
$\partial_\xi J_{(\Psi)}^\xi (x,\varsigma)=0$, $\xi=\pm$, we come to the same 
linearization, renormalization and bosonization conditions in the sense of weak equality. 
We reproduce briefly all these steps in the next subsections to outline the main 
differences. The first one is due to appearance of tilde-conjugate fields. 
By virtue of (\ref{OJ_5}), (\ref{c_tc_V}), the tilde-conjugation rule for corresponding 
DM (\ref{cold_Psi_-}), (\ref{hot_Psi_+}) takes the most simple form in momentum 
representation \cite{mtu}, in terms of initial annihilation/creation operators $c(k^1), 
{\rm b}(k^1), {\rm f}(k^1)$, etc. for zero temperature. 

\subsection{Linearization of the Heisenberg equation}

From the equation (\ref{T_45bn6l4}) and anticommutation relations
for the field operators (\ref{vbnmei2})--(\ref{vbnmei4}) it
follows again that in the canonical equation of motion for the
``total current'' operator (\ref{vbnmei5}), (\ref{T_vbnmei8}) from the
right hand side of HEq. (\ref{T_45bn6l4}):
\begin{eqnarray}
&&\!\!\!\!\!\!\!\!\!\!\!\!\!\!\!\!\!\!\!\!
i\partial_0\gamma^0\gamma_\nu J_{(\Psi)}^\nu(x,\varsigma)-
\left[\gamma^0\gamma_\nu J_{(\Psi)}^\nu (x,\varsigma),
H_{0[\Psi]}(x^0,\varsigma)\right]=
iI\,\partial_\mu J_{(\Psi)}^\mu(x,\varsigma)+
i\gamma^5\,\epsilon_{\mu\nu}\partial^\mu J^\nu_{(\Psi)}(x,\varsigma)=0, 
\label{T_bnemti55} \\
&&\!\!\!\!\!\!\!\!\!\!\!\!\!\!\!\!\!\!\!\!
i\partial_0\gamma^0\gamma_\nu J_{(\Psi)}^\nu(x,\varsigma)-
\left[\gamma^0\gamma_\nu J_{(\Psi)}^\nu (x,\varsigma),
H_{0[\Psi]}(x^0,\varsigma)\right]
=\left[\gamma^0\gamma_\nu J_{(\Psi)}^\nu (x,\varsigma),
H_{I[\Psi]}(x^0,\varsigma)\right] = 0,
\label{T_bnemti5} 
\end{eqnarray}
due to currents conservation $\partial_\xi J_{(\Psi)}^\xi (x,\varsigma)=0$, again 
vanishes the contribution of the commutator with the
interaction Hamiltonian $H_{I(\Psi)} \left(x^0,\varsigma\right)$. Therefore, the temporal 
evolution of this ``total current'' will be again described by a free Hamiltonian
$H_{0(\chi)}\left(x^0,\varsigma\right)$, quadratic on some free trial physical Dirac 
fields $\chi(x,\varsigma)$, furnished by the same anti-commutation relations 
(\ref{vbnmei2})--(\ref{vbnmei4}), (\ref{T_vbnmei4}), (\ref{tT_vbnmei4}), and by the same 
conservation laws for corresponding currents $J_{(\chi)}^\nu (x,\varsigma)$, 
$J_{(\chi)}^{5\nu}(x,\varsigma)$, given by Eqs.  (\ref{vbnmei5}), (\ref{T_vbnmei8}) with 
$\Psi(x,\varsigma)\mapsto\chi(x,\varsigma)$:
\begin{eqnarray}
i\partial_0\gamma^0\gamma_\nu J_{(\chi)}^\nu (x,\varsigma)-
\left[\gamma^0\gamma_\nu J_{(\chi)}^\nu (x,\varsigma),
H_{0[\chi]}(x^0,\varsigma)\right]=
iI\,\partial_\mu J_{(\chi)}^\mu(x,\varsigma)+
i\gamma^5\,\epsilon_{\mu\nu}\partial^\mu J^\nu_{(\chi)}(x,\varsigma)=0.
\label{T_bnemti6}
\end{eqnarray}
So, as in above section, the allowance of a possible contribution into (\ref{T_bnemti5}) 
of the Schwinger terms would be premature, leading to contradiction 
with the vector and pseudovector currents conservation conditions.
As above, the Heisenberg current operators appearing in (\ref{T_45bn6l4}),
(\ref{T_45bn6l5}) acquire precise operator meaning -- with non-vanishing Schwinger
term -- again only after the choice of the representation space \cite{Leut}, \cite{hep},
\cite{Vlad} for anticommutation relations (\ref{vbnmei2})--(\ref{vbnmei4}), 
(\ref{T_vbnmei4}),  (\ref{tT_vbnmei4}), and subsequent reduction in this representation 
to the normal ordered form by means of renormalization, for example, again via 
point-splitting and subtraction of the VEV \cite{blot}, but taken now with respect to 
the initial ``cold'' vacuum $|0\widetilde{0}\rangle$: 
\begin{eqnarray}
&&\!\!\!\!\!\!\!\!\!\!\!\!\!\!\!\!\!\!
J^0_{(\Psi)} (x,\varsigma) \longmapsto
\lim\limits_{\widetilde{\varepsilon} \rightarrow 0}
\widehat{J}^0_{(\Psi)}(x;\widetilde{\varepsilon},\varsigma)
=\widehat{J}^0_{(\Psi)}(x,\varsigma),
\quad J^1_{(\Psi)} (x,\varsigma) \longmapsto
\lim\limits_{\varepsilon \rightarrow 0}
\widehat{J}^1_{(\Psi)}(x;\varepsilon,\varsigma)=
\widehat{J}^1_{(\Psi)}(x,\varsigma),
\label{T_bos-111}\\
&&\!\!\!\!\!\!\!\!\!\!\!\!\!\!\!\!\!\!
\mbox{where at first: }\; \widetilde{\varepsilon}^0 =
\varepsilon^1\rightarrow 0,\;
\mbox{ when: }\; \widetilde{\varepsilon}^1 =\varepsilon^0,\;\;
\varepsilon^2=-\widetilde{\varepsilon}^2>0,
\label{T_K_E} \\
&&\!\!\!\!\!\!\!\!\!\!\!\!\!\!\!\!\!\!
\mbox{for: }\;
\widehat{J}^\nu_{(\Psi)}(x;a,\varsigma)=
Z^{-1}_{(\Psi)}(a)\left[\overline{\Psi}(x + a,\varsigma)\gamma^\nu \Psi (x,\varsigma)- 
\langle 0\widetilde{0}|\overline{\Psi}(x + a,\varsigma)\gamma^\nu 
\Psi(x,\varsigma)|0\widetilde{0}\rangle\right],
\label{T_K_Z}
\end{eqnarray}
and accordingly for every $\xi$- component (\ref{T_vbnmei8}). The
renormalization ``constant'' $Z_{(\Psi)}(a)$ is defined below in Eq.
(\ref{T_K_Za}). With these remarks, observations (\ref{T_bnemti5}),
(\ref{T_bnemti6}) again allow to identify at least in a weak sense, the 
Heisenberg operator of ``total current'' on the r.h.s. of Eq. (\ref{T_45bn6l4}), 
defined by Eqs. (\ref{vbnmei5}), (\ref{T_bnemti5}), with that operator, defined by Eqs.  
(\ref{vbnmei5}), (\ref{T_bnemti6}) for the free massless trial physical Dirac fields 
$\chi(x,\varsigma)$ and renormalized in the sense of normal form 
(\ref{T_bos-111})--(\ref{T_K_Z}) up to an unknown yet constant $\beta$:
\begin{eqnarray}
&&\!\!\!\!\!\!\!\!\!\!\!\!\!\!\!\!\!\!\!\!
\gamma^0\gamma_\nu J_{(\Psi)}^\nu(x,\varsigma)
\stackrel{\rm w}{\longmapsto}
\frac{\beta}{2\sqrt{\pi}}\gamma^0\gamma_\nu
\widehat{J}_{(\chi)}^\nu(x,\varsigma),\;\;\mbox{ where again:}
\label{T_ns94m61} \\
&&\!\!\!\!\!\!\!\!\!\!\!\!\!\!\!\!\!\!\!\!
\widehat{J}_{(\chi)}^\nu(x,\varsigma)=
\lim\limits_{\varepsilon,(\widetilde{\varepsilon})\rightarrow 0}
\widehat{J}_{(\chi)}^\nu\left(x;\varepsilon
(\widetilde{\varepsilon}),\varsigma\right)\,
\equiv\, :J_{(\chi)}^\nu(x,\varsigma):\,,\;\mbox{ for: }\;Z_{(\chi)}(a)=1,
\label{T_ns94m62}
\end{eqnarray}
what leads again to the linearization of both equations (\ref{T_45bn6l4}), 
(\ref{T_45bn6l5}) in the representation of these free trial physical massless Dirac 
fields $\chi(x,\varsigma)$. 

\subsection{Thermal bosonization and scalar fields}

Again, the use of BP also simplifies integration of linearized HEqs (\ref{T_45bn6l4}). 
Being again a formal consequence of the current conservation conditions
(\ref{T_bnemti55}) only, the bosonization rules have, generally
speaking, the sense of weak equalities only for the current
operator in the normal-ordered form (\ref{T_bos-111})--(\ref{T_K_Z}), that already 
implies a choice of certain representations of (anti-) commutation relations
(\ref{vbnmei2})--(\ref{vbnmei4}), (\ref{T_vbnmei4}),  (\ref{tT_vbnmei4}), and 
(\ref{T_K_8_1}) below. 
However, for the free massless fields $\chi(x,\varsigma)$, $\varphi(x,\varsigma)$,
$\phi(x,\varsigma)$, this choice is again carried out ``almost automatically''.
This, due to the linearization condition (\ref{T_ns94m61}),
(\ref{T_ns94m62}), again becomes enough for our purposes, since for the
free fields these relationships appear as operator equalities \cite{blot}:
\begin{eqnarray}
\widehat{J}_{(\chi)}^\mu (x,\varsigma)=\frac{1}{\sqrt{\pi}}
\partial^\mu\varphi(x,\varsigma)=
- \frac{1}{\sqrt{\pi}}\epsilon^{\mu\nu}\partial_\nu \phi
(x,\varsigma), \quad
\widehat{J}_{(\chi)}^{-\xi} (x,\varsigma) = \frac{2}{\sqrt{\pi}}
\partial_{\xi} \varphi^{\xi}\left(x^{\xi},\varsigma\right)\,, 
\label{T_nweyi19}
\end{eqnarray}
where the thermofields $\varphi (x,\varsigma)$ and $\phi(x,\varsigma)$ are 
defined in (\ref{T_ppp}) below  as unitarily inequivalent representations of the 
massless scalar and pseudoscalar Klein-Gordon fields: 
$\partial_\mu\partial^\mu\varphi(x,\varsigma)=0$, and 
$\partial_\mu\partial^\mu\phi(x,\varsigma)=0$, and are taken again mutually
dual and coupled by the symmetric integral relations: 
\begin{eqnarray}
&&\!\!\!\!\!\!\!\!\!\!\!\!\!\!\!\!\!\!\!\!
\left. \begin{array}{c}\phi(x,\varsigma) \\
\varphi(x,\varsigma)\end{array}\right\}
=-\frac{1}{2}\int\limits_{-\infty}^\infty dy^1
\varepsilon \left(x^1-y^1\right)\partial_0
\left\{\begin{array}{c}\varphi\left(y^1,x^0,\varsigma\right), \\
\phi\left(y^1,x^0,\varsigma \right), \end{array}\right.
\label{T_K_5}
\end{eqnarray}
that again implies the solitonic type of asymptotical conditions:  
\begin{eqnarray}
&&\!\!\!\!\!\!\!\!\!\!\!\!\!\!\!\!\!\!\!\!
\varphi(-\infty,x^0,\varsigma)+\varphi(\infty,x^0,\varsigma)=0, \quad 
\phi(-\infty,x^0,\varsigma)+\phi(\infty,x^0,\varsigma)=0, 
\label{T_as_cond}
\end{eqnarray}
with the conserved charges corresponding to these fields, as: 
\begin{eqnarray}
&&\!\!\!\!\!\!\!\!\!\!\!\!\!\!\!\!\!\!\!\!
\left. \begin{array}{c} O(\varsigma) \\ O_5(\varsigma)\end{array}\right\}=
\lim_{L\to\infty}
\int\limits_{-\infty}^\infty dy^1\Delta\left(\frac{y^1}L\right)\partial_0
\left\{\begin{array}{c}\varphi\left(y^1,x^0,\varsigma\right) \\
\phi\left(y^1,x^0,\varsigma\right) \end{array}\right\}
\stackunder{\Delta=1}{\Longrightarrow}
\left\{\begin{array}{c}\phi(-\infty, x^0,\varsigma)-\phi(\infty,x^0,\varsigma) \\
\varphi(-\infty,x^0,\varsigma)-\varphi(\infty,x^0,\varsigma).\end{array}\right. 
\label{T_K_O} 
\end{eqnarray}
The right and left thermofields $\varphi^{\xi}\left(x^{\xi},\varsigma\right)$ and their 
charges $ Q^\xi(\varsigma)$ are defined again by the same linear combinations 
(\ref{K_7})--(\ref{sum_xi}), \cite{blot}:
\begin{eqnarray}
&&\!\!\!\!\!\!\!\!\!\!\!\!\!\!\!\!\!\!\!\!
\varphi^\xi\left(x^\xi,\varsigma\right)=
\frac{1}{2}\left[\varphi(x,\varsigma)-\xi\phi(x,\varsigma)\right], \;
\mbox { for: }\; \xi=\pm ,
\label{T_K_7} \\
&&\!\!\!\!\!\!\!\!\!\!\!\!\!\!\!\!\!\!\!\!
Q^\xi (\varsigma)=\frac{1}{2}\left[O(\varsigma)-
\xi{O}_5(\varsigma)\right]=
\xi\varphi^\xi(x^0+\xi\infty,\varsigma)-\xi\varphi^\xi(x^0-\xi\infty,\varsigma)
=\pm 2\varphi^\xi(x^0\pm\infty,\varsigma),
\label{T_K_7_Q}
\end{eqnarray}
The fields $\varphi(x,\varsigma)$, $\phi(x,\varsigma)$,
$\varphi^\xi\left(x^\xi,\varsigma\right)$ and their charges 
obey the same commutation relations (\ref{K_8_1})--(\ref{K_9}), that 
are not depended on temperature, for example:
\begin{eqnarray}
&&\!\!\!\!\!\!\!\!\!\!\!\!\!\!\!\!\!\!\!\!
\left[\varphi(x,\varsigma),\partial_0 \varphi (y,\varsigma)\right]
\bigr|_{x^0=y^0}=
\left[\phi(x,\varsigma),\partial_0\phi(y,\varsigma)\right]
\bigr|_{x^0=y^0}=i\delta(x^1-y^1),
\label{T_K_8_1} \\
&&\!\!\!\!\!\!\!\!\!\!\!\!\!\!\!\!\!\!\!\!
\left[\varphi(x,\varsigma), \varphi (y,\varsigma)\right]=
\left[\phi(x,\varsigma),\phi(y,\varsigma)\right]=
-i\frac{\varepsilon(x^0-y^0)}2\theta\left((x-y)^2\right),
\label{T_K_8} \\
&&\!\!\!\!\!\!\!\!\!\!\!\!\!\!\!\!\!\!\!\!
\left[\varphi^\xi\left(s,\varsigma\right),\varphi^{\xi'}
\left(\tau,\varsigma\right)\right]=
-\frac{i}{4}\varepsilon(s - \tau)\delta_{\xi, \xi'}, \quad
\left[\varphi^\xi(s,\varsigma), Q^{\xi'}(\varsigma)\right]=
\frac{i}{2}\delta_{\xi,\xi'}.
\label{T_K_9}
\end{eqnarray}
Moreover, the similar commutation relations (but not the same!) take place for their 
tilde-partners:
\begin{eqnarray}
&&
\left[ \widetilde{\varphi}(x,\varsigma),\partial_0\widetilde{\varphi}(y,\varsigma)\right]
\bigr|_{x^0=y^0}=
\left[\widetilde{\phi}(x,\varsigma),\partial_0 \widetilde{\phi}(y,\varsigma)\right]
\bigr|_{x^0=y^0}=- i\delta(x^1-y^1),
\label{T_K_8_1_t} \\
&&
\left[\widetilde{\varphi}(x,\varsigma),\widetilde{\varphi}(y,\varsigma)\right]=
\left[\widetilde{\phi}(x,\varsigma),\widetilde{\phi}(y,\varsigma)\right]=
+ i\frac{\varepsilon(x^0-y^0)}2\theta\left((x-y)^2\right),
\label{T_K_8_t} \\
&&
\left[\widetilde{\varphi}^\xi\left(s,\varsigma\right),\widetilde{\varphi}^{\xi'}
\left(\tau,\varsigma\right)\right]=
+ \frac{i}{4}\varepsilon(s - \tau)\delta_{\xi, \xi'}, \quad
\left[\widetilde{\varphi}^\xi(s,\varsigma),\widetilde{Q}^{\xi'}(\varsigma)\right]=
-\,\frac{i}{2}\delta_{\xi,\xi'}, 
\label{T_K_9_t}
\end{eqnarray}
that remain kinematically independent in the sense of Eqs. (\ref{T_vbnmei4}),  
(\ref{tT_vbnmei4}), also at finite temperature:  
\begin{eqnarray}
&&\!\!\!\!\!\!\!\!\!\!\!\!\!\!\!\!\!\!
\left[\varphi^\xi\left(s,\varsigma\right),\widetilde{\varphi}^{\xi'}
\left(\tau,\varsigma\right)\right]=0, \quad
\left[\varphi^\xi(s,\varsigma), \widetilde{Q}^{\xi'}(\varsigma)\right]=0, \quad
\left[ Q^{\xi}(\varsigma), \widetilde{Q}^{\xi'}(\varsigma)\right]=0. 
\label{T_C_til}
\end{eqnarray}
So, up to now we cannot distinguish the ``hot'' and ``cold'' physical thermofields. 


The kinematic independence of the tilde-partners fails and the difference between 
the ``hot'' and ``cold'' physical thermofields appears on going to the ``frequency'' 
parts of 
corresponding physical fields $\varphi^{\xi(\pm)}\left(x^\xi,\varsigma\right)$, and their 
charges $ Q^{\xi(\pm)}(\varsigma)$ with respect to any of chosen vacuum state. In 
particular, it 
manifests itself in the commutators of annihilation $(+)$ and creation $(-)$ (frequency) 
parts, defined according to (\ref{cc_k0}), by annihilation and creation operators for 
the one and the same initial ``cold'' vacuum $|0\widetilde{0}\rangle$: 
$c(k^1)|0\widetilde{0}\rangle=\widetilde{c}(k^1)|0\widetilde{0}\rangle=0$, 
for both the ``hot'' $[+]$, and ``cold'' $[-]$ physical thermofields by making use of 
Eqs. (\ref{X_X})--(\ref{phi_t_c+}), in the form:
\begin{eqnarray}
&&\!\!\!\!\!\!\!\!\!\!\!\!\!\!\!\!\!\!
| 0 (\varsigma) \rangle = {\cal V}_{\vartheta(B)}^{-1}
| 0\widetilde{0} \rangle \equiv {\cal V}_{(B)} [-\vartheta]
| 0\widetilde{0} \rangle, \quad  \vartheta=\vartheta(k^1;\varsigma), 
\quad  \tanh^2\vartheta(k^1;\varsigma)=e^{-\varsigma k^0},
\label{vac_vac} \\
&&\!\!\!\!\!\!\!\!\!\!\!\!\!\!\!\!\!\!
\varphi(x;[\pm]\varsigma)=
{\cal V}_{\vartheta(B)}^{\mp 1}\varphi(x){\cal V}_{\vartheta(B)}^{\pm 1}
\Longrightarrow \varphi^{(+)} (x; [\pm] \varsigma) +
\varphi^{(-)} (x; [\pm] \varsigma),
\label{T_ppp}
\end{eqnarray}
and so on for all other free physical (pseudo) scalar fields and charges 
$\phi(x),\omega(x),\Omega(x), O, O_5,{\cal W}^\xi,...$, with corresponding Fourier 
expansions and commutators. Below we put corresponding $\pm$ into respective 
brackets, and $k^0 = |k^1|$: 
\begin{eqnarray}
&&\!\!\!\!\!\!\!\!\!\!\!\!\!\!\!\!\!\!
\varphi^{\xi(+)} \left(x^\xi; [\pm]\varsigma\right) = -
\frac{\xi}{2\pi} \int\limits_{-\infty}^\infty \frac{d k^1}{2 k^0}
\theta \left(- \xi k^1\right) \left[\cosh \vartheta c
\left(k^1\right) e^{- i k^0 x^\xi} \mp \sinh \vartheta
\widetilde{c} \left(k^1\right) e^{i k^0 x^\xi} \right],
\label{phi_T_p} \\
&&\!\!\!\!\!\!\!\!\!\!\!\!\!\!\!\!\!\!
\varphi^{\xi (-)} \left(x^\xi; [\pm]\varsigma\right) =
\left\{\varphi^{\xi (+)} \left(x^\xi;[\pm]
\varsigma\right)\right\}^\dagger,
\label{phi_T_m} \\ 
&&\!\!\!\!\!\!\!\!\!\!\!\!\!\!\!\!\!\!
\widetilde{\varphi}^{\xi (+)} \left(x^\xi; [\pm]\varsigma\right) = -
\frac{\xi}{2\pi} \int\limits_{-\infty}^\infty \frac{d k^1}{2 k^0}
\theta \left(- \xi k^1\right) \left[\cosh \vartheta \widetilde{c}
\left(k^1\right) e^{i k^0 x^\xi} \mp \sinh \vartheta
c \left(k^1\right) e^{- i k^0 x^\xi} \right],
\label{phi_Tt_p} \\
&&\!\!\!\!\!\!\!\!\!\!\!\!\!\!\!\!\!\!
\widetilde{\varphi}^{\xi (-)} \left(x^\xi; [\pm]\varsigma\right) =
\left\{\widetilde{\varphi}^{\xi (+)} \left(x^\xi;[\pm]
\varsigma\right)\right\}^\dagger, 
\label{phi_Tt_m} \\
&&\!\!\!\!\!\!\!\!\!\!\!\!\!\!\!\!\!\!
Q^{\xi (+)} ([\pm]\varsigma) = \lim_{L \rightarrow \infty} i
\frac{\xi}{2} \int\limits_{-\infty}^\infty d k^1
\theta \left(-\xi k^1\right) \left[\cosh \vartheta c
\left(k^1\right) e^{- i k^0 \widehat{x}^0} \pm \sinh \vartheta \widetilde{c}
\left(k^1\right) e^{i k^0\widehat{x}^0}\right]\delta_L \left(k^1\right),
\label{QQ_T_p} \\
&&\!\!\!\!\!\!\!\!\!\!\!\!\!\!\!\!\!\!
Q^{\xi (-)} ([\pm]\varsigma) = \left\{Q^{\xi (+)}
([\pm]\varsigma)\right\}^\dagger, 
\label{QQ_T_m} \\
&&\!\!\!\!\!\!\!\!\!\!\!\!\!\!\!\!\!\!
\widetilde{Q}^{\xi (+)} ([\pm]\varsigma) = \lim_{L \rightarrow \infty} - i
\frac{\xi}{2} \int\limits_{-\infty}^\infty d k^1
\theta \left(-\xi k^1\right) \left[\cosh \vartheta \widetilde{c}
\left(k^1\right) e^{i k^0 \widehat{x}^0} \pm \sinh \vartheta c
\left(k^1\right) e^{- i k^0 \widehat{x}^0}\right]\delta_L \left(k^1\right),
\label{QQ_Tt_p} \\
&&\!\!\!\!\!\!\!\!\!\!\!\!\!\!\!\!\!\!
\widetilde{Q}^{\xi (-)} ([\pm]\varsigma) = \left\{\widetilde{Q}^{\xi (+)}
([\pm]\varsigma)\right\}^\dagger.
\label{QQ_Tt_m} 
\end{eqnarray}
Here, as before in (\ref{Q_pm}), the $\widehat{x}^0$ -- dependence of charge frequency 
parts (\ref{QQ_T_p}), (\ref{QQ_Tt_p}) is fictitious and non-physical. 
It is an artifact of space regularization (\ref{T_K_O}) and should be eliminated at 
the end of calculation.  

Only for ``hot'' $[+]$ thermofields one has:
\begin{eqnarray}
&&\!\!\!\!\!\!\!\!\!\!\!\!\!\!\!\!\!\!
\langle 0 (\varsigma)|\varphi^\xi(s;[+]\varsigma)\varphi^{\xi'}(\tau;[+]\varsigma)
|0(\varsigma)\rangle=\langle 0|\varphi^{\xi}(s)\varphi^{\xi'} (\tau)|0\rangle=
\label{hot_DW0} \\
&&\!\!\!\!\!\!\!\!\!\!\!\!\!\!\!\!\!\!
= \left[\varphi^{\xi(+)}(s),\varphi^{\xi'(-)}(\tau)\right]
= \frac{\delta_{\xi,\xi'}}{i} D^{(-)} (s-\tau),
\label{hot_DW} 
\end{eqnarray}
(here $D^{(-)}(s)=\lim\limits_{\varsigma \rightarrow \infty}
{\cal D}^{(-)}(s,\varsigma;\mu_1)$, see Appendix B), but for both of them: 
\begin{eqnarray}
&&\!\!\!\!\!\!\!\!\!\!\!\!\!\!\!\!\!\!
\langle 0\widetilde{0} |
\varphi^\xi (s; [\pm] \varsigma)\varphi^{\xi'} (\tau; [\pm] \varsigma)
|0\widetilde{0} \rangle
= \left[\varphi^{\xi(+)}(s;[\pm]\varsigma),\varphi^{\xi'(-)}(\tau;[\pm]\varsigma)\right],
\label{p_DW} \\
&&\!\!\!\!\!\!\!\!\!\!\!\!\!\!\!\!\!\!
\left[\varphi^{\xi (\pm)} \left(s; [\pm]\varsigma\right), \varphi^{\xi'
(\mp)} \left(\tau; [\pm]\varsigma\right)\right]
= (\pm 1) \frac{\delta_{\xi,\xi'}}{i}{\cal D}^{(-)} (\pm(s-\tau),\varsigma;\mu_1)=
\nonumber \\
&&\!\!\!\!\!\!\!\!\!\!\!\!\!\!\!\!\!\!
= (\mp 1) \frac{1}{4\pi} \delta_{\xi, \xi'} \left\{\ln \left(i\overline{\mu} 
\frac{\varsigma}{\pi} \sinh \left(\frac{\pi}{\varsigma}
(\pm(s-\tau)-i0)\right)\right) -
g \left(\varsigma, \mu_1\right)\right\},
\label{pm_DW} \\
&&\!\!\!\!\!\!\!\!\!\!\!\!\!\!\!\!\!\!
\left[\widetilde{\varphi}^{\xi (\pm)} \left(s; [\pm]\varsigma\right),
\widetilde{\varphi}^{\xi'(\mp)} \left(\tau; [\pm]\varsigma\right)\right]= (\mp 1)
\frac{\delta_{\xi,\xi'}}{i} \widetilde{\cal D}^{(-)} (\pm(s-\tau),\varsigma;\mu_1)=
\nonumber \\
&&\!\!\!\!\!\!\!\!\!\!\!\!\!\!\!\!\!\!
= (\mp 1) \frac{1}{4\pi} \delta_{\xi, \xi'} \left\{\ln \left(i \overline{\mu}
\frac{\varsigma}{\pi} \sinh \left(\frac{\pi}{\varsigma}
(\mp(s-\tau)-i0)\right)\right) -
g \left(\varsigma, \mu_1\right)\right\},
\label{pmtt_DW} \\
&&\!\!\!\!\!\!\!\!\!\!\!\!\!\!\!\!\!\!
\left[\varphi^{\xi(\pm)}\left(s;[\pm]\varsigma\right),
\widetilde{\varphi}^{\xi'(\mp)}\left(\tau;[\pm]\varsigma\right)\right]=
(\pm 1)[\pm 1] \frac{1}{4\pi}\delta_{\xi, \xi'}
\left\{\ln \left(\cosh \left(\frac{\pi}{\varsigma}(s-\tau)\right)\right)
-f(\varsigma, \mu_2)\right\},
\label{pmt_DW} \\
&&\!\!\!\!\!\!\!\!\!\!\!\!\!\!\!\!\!\!
\left[\varphi^{\xi (\pm)} (s; [\pm]\varsigma), Q^{\xi' (\mp)}([\pm]\varsigma)\right]=
\delta_{\xi, \xi'}\left[\frac{i}{4}-
(\pm 1)\left(\frac{\widehat{x}^0-s}{2\varsigma}\right)\right],
\label{pm_ph_Q} \\
&&\!\!\!\!\!\!\!\!\!\!\!\!\!\!\!\!\!\!
\left[\widetilde{\varphi}^{\xi (\pm)} (s; [\pm]\varsigma),
\widetilde{Q}^{\xi'(\mp)}([\pm]\varsigma)\right] =
\delta_{\xi, \xi'}\left[-\,\frac{i}{4}-
(\pm 1)\left(\frac{\widehat{x}^0-s}{2\varsigma}\right)\right],
\label{pm_ph_Q_tt} \\
&&\!\!\!\!\!\!\!\!\!\!\!\!\!\!\!\!\!\!
\left[\varphi^{\xi(\pm)}(s;[\pm]\varsigma),
\widetilde{Q}^{\xi'(\mp)}([\pm]\varsigma)\right] = (\pm 1)[\pm 1]
\delta_{\xi, \xi'}\left(\frac{\widehat{x}^0-s}{2\varsigma}\right)=
\left[\widetilde{\varphi}^{\xi(\pm)}(s;[\pm]\varsigma),
Q^{\xi'(\mp)}([\pm]\varsigma)\right],
\label{pm__ph_Qt} \\
&&\!\!\!\!\!\!\!\!\!\!\!\!\!\!\!\!\!\!
\left[Q^{\xi (\pm)} ([\pm]\varsigma), Q^{\xi'
(\mp)}([\pm]\varsigma) \right] = (\pm 1) a_1 \delta_{\xi, \xi'}
= \left[\widetilde{Q}^{\xi (\pm)} ([\pm]\varsigma), \widetilde{Q}^{\xi'
(\mp)}([\pm]\varsigma) \right],
\label{QQ_QtQt} \\
&&\!\!\!\!\!\!\!\!\!\!\!\!\!\!\!\!\!\!
\left[Q^{\xi (\pm)} ([\pm]\varsigma), \widetilde{Q}^{\xi'
(\mp)}([\pm]\varsigma) \right] = (\pm 1)[\mp 1] a_2 \delta_{\xi, \xi'}
= \left[\widetilde{Q}^{\xi (\pm)} ([\pm]\varsigma), Q^{\xi'
(\mp)}([\pm]\varsigma) \right].
\label{QQ_Qt} 
\end{eqnarray}
Here the following quantities are defined (see Appendix B and C): 
\begin{eqnarray}
&&\!\!\!\!\!\!\!\!\!\!\!\!\!\!\!\!\!\!
g \left(\varsigma, \mu_1\right) = \int\limits_{\mu_1}^\infty
\frac{d k^1}{k^0} \frac{2}{e^{\varsigma k^0} - 1} \Longrightarrow
\frac{2}{\varsigma \mu_1} - \ln
\left(\frac{2\pi}{\varsigma \overline{\mu}_1}\right), \quad \overline{\mu}_1
= \mu_1 e^{C_\ni} \rightarrow 0, \quad \lim_{\varsigma \rightarrow \infty}
g \left(\varsigma, \mu_1\right) = 0,
\label{ggg} \\
&&\!\!\!\!\!\!\!\!\!\!\!\!\!\!\!\!\!\!
f (\varsigma, \mu_2) = \int\limits_{\mu_2}^\infty \frac{dk^1}{k^0}
\frac{1}{\sinh(\varsigma k^0/2)} \Longrightarrow \frac{2}{\varsigma\mu_2}-\ln 2,
\quad \mu_2 \rightarrow 0, \quad \lim_{\varsigma \rightarrow \infty}
f(\varsigma, \mu_2) = 0,
\label{fff} \\
&&\!\!\!\!\!\!\!\!\!\!\!\!\!\!\!\!\!\!
\delta_L \left(k^1\right)=
\int\limits_{-\infty}^\infty\frac{dx^1}{2\pi}\Delta\left(\frac{x^1}L\right)
e^{\pm ik^1x^1} \equiv L\overline{\Delta} \left(k^1 L\right),
\quad L\rightarrow\infty, \quad \lim_{L \rightarrow \infty} \delta_L \left(k^1\right)
= \delta \left(k^1\right), 
\label{DDddDD} \\
&&\!\!\!\!\!\!\!\!\!\!\!\!\!\!\!\!\!\!
a_0=a_0(L)=\pi\int\limits_0^\infty d k^1 k^1\left(\delta_L(k^1)\right)^2\Longrightarrow 
\pi \int\limits_0^\infty dt\, t\left(\overline{\Delta}(t)\right)^2\equiv
\pi I_1^\Delta, \quad 
I_n^\Delta\equiv\int\limits_0^\infty dt\, t^n \left(\overline{\Delta}(t)\right)^2,
\label{a_0_I_n} \\
&&\!\!\!\!\!\!\!\!\!\!\!\!\!\!\!\!\!\!
a_1=a_1(L,\varsigma)= a_0(L)+ 2\pi \int\limits_0^\infty d k^1 k^1 
\frac{\left(\delta_L(k^1)\right)^2}{e^{\varsigma k^0}-1}
=a_0(L) + 2 \pi \int\limits_0^\infty dt t
\frac{(\overline{\Delta}(t))^2}{e^{\varsigma t/L}-1}
\Longrightarrow
\label{a_1_00} \\
&&\!\!\!\!\!\!\!\!\!\!\!\!\!\!\!\!\!\!
\stackunder{L\to\infty}{\Longrightarrow}
2 \pi I_0^\Delta \frac{L}{\varsigma}
+ \frac{\pi}{6} I_2^\Delta \frac{\varsigma}{L}
+ O \left(\left(\frac{\varsigma}{L}\right)^3\right), \quad 
\lim_{\varsigma \rightarrow \infty} a_1(L,\varsigma)=a_0(L), 
\label{a_1_0} \\
&&\!\!\!\!\!\!\!\!\!\!\!\!\!\!\!\!\!\!
a_2 =a_2(L,\varsigma)= \pi \int\limits_0^\infty d k^1 k^1
\frac{\left(\delta_L(k^1)\right)^2}{\sinh(\varsigma k^0/2)}
= \pi \int\limits_0^\infty dt t
\frac{(\overline{\Delta}(t))^2}{\sinh \left(t\varsigma/2L\right)}\Longrightarrow
\label{a_2_0} \\
&&\!\!\!\!\!\!\!\!\!\!\!\!\!\!\!\!\!\!
\stackunder{L\to\infty}{\Longrightarrow} 
2\pi I_0^\Delta \frac{L}{\varsigma}
+ \left(\frac{\pi}{6}-\frac{\pi}{4}\right) I_2^\Delta \frac{\varsigma}{L} + O
\left(\left(\frac{\varsigma}{L}\right)^3\right), \quad 
\lim_{\varsigma \rightarrow \infty} a_2(L,\varsigma)=0, 
\label{a_2_}
\end{eqnarray}
where $C_\ni$ is again the Euler-Mascheroni constant, and appearance of additional 
infrared regulators $\mu_1, \mu_2$ \cite{abr,ks_ttt} is clarified in Appendix B. It is 
worth to note that for chosen general type of volume cut-off regularization 
(\ref{K_O}), (\ref{T_K_O}) with arbitrary appropriate even function $\Delta(x^1/L)$ 
(\ref{DDddDD}) the value of $a_0$ (\ref{D_a0}), (\ref{a_0_I_n}), if it exists 
(is finite), does not depend on $L$ at all, while the $a_1$ (\ref{a_1_00}) and $a_2$ 
(\ref{a_2_0}) in any case have the same divergent asymptotic behaviour (\ref{a_1_0}), 
(\ref{a_2_}) for $L\to\infty$, but, in general, have different behaviour at 
$\varsigma \rightarrow \infty$. Here the existence of $I_n^\Delta$ for $n=0,1,2$ is 
assumed, that for different regularizations are displayed in the Table of Appendix C. 
It is important to note that in any case the difference $a_1-a_2$ becomes $L$ - 
independent at $L\to\infty$, and if $a_0$ is finite, then $a_1-a_2\rightarrow 0$ at 
$L\to\infty$ (see Appendix C). 

The $(\widehat{x}^0-s)$ -dependence of commutators (\ref{pm_ph_Q})--(\ref{pm__ph_Qt}) 
has no physical meaning and below will be eliminated automatically. But it seems 
convenient to retain it for additional control up to the end of calculation, because 
further it suggests the way of correct doubling of the number of degrees of freedom.

Following \cite{blot}, by the use of the (pseudo) scalar fields given above, one
can again construct a representation of solutions of the Dirac equation for a free 
massless trial field at finite temperature, 
$\partial_\xi\chi_\xi \left(x^{-\xi},\varsigma\right)= 0$, in the form of local 
normal-ordered exponentials of the left and right bosonic thermofields 
$\varphi^{-\xi}(x^{-\xi},\varsigma)$, and their charges $ Q^\xi(\varsigma)$ 
(\ref{T_K_7}), (\ref{T_K_7_Q}). The naive expression, which implies the bosonization 
relations (\ref{T_nweyi19}) for the currents (\ref{T_bos-111})--(\ref{T_K_Z}) with 
$Z_{(\chi)}(a)=1$, is obtained from (\ref{nblaie12}), (\ref{u_xi}), 
(\ref{UAU})--(\ref{w_2}), (\ref{T_ppp}) as: 
\begin{eqnarray}
&&\!\!\!\!\!\!\!\!\!\!\!\!\!\!\!\!\!\!
{\cal V}_{\vartheta(B)}^{\mp 1}\chi_\xi(x^{-\xi}){\cal V}_{\vartheta(B)}^{\pm 1}=
{\cal N}_\varphi\left\{\exp\left(- i2\sqrt{\pi}\left[
\varphi^{-\xi} \left(x^{-\xi}; [\pm] \varsigma\right)+
\frac{\xi}{4}Q ^{\xi}([\pm]\varsigma)\right]\right)\right\}
u_\xi\left(\mu_1, \varsigma\right), 
\label{naiv_chi} \\
&&\!\!\!\!\!\!\!\!\!\!\!\!\!\!\!\!\!\!
u_\xi \left(\mu_1, \varsigma\right)=
\left(\frac{\overline{\mu}}{2\pi}\right)^{1/2}e^{i\varpi - i\xi\Theta/4}
\exp\left\{-\,\frac{g(\varsigma, \mu_1)}{2}\right\}
\exp\left\{-a_1(L)\frac{\pi}{8}\right\}, 
\label{naiv_u_xi}
\end{eqnarray}
where $\varpi$ and $\Theta$ are again arbitrary initial overall and relative phases. 
It is worth to note that to have a correct zero temperature limit $\varsigma\to\infty$ 
for this field, here it is necessary to keep finite the values of all infrared 
regulators: $\overline{\mu}, \mu_1,L, a_1(L)$.
However, the kinematic independence (\ref{T_vbnmei4}),  (\ref{tT_vbnmei4}), of the 
tilde-partners can be achieved now only by ``admixing'' the Klein factors of both the 
charges $\widetilde{Q}^\xi(\varsigma)$ and $\widetilde{Q}^{-\xi}(\varsigma)$ to the same 
field. Moreover, according to the meaning of $L$ as 
macroscopic parameter explained above, the wanted thermofield should have a correct 
thermodynamic limit $L\to\infty$ for the finite temperature $T>0$. To this end 
we define the new charges with simple commutation relations following from 
(\ref{pm_ph_Q})--(\ref{QQ_Qt}): 
\begin{eqnarray}
&&\!\!\!\!\!\!\!\!\!\!\!\!\!\!\!\!\!\!
{\rm G}^{\xi}([\pm]\varsigma)=Q^{\xi}([\pm]\varsigma)+[\pm 1]
\widetilde{Q}^{\xi}([\pm]\varsigma), \quad 
\widetilde{\rm G}^{\xi}([\pm]\varsigma)=\widetilde{Q}^{\xi}([\pm]\varsigma)+[\pm 1]
Q^{\xi}([\pm]\varsigma),\; \mbox { with:}
\label{G_Gt} \\
&&\!\!\!\!\!\!\!\!\!\!\!\!\!\!\!\!\!\!
\left[\varphi^{\xi(\pm)}(s;[\pm]\varsigma),{\rm G}^{\xi'(\mp)}([\pm]\varsigma)\right]=
\frac{i}{4}\,\delta_{\xi, \xi'}, \quad 
\left[\widetilde{\varphi}^{\xi(\pm)}(s;[\pm]\varsigma),
\widetilde{\rm G}^{\xi'(\mp)}([\pm]\varsigma)\right]=-\,\frac{i}{4}\,\delta_{\xi, \xi'}, 
\label{cc_phi_G_G} \\
&&\!\!\!\!\!\!\!\!\!\!\!\!\!\!\!\!\!\!
\left[\varphi^{\xi(\pm)}(s;[\pm]\varsigma),
\widetilde{\rm G}^{\xi'(\mp)}([\pm]\varsigma)\right]=
\frac{i}{4}\,[\pm 1]\delta_{\xi,\xi'}, 
\label{cc_phi_Gt} \\
&&\!\!\!\!\!\!\!\!\!\!\!\!\!\!\!\!\!\!
\left[{\rm G}^{\xi(\pm)}([\pm]\varsigma),{\rm G}^{\xi'(\mp)}([\pm]\varsigma)\right]=
(\pm 1)2(a_1-a_2)\delta_{\xi,\xi'}, 
\label{cc_phi_G_Gt} \\
&&\!\!\!\!\!\!\!\!\!\!\!\!\!\!\!\!\!\!
\left[{\rm G}^{\xi(\pm)}([\pm]\varsigma),
\widetilde{\rm G}^{\xi'(\mp)}([\pm]\varsigma)\right]=
(\pm 1)[\pm 1]2(a_1-a_2)\delta_{\xi,\xi'}. 
\label{cc_G_Gt}
\end{eqnarray}
Thus induced natural generalization, which, due to (\ref{d_xi_phi}), again gives 
nevertheless the bosonization relations (\ref{T_nweyi19}) for the currents 
(\ref{T_bos-111})--(\ref{T_K_Z}) of the trial physical fields $\chi(x; [\pm]\varsigma)$ 
with $Z_{(\chi)}(a)=1$, reads:
\begin{eqnarray}
&&\!\!\!\!\!\!\!\!\!\!\!\!\!\!\!\!\!\!
\chi_\xi (x^{-\xi}; [\pm]\varsigma)={\cal N}_\varphi
\left(\exp\left\{R_\xi (x^{-\xi};[\pm]\varsigma)\right\}\right) 
\widehat{\rm u}_\xi\left(\mu_1,[\pm ]\varsigma\right), 
\label{T_nblaie12} \\
&&\!\!\!\!\!\!\!\!\!\!\!\!\!\!\!\!\!\!
R_\xi (x^{-\xi}; [\pm]\varsigma)= - i2\sqrt{\pi}\left[
\varphi^{-\xi} \left(x^{-\xi}; [\pm] \varsigma\right)
+ \frac{1}{4} \sigma_0^\xi{\rm G}^{-\xi}([\pm] \varsigma)
+ \frac{1}{4} \sigma_1^\xi {\rm G}^{\xi}([\pm] \varsigma)\right],
\label{chi_chi} \\
&&\!\!\!\!\!\!\!\!\!\!\!\!\!\!\!\!\!\!
\widehat{\rm u}_\xi\left(\mu_1,[\pm]\varsigma\right)=
\left(\frac{\overline{\mu}}{2\pi}\right)^{1/2}e^{i\varpi - i\xi\Theta/4}
\exp\left\{-\,\frac{g(\varsigma,\mu_1)}{2}\right\}
\exp\left\{-\,\frac{\pi}4(a_1-a_2)\left[(\sigma^\xi_0)^2+(\sigma^\xi_1)^2\right]\right\}.
\label{T_nblaie13}
\end{eqnarray}
Independently the same expression for $\widehat{\rm u}_\xi$ together with 
admissible values of $\sigma_{0,1}^\xi$ follows from the anticommutation relations 
(\ref{vbnmei2})--(\ref{vbnmei4}), kinematic independence conditions (\ref{T_vbnmei4}), 
(\ref{tT_vbnmei4}), and symmetry under the above tilde - operation (\ref{OJ_5}), 
\cite{ojima}, as: 
\begin{eqnarray}
&&\!\!\!\!\!\!\!\!\!\!\!\!\!\!\!\!\!\!
\mbox{from }\;\left\{\chi,\chi^{\#}\right\}:\;\;
\frac{\sigma_1^\xi-\sigma_1^{-\xi}}2=\xi(2n_1+1), \;\;
\mbox { with arbitrary integer }\; n_0,\,n_1,\,n_2,\, n_3,  
\label{sigm_cond_chi} \\
&&\!\!\!\!\!\!\!\!\!\!\!\!\!\!\!\!\!\!
\mbox{from }\;\left\{\chi,\widetilde{\chi}^{\#}\right\}:\;\;
\frac{\sigma_1^\xi+\sigma_1^{-\xi}}2=2n_2+1,\;\mbox { and: }\;
\sigma_0^\xi=n_0-n_3+\xi(n_0+n_3+1).  
\label{sigm_cond_t_chi}
\end{eqnarray}
Contrary to (\ref{naiv_chi}) and \cite{abr}, thus chosen free field is a simple 
transformation of its zero temperature case furnished by all necessary tilde Klein 
factors and connected with the solution (\ref{nblaie12_0})--(\ref{u_xi}) by the 
following steps. The first one is the generalization to the two-parametric family 
(\ref{chi_r_sig}) of zero temperature solutions of free Dirac equation. Then, the 
second one is prompted by Eqs. (\ref{nblaie12_0}), (\ref{nblaie12}) as following. 

By virtue of the relations (\ref{VEV_0}), (\ref{UAU})--(\ref{w_2}), (\ref{sgm_Sgm_01}), 
(\ref{chi_r_sig})--(\ref{u_rho_sigm}), (\ref{pm_ph_Q})--(\ref{QQ_Qt}), (\ref{a_2_}), 
(\ref{G_Gt}), this suggests the simple admixing of the all necessary tilde Klein factors 
at zero temperature and directly leads to the field (\ref{T_nblaie12}) for zero 
temperature ($\varsigma=\infty$): 
\begin{eqnarray}
&&\!\!\!\!\!\!\!\!\!\!\!\!\!\!\!\!\!\!
\chi_\xi (x^{-\xi};[\pm]\infty)=\chi_\xi(x^{-\xi},\sigma,\rho)
\exp\left[-i[\pm 1]\frac{\sqrt{\pi}}{2}\sigma^\xi_0\widetilde{Q}^{-\xi}\right]
\exp\left[-i[\pm 1]\frac{\sqrt{\pi}}{2}\sigma^\xi_1\widetilde{Q}^{\xi}\right],\;
\mbox{ or:}
\label{xi_inf} \\
&&\!\!\!\!\!\!\!\!\!\!\!\!\!\!\!\!\!\!
\chi_\xi (x^{-\xi};[\pm]\infty)={\cal N}_\varphi
\left(\exp\left\{R_\xi (x^{-\xi};[\pm]\infty)\right\}\right) 
\widehat{\rm u}_\xi([\pm]\infty), \quad 
\sigma^\xi_0=- \xi\sigma, \quad \sigma^\xi_1=\xi 1+\rho,
\label{xi_inf_1} \\
&&\!\!\!\!\!\!\!\!\!\!\!\!\!\!\!\!\!\!
R_\xi (x^{-\xi};[\pm]\infty)=B_\xi(x^{-\xi},\sigma,\rho)
-i[\pm 1]\frac{\sqrt{\pi}}{2}\sigma^\xi_0\widetilde{Q}^{-\xi}
-i[\pm 1]\frac{\sqrt{\pi}}{2}\sigma^\xi_1\widetilde{Q}^{\xi}, 
\label{B_R_r_s} \\
&&\!\!\!\!\!\!\!\!\!\!\!\!\!\!\!\!\!\!
\widehat{\rm u}_\xi([\pm]\infty)=
\left(\frac{\overline{\mu}}{2\pi}\right)^{1/2}e^{i\varpi - i\xi\Theta/4}
\exp\left\{-a_0\frac{\pi}{4}\left[\left(\sigma^\xi_0\right)^2+
\left(\sigma^\xi_1\right)^2\right]\right\}=\lim\limits_{\varsigma\to\infty}
\widehat{\rm u}_\xi\left(\mu_1,[\pm]\varsigma\right),\;\mbox{so that:}
\label{hu_inf} \\
&&\!\!\!\!\!\!\!\!\!\!\!\!\!\!\!\!\!\!
R_\xi (x^{-\xi};[\pm]\varsigma)={\cal V}_{\vartheta(B)}^{\mp 1}
R_\xi (x^{-\xi};[\pm]\infty){\cal V}_{\vartheta(B)}^{\pm 1},\quad 
\chi_\xi (x^{-\xi};[\pm]\varsigma)={\cal V}_{\vartheta(B)}^{\mp 1}
\chi_\xi (x^{-\xi};[\pm]\infty){\cal V}_{\vartheta(B)}^{\pm 1}.
\label{chi_T_chi_inf}
\end{eqnarray}
This means, that in accordance with the fermionic thermofield transformation, which due 
to Eq. (\ref{OJ_5}), of course, gives automatically the field with all necessary 
anticommutation properties uniformly with continuous limit to zero temperature case 
(\ref{chi_xi_0}), (\ref{bar_chi_xi_0}) of Appendix E: 
\begin{eqnarray}
&&\!\!\!\!\!\!\!\!\!\!\!\!\!\!\!\!\!\!\!\!\!\!\!\!
\chi_\xi(x^{-\xi}; [\pm]\varsigma)=   
{\cal V}_{\vartheta(F)}^{\mp 1}\chi_\xi(x^{-\xi};[\pm]\infty)
{\cal V}_{\vartheta(F)}^{\pm 1}=
\label{X_T_pm} \\
&&\!\!\!\!\!\!\!\!\!\!\!\!\!\!\!\!\!\!\!\!\!\!\!\!
=\int\limits^\infty_{-\infty}\frac{dp^1}{\sqrt{2\pi}}
\left[\theta(\xi p^1){\rm b}(p^1; [\pm]\varsigma)e^{-i(px)}+
\xi\theta(\xi p^1){\rm f}^\dagger(p^1; [\pm]\varsigma)e^{i(px)}\right]
e^{i\varpi-i\xi\Theta/4}, 
\nonumber 
\end{eqnarray}
the bosonic field representation (\ref{T_nblaie12})--(\ref{T_nblaie13}) of fermi field 
also reproduce such commutation and continuity properties at $T\to 0$ due to 
the introduced Klein factors in Eqs. (\ref{xi_inf})--(\ref{B_R_r_s}), that accords with 
the necessity of doubling the number of degrees of freedom already at $T=0$ 
\cite{mtu,ojima}. 
From the conditions (\ref{sgm_Sgm_01}), (\ref{sigm_cond_chi}), (\ref{sigm_cond_t_chi}), 
(\ref{xi_inf_1}), we conclude that it is enough without loss of generality to take 
for $\xi=\pm $:
\begin{eqnarray}
&&\!\!\!\!\!\!\!\!\!\!\!\!\!\!\!\!\!\!
n_3=n_0=\ell,\quad \sigma=-(2\ell+1), \quad \sigma^\xi_0=-\xi\sigma;  \quad
n_1=0,\quad n_2=n, \quad \rho=2n+1, \quad \sigma^\xi_1=\xi 1+\rho.
\label{s_l_i_n_i} 
\end{eqnarray}
This induce inevitable additional $\xi$- dependence for c -number spinor 
$\widehat{\rm u}_\xi\left(\mu_1,[\pm]\varsigma\right)$ (\ref{T_nblaie13}), if 
$a_1-a_2\neq 0$. Indeed, the last exponential factor in (\ref{T_nblaie13}), if it differ 
from unity (as for the usual box, see (\ref{a1_a2}) and the Table in Appendix C), leads 
to non-physical regularization dependent ``temperature induced anomalous dimensions'', 
different for various components of this field. 

Nevertheless, since due to (\ref{a_1_0}), (\ref{a_2_}) and the discussion in Appendix C, 
excluding the case of usual box, for any continuous regularizations $\Delta(y^1/L)$: 
$a_0$ is finite and $a_1-a_2\Rightarrow 0$, for $L\to\infty$, then the c - number spinor 
(\ref{T_nblaie13}) is reduced to its most 
simple ``Oksak's'' form, which directly has a correct zero temperature behavior for Oksak 
case $a_0=0$ in (\ref{u_xi}), \cite{blot,oksak} (see Appendix C): 
\begin{eqnarray} 
&&\!\!\!\!\!\!\!\!\!\!\!\!\!\!\!\!\!\!\!\!\!\!\!\!
\widehat{\rm u}_\xi\left(\mu_1,[\pm]\varsigma\right)\Longrightarrow 
\left(\frac{\overline{\mu}}{2\pi}\right)^{1/2}e^{i\varpi - i\xi\Theta/4}
\exp\left\{-\,\frac{g(\varsigma,\mu_1)}{2}\right\}=
\widehat{\rm u}^{Ok}_\xi\left(\mu_1,[\pm]\varsigma\right),\quad 
\lim_{\varsigma\to\infty}\widehat{\rm u}^{Ok}_\xi\left(\mu_1,[\pm]\varsigma\right)=
u^{Ok}_\xi. 
\label{u_u} 
\end{eqnarray}

\subsection{Integration of the Heisenberg equation}

For the chosen representation (\ref{T_nweyi19})--(\ref{T_nblaie13})
the operator product in the linearized by means of
(\ref{T_ns94m61}), (\ref{T_ns94m62}) HEq (\ref{T_45bn6l4}) or
(\ref{T_45bn6l5}) is naturally redefined again into the normal-ordered
form \cite{blot} with respect to the fields $\varphi^\xi(x^\xi,\varsigma)$:
\begin{eqnarray}
\partial_0 \Psi_\xi (x,\varsigma)=\left(-\xi
\partial_1-i\frac{\beta g}{2\sqrt{\pi}}
\widehat{J}_{(\chi)}^{-\xi(-)} (x,\varsigma)\right)
\Psi_\xi (x,\varsigma)-
\Psi_\xi(x,\varsigma)\left(i\frac{\beta g}{2\sqrt{\pi}}
\widehat{J}_{(\chi)}^{-\xi(+)}(x,\varsigma)\right).
\label{T_nweyi13}
\end{eqnarray}
As above, the famous expression for the derivative of function 
$F\left(x^1\right)$ in terms of the operator $P^1$: 
$-i\partial_1 F(x^1)=\left[P^1,F(x^1)\right]$ and its finite-shift equivalent: 
$e^{i a P^1}F(x^1)e^{-iaP^1}=F(x^1+a)$ allows to transcribe the equation
(\ref{T_nweyi13}) for $x^0=t$, $\Psi_\xi(x,\varsigma)\longleftrightarrow Y(t)$, as 
follows:
\begin{eqnarray}
\frac {d}{dt}Y(t) = A(t)Y(t)-Y(t)B(t),
\label{T_nweyi15}  
\end{eqnarray}
and to have again the following formal solution in terms of the time-ordered exponential:
\begin{eqnarray}
Y(t) = {T}_A \left\{\exp\left(\int\limits_0^t d\tau
A(\tau)\right)\right\}Y(0)
\left[{T}_B \left\{\exp\left(\int\limits_0^t d\tau
B(\tau)\right) \right\}\right]^{-1},
\label{T_nweyi16}
\end{eqnarray}
which in this case is immediately transformed to the usual one for 
$Y(t)\rightarrow\Psi_\xi (x,\varsigma)$: 
\begin{eqnarray}
\Psi_\xi (x,\varsigma)
= e^{C^{\xi(-)}(x,\varsigma)}\Psi_\xi\left(x^1-\xi x^0, 0,\varsigma\right)
e^{C^{\xi(+)}(x,\varsigma)}.
\label{T_nweyi21}
\end{eqnarray}
Using the operator bosonization (\ref{T_nweyi19}) for the vector
current of the trial physical field (\ref{T_nblaie12}), we find as above:
\begin{eqnarray}
&&\!\!\!\!\!\!\!\!\!\!\!\!\!\!\!\!\!\!\!\!
C^{\xi(\pm)} (x,\varsigma)=-i\frac{\beta g}{2\sqrt{\pi}}\int\limits_0^{x^0}d y^0
\widehat{J}_{(\chi)}^{-\xi(\pm)}\left(x^1+\xi y^0 -\xi x^0, y^0,\varsigma\right)=
\label{T_nweyi22}  \\
&&\!\!\!\!\!\!\!\!\!\!\!\!\!\!\!\!\!\!\!\!
=-i\frac{\beta g}{2\pi}\left[\varphi^{(\pm)}\left(x^1,x^0,\varsigma\right)-
\varphi^{(\pm)}\left(x^1-\xi x^0,0,\varsigma\right)\right]
= -i\frac{\beta g}{2\pi}\left[\varphi^{\xi(\pm)}\left(x^\xi,\varsigma\right)-
\varphi^{\xi(\pm)}\left(-x^{-\xi},\varsigma\right)\right].
\nonumber
\end{eqnarray}
Remarkably, that the completely unknown ``initial'' HF 
$\Psi_\xi(x^1-\xi x^0,0,\varsigma)=\lambda_\xi(x^{-\xi},\varsigma)$ appears here
also as a solution of free massless Dirac equation,
$\partial_\xi\lambda_\xi(x^{-\xi},\varsigma)= 0$, but certainly again 
unitarily inequivalent to the free field $\chi(x,\varsigma)$
(\ref{T_nblaie12}). The expressions  (\ref{T_nweyi21}),
(\ref{T_nweyi22}) suggest to choose it also in the normal-ordered
form with respect to the field $\varphi$, using appropriate
``bosonic canonical transformation'' of this field with
parameters $\overline{\alpha}=2\sqrt{\pi}\cosh\eta$,
$\overline{\beta}=2\sqrt{\pi}\sinh\eta$, connected by
$\overline{\alpha}^2-\overline{\beta}^2 = 4\pi$, which is generated now by the operator 
$F_\eta(\varsigma)$ (for $y^0=x^0$) in the form 
$U_\eta(\varsigma) =\exp F_\eta (\varsigma)$, 
which as above (\ref{intro_009}), in fact does not depend on $\xi$ and $x^0$ at all:
\begin{eqnarray}
&&\!\!\!\!\!\!\!\!\!\!\!\!\!\!\!\!\!\!\!\!
U^{-1}_\eta(\varsigma)\varphi(x,\varsigma)U_\eta(\varsigma)
=\omega(x,\varsigma)\equiv
\omega^\xi(x^\xi,\varsigma)+\omega^{-\xi}(x^{-\xi},\varsigma)=
\frac{1}{2\sqrt{\pi}}\Bigl[\overline{\alpha}\varphi(x^1,x^0,\varsigma)+
\overline{\beta}\varphi(x^1,-x^0,\varsigma)\Bigr],
\label{T_om_U} \\
&&\!\!\!\!\!\!\!\!\!\!\!\!\!\!\!\!\!\!\!\!
U^{-1}_\eta(\varsigma)\phi(x,\varsigma)U_\eta(\varsigma)=\Omega(x,\varsigma)\equiv
\xi\left(\omega^{-\xi}(x^{-\xi},\varsigma)-\omega^\xi(x^\xi,\varsigma)\right)=
\frac{1}{2\sqrt{\pi}}\Bigl[\overline{\alpha}\phi(x^1,x^0,\varsigma)-
\overline{\beta}\phi(x^1,-x^0,\varsigma)\Bigr], 
\label{T_oOm_U} \\
&&\!\!\!\!\!\!\!\!\!\!\!\!\!\!\!\!\!\!\!\!
U^{-1}_\eta(\varsigma)\varphi^\xi(x^\xi,\varsigma)U_\eta(\varsigma)
=\omega^\xi(x^\xi,\varsigma)=
\frac{1}{2\sqrt{\pi}}\left[\overline{\alpha}\varphi^\xi(x^\xi,\varsigma)+
\overline{\beta}\varphi^{-\xi}(-x^\xi,\varsigma)\right]\equiv
{\cal V}_{\vartheta(B)}^{\mp 1}\omega^\xi(x^\xi){\cal V}_{\vartheta(B)}^{\pm 1},
\label{T_intro_007} \\
&&\!\!\!\!\!\!\!\!\!\!\!\!\!\!\!\!\!\!\!\!
U^{-1}_\eta(\varsigma)Q^\xi([\pm]\varsigma) U_\eta (\varsigma)=
{\cal W}^\xi([\pm]\varsigma)=
\frac{1}{2\sqrt{\pi}}\left[\overline{\alpha}Q^\xi([\pm]\varsigma)-
\overline{\beta}Q^{-\xi}([\pm]\varsigma)\right]\equiv
{\cal V}_{\vartheta(B)}^{\mp 1}{\cal W}^\xi{\cal V}_{\vartheta(B)}^{\pm 1},
\label{T_intro_008}\\
&&\!\!\!\!\!\!\!\!\!\!\!\!\!\!\!\!\!\!\!\!
U^{- 1}_\eta (\varsigma){\rm G}^\xi([\pm]\varsigma)U_\eta (\varsigma)= 
{\cal G}^\xi([\pm]\varsigma)=
\frac{1}{2\sqrt{\pi}}\left[\overline{\alpha}{\rm G}^\xi([\pm]\varsigma)-
\overline{\beta}{\rm G}^{-\xi}([\pm]\varsigma)\right]=
{\cal W}^\xi([\pm]\varsigma)+[\pm 1]\widetilde{\cal W}^\xi([\pm]\varsigma),
\label{cG_G_W}\\
&&\!\!\!\!\!\!\!\!\!\!\!\!\!\!\!\!\!\!\!\!
F_\eta ([\pm]\varsigma)=
{\cal V}_{\vartheta(B)}^{\mp 1}\vec{\rm F}_\eta[c(k^1)]{\cal V}_{\vartheta(B)}^{\pm 1}+
{\cal V}_{\vartheta(B)}^{\mp 1}\widetilde{\vec{\rm F}}_{\eta}[\widetilde{c}(k^1)]
{\cal V}_{\vartheta(B)}^{\pm 1}=
\vec{\rm F}_\eta\left[c\left(k^1;[\pm]\varsigma\right)\right]+
\widetilde{\vec{\rm F}}_{\eta}\left[\widetilde{c}\left(k^1;[\pm]\varsigma\right)\right]
\equiv
\label{T_intro_0_c09} \\
&&\!\!\!\!\!\!\!\!\!\!\!\!\!\!\!\!\!\!\!\!
\equiv\eta \int\limits_{-\infty}^\infty \frac{d k^1 \theta(k^1)}{2\pi 2 k^0}
\left[c\left(k^1;[\pm]\varsigma\right) c\left(-k^1; [\pm]\varsigma\right)-
c^\dagger\left(k^1;[\pm]\varsigma\right) 
c^\dagger\left(-k^1;[\pm]\varsigma\right)\right]+
\nonumber  \\
&&\!\!\!\!\!\!\!\!\!\!\!\!\!\!\!\!\!\!\!\!
+\eta\int\limits_{-\infty}^\infty\frac{d k^1 \theta(k^1)}{2\pi 2k^0}
\left[\widetilde{c}\left(k^1;[\pm]\varsigma\right)
\widetilde{c}\left(-k^1;[\pm]\varsigma\right)-
\widetilde{c}^\dagger\left(k^1;[\pm]\varsigma\right)
\widetilde{c}^\dagger\left(-k^1;[\pm]\varsigma\right)\right],\; \mbox{ or:}
\label{T_intro_0_cc09} \\
&&\!\!\!\!\!\!\!\!\!\!\!\!\!\!\!\!\!\!\!\!
F_\eta (\varsigma) =
2i\eta\int\limits_{-\infty}^\infty d y^1
\varphi^\xi\left(y^\xi,\varsigma\right)
\partial_0\varphi^{-\xi}\left(-y^\xi,\varsigma\right)-
2i\eta\int\limits_{-\infty}^\infty d y^1
\widetilde\varphi^\xi\left(y^\xi,\varsigma\right)
\partial_0\widetilde\varphi^{-\xi}\left(-y^\xi,\varsigma\right)=
\label{T_in_009}\\
&&\!\!\!\!\!\!\!\!\!\!\!\!\!\!\!\!\!\!\!\!
= 2i\eta\int\limits_{-\infty}^\infty d y^1
\omega^\xi\left(y^\xi,\varsigma\right)
\partial_0\omega^{-\xi}\left(-y^\xi,\varsigma\right)-
2i\eta\int\limits_{-\infty}^\infty d y^1
\widetilde\omega^\xi\left(y^\xi,\varsigma\right)
\partial_0\widetilde\omega^{-\xi}\left(-y^\xi,\varsigma\right), \; \mbox{ with:}
\label{T_intro_009} \\
&&\!\!\!\!\!\!\!\!\!\!\!\!\!\!\!\!\!\!\!\!
\left[\varphi^{\xi(\pm)}(s,\varsigma),F_\eta(\varsigma)\right]
=\eta\,\varphi^{-\xi(\mp)}(-s,\varsigma), \quad 
\left[Q^{\xi(\pm)}(\widehat{x}^0,\varsigma),F_\eta(\varsigma)\right]=
-\eta\,Q^{-\xi(\mp)}(-\widehat{x}^0,\varsigma),\quad \widehat{x}^0\Rightarrow 0. 
\label{ph_T_F}
\end{eqnarray}
As shown in Appendix D, here the second tilde-conjugate term in formulas 
(\ref{T_intro_0_c09})--(\ref{T_intro_009}) is very important, providing 
$\widetilde{F}_\eta(\varsigma)=F_\eta(\varsigma)$. 
For the field $\lambda_\xi(x^{-\xi},\varsigma)$ by the same way as above 
(\ref{UAU})--(\ref{M_2}), keeping in mind the relations (\ref{ID_ID}), 
(\ref{v_xi_r_S})--(\ref{sgm_Sgm_01}), one obtains, with: 
$\Sigma^\xi_0=\overline{\alpha}\sigma^\xi_0-\overline{\beta}\sigma^\xi_1$,  
$\Sigma^\xi_1=\overline{\alpha}\sigma^\xi_1-\overline{\beta}\sigma^\xi_0$, that:   
\begin{eqnarray}
&&\!\!\!\!\!\!\!\!\!\!\!\!\!\!\!\!\!\!\!\!
\lambda_\xi(x^{-\xi},\varsigma)=
U^{-1}_\eta(\varsigma)\chi_\xi(x^{- \xi};\varsigma)U_\eta (\varsigma)=
\lambda_\xi (x^{-\xi}; [\pm]\varsigma)=
{\cal N}_\varphi \left(\exp\left\{ {\cal R}_\xi(x^{-\xi};[\pm]\varsigma)\right\}\right) 
w_\xi(\mu_1,\varsigma), 
\label{T_M_1_2} \\
&&\!\!\!\!\!\!\!\!\!\!\!\!\!\!\!\!\!\!\!\!
{\cal R}_\xi (x^{-\xi};[\pm]\varsigma)=- i2\sqrt{\pi}
\left[\omega^{-\xi}(x^{-\xi}; [\pm]\varsigma)+ 
\frac{\sigma_0^\xi}{4}{\cal G}^{-\xi}([\pm]\varsigma)+
\frac{\sigma_1^\xi}{4}{\cal G}^{\xi}([\pm]\varsigma)\right], \;\mbox{ or:}
\label{T_lambd_W} \\
&&\!\!\!\!\!\!\!\!\!\!\!\!\!\!\!\!\!\!\!\!
{\cal R}_\xi (x^{-\xi};[\pm]\varsigma)=
- i\left[2\sqrt{\pi}\omega^{-\xi}(x^{-\xi};[\pm]\varsigma)+ 
\frac{\Sigma_0^\xi}{4}{\rm G}^{-\xi}([\pm]\varsigma)+
\frac{\Sigma_1^\xi}{4}{\rm G}^{\xi}([\pm]\varsigma)\right],
\label{T_lambd_W_1} \\
&&\!\!\!\!\!\!\!\!\!\!\!\!\!\!\!\!\!\!\!\!
w_\xi(\mu_1,\varsigma)=\left(\frac{\overline{\mu}}{2\pi}\right)^{1/2}
\left(\frac{\overline{\mu}}{\Lambda}\right)^{\overline{\beta}^2/4\pi }
e^{i\varpi- i \xi\Theta/4}
\exp\left\{-g(\varsigma,\mu_1)
\left(\frac{1}{2}+\frac{\overline{\beta}^2}{4\pi}\right)\right\}
\nonumber \\
&&\!\!\!\!\!\!\!\!\!\!\!\!\!\!\!\!\!\!\!\!
\cdot \exp\left\{-(a_1-a_2)\frac{\pi}{4}\left(
\left[(\sigma^\xi_0)^2+(\sigma^\xi_1)^2\right]\cosh 2\eta-
2\sigma^\xi_0\sigma^\xi_1\sinh 2\eta\right)\right\}.
\label{exp_s_s}
\end{eqnarray}
The CAR (\ref{Ps_d_Z}) may be easy verified for this field with 
$Z_{(\lambda)}(a)$ (\ref{J_L}). 
For the corresponding current $\widehat{J}_{(\lambda)}^\mu(x,\varsigma)$, defined by 
Eqs. (\ref{T_bos-111})--(\ref{T_K_Z}), or by the Johnson definition \cite{Jon,s_w,wai}, 
with the same zero temperature renormalization constant 
$Z_{(\lambda)}(a)$ (\ref{J_L}), one finds again the previous bosonization rules 
(\ref{T_nweyi19}):
\begin{eqnarray}
\widehat{J}_{(\lambda)}^\mu(x,\varsigma)=
\frac{1}{\sqrt{\pi}}\partial^\mu\omega(x,\varsigma)=
-\,\frac{1}{\sqrt{\pi}}\epsilon^{\mu\nu}\partial_\nu\Omega(x,\varsigma),
\label{T_J_L}
\end{eqnarray}
onto the new scalar fields $\omega(x,\varsigma)$, $\Omega(x,\varsigma)$, 
$\omega^\xi(x^\xi,\varsigma)$, ${\cal W}^{\xi} ([\pm] \varsigma)$, obey the same 
commutation relations, as initial fields $\varphi^\xi(x^\xi,\varsigma)$ 
(\ref{T_K_8_1})--(\ref{T_K_9}), (\ref{T_C_til}):
\begin{eqnarray}
&&\!\!\!\!\!\!\!\!\!\!\!\!\!\!\!\!\!\!
\left[\omega^\xi\left(s,\varsigma\right),\omega^{\xi'}
\left(\tau,\varsigma\right)\right]=
-\frac{i}{4}\varepsilon(s - \tau)\delta_{\xi, \xi'}, \quad
\left[\omega^\xi(s,\varsigma), {\cal W}^{\xi'}(\varsigma)\right]=
\frac{i}{2}\delta_{\xi,\xi'},
\label{T_K_om_9} \\
&&\!\!\!\!\!\!\!\!\!\!\!\!\!\!\!\!\!\!
\left[\omega^\xi\left(s,\varsigma\right),\widetilde{\omega}^{\xi'}
\left(\tau,\varsigma\right)\right]=0, \quad
\left[\omega^\xi(s,\varsigma), \widetilde{\cal W}^{\xi'}(\varsigma)\right]=0, \quad
\left[{\cal W}^{\xi}(\varsigma), \widetilde{\cal W}^{\xi'}(\varsigma)\right]=0. 
\label{T_C_om_til}
\end{eqnarray}
Substituting the normal form (\ref{T_M_1_2}) into the solution (\ref{T_nweyi21}), 
by imposing the same conditions (\ref{nweyi32}) onto the parameters 
$\overline{\alpha},\overline{\beta}$,  that are
necessary to have correct Lorentz-transformation properties
corresponding to the spin $1/2$, and correct canonical
anticommutation relations, we again immediately obtain the normal exponential of
the DM for Thirring field in the form, analogous to Oksak solution 
(\ref{Psi_varho}), \cite{blot,oksak}, where the condition 
$\overline{\beta}=\beta g/(2\pi)$ from (\ref{nweyi32}) again replaces the thermofield 
$\omega^{-\xi}(x^{-\xi},\varsigma)$ by thermofield $\varrho^{-\xi}(x;\varsigma)$, 
and the use of the charges (\ref{G_Gt})--(\ref{cc_G_Gt}) assures a correct doubling of 
the number of degrees of freedom:
\begin{eqnarray}
&&\!\!\!\!\!\!\!\!\!\!\!\!\!\!\!\!\!\!
\Psi_\xi (x; [\pm]\varsigma)={\cal N}_\varphi\left(\exp\left\{
\Re_\xi(x;[\pm]\varsigma)\right\}\right) w_\xi\left(\mu_1,\varsigma\right), 
\label{T_nweyi29} \\
&&\!\!\!\!\!\!\!\!\!\!\!\!\!\!\!\!\!\!
\Re_\xi(x;[\pm]\varsigma)=-i2\sqrt{\pi}\left[\varrho^{-\xi}(x;[\pm]\varsigma)+ 
\frac{\sigma_0^\xi}{4}{\cal G}^{-\xi}([\pm]\varsigma)+
\frac{\sigma_1^\xi}{4}{\cal G}^{\xi}([\pm]\varsigma)\right],\; \mbox{ or:}
\label{T_Psi_Psi} \\
&&\!\!\!\!\!\!\!\!\!\!\!\!\!\!\!\!\!\!
\Re_\xi(x;[\pm]\varsigma)=-i\left[2\sqrt{\pi}\,\varrho^{-\xi}(x;[\pm]\varsigma)+ 
\frac{\Sigma_0^\xi}{4}{\rm G}^{-\xi}([\pm]\varsigma)+
\frac{\Sigma_1^\xi}{4}{\rm G}^{\xi}([\pm]\varsigma)\right], 
\label{T_Psi_Psi_1} \\
&&\!\!\!\!\!\!\!\!\!\!\!\!\!\!\!\!\!\!
2\sqrt{\pi}\,\varrho^{-\xi}(x;[\pm]\varsigma)=
\overline{\alpha}\varphi^{-\xi}(x^{-\xi};[\pm]\varsigma))+
\overline{\beta}\varphi^{\xi}(x^\xi;[\pm]\varsigma)), 
\label{T_Psi_vrho} \\
&&\!\!\!\!\!\!\!\!\!\!\!\!\!\!\!\!\!\!
\mbox{with: }\;
\Sigma^\xi_0=\overline{\alpha}\sigma^\xi_0-\overline{\beta}\sigma^\xi_1, \quad 
\Sigma^\xi_1=\overline{\alpha}\sigma^\xi_1-\overline{\beta}\sigma^\xi_0, 
\label{Sgm_nn} \\
&&\!\!\!\!\!\!\!\!\!\!\!\!\!\!\!\!\!\!
\mbox {and: }\;
\sigma^\xi_0=-\xi\sigma \Rightarrow \xi(2\ell+1), \quad 
\sigma^\xi_1=\xi 1+\rho \Rightarrow \xi 1+(2n+1).
\label{sgm_n_n} 
\end{eqnarray}
Straightforward calculation of the current operators (\ref{T_bos-111})--(\ref{T_K_Z}) 
with the same current's (\ref{K_ZaZx}) and field's (\ref{ZLa_Z}) renormalization 
constant $Z_{(\Psi)}(a)$, arisen from the same short-distance behavior of Wightman 
functions (\ref{A_d_mu-}), reproduces the CAR (\ref{Ps_d_Z}) and bosonization 
relations (\ref{T_ns94m61})--(\ref{T_nweyi19}) as following weak equalities:
\begin{eqnarray}
&&\!\!\!\!\!\!\!\!\!\!\!\!\!\!\!\!\!\!
\widehat{J}_{(\Psi)}^\nu(x,\varsigma)\stackrel{\rm w}{=}
-\,\frac{\beta}{2\pi}\,\epsilon^{\mu\nu}\partial_\nu\phi(x, \varsigma)=
\frac{\beta}{2\sqrt{\pi}}\widehat{J}_{(\chi)}^\nu(x,\varsigma),\;
\mbox{ for: }\; Z_{(\chi)}(a)=1,\quad 
Z_{(\Psi)}(a)=(-\Lambda^2 a^2)^{-{\overline{\beta}^2}/{4\pi}},  
\label{T_K_Za} 
\end{eqnarray}
where $\beta$ is again defined by Eqs. (\ref{Kab}). And vice versa, this bosonization 
relation and CAR (\ref{Ps_d_Z}) separately imply, similarly to the free case 
(\ref{T_nblaie12})--(\ref{T_nblaie13}), the one and the same expression (\ref{exp_s_s}) 
for the c - number spinor $w_\xi\left(\mu_1,\varsigma\right)$ in terms of arbitrary 
parameters $\sigma,\, \rho$ and $a_{1,2}$, with the same $Z_{(\Psi)}(a)$ (\ref{K_ZaZx}), 
(\ref{T_K_Za}). 

Remarkably, that the CAR (\ref{vbnmei2}), (\ref{Ps_d_Z}), jointly with locality 
(\ref{vbnmei3}), (\ref{vbnmei4}) and kinematic independence conditions (\ref{tT_vbnmei4}) 
give for the fermionic fields $\chi(x,[\pm]\varsigma)$, $\lambda(x,[\pm]\varsigma)$, 
$\Psi(x,[\pm]\varsigma)$ separately the above one and the same relations 
(\ref{sigm_cond_chi}), (\ref{sigm_cond_t_chi}) for the parameters $\sigma_{0,1}^\xi$ 
simultaneously, leading to their final values (\ref{sgm_n_n}). 

The same calculation as above leads to the same values of the Johnson commutators 
(\ref{J_Psi_D})--(\ref{J_33}) and the corresponding charge algebras 
(\ref{Q_J_Pchi})--(\ref{Q5_J_Pchi}). 

For the non-mixed VEV of the strings of these fields (\ref{T_nweyi29}), following to 
(\ref{Psi_ppp})--(\ref{LL_LL}), by means of the formulae (\ref{x_xi_AB})--(\ref{A_D_-}) 
from Appendix B, one can obtain, again denoting here 
$l_i\Re_{\xi_i}(x_i;[\pm]\varsigma)\longmapsto {\cal R}_i$: 
\begin{eqnarray}
&&\!\!\!\!\!\!\!\!\!\!\!\!\!\!\!\!\!\!
\left(\Lambda^{\overline{\beta}^2/4\pi}\sqrt{2\pi}\right)^p
\left\langle 0\left|\prod\limits_{i=1}^p
\Psi_{\xi_i}^{(l_i)}\left(x_i;[\pm]\varsigma\right)\right|0 \right\rangle=
\left\langle 0\left|{\cal N}_\varphi\left\{\exp\left(\sum\limits_{j=1}^p
{\cal R}_j\right)\right\}\right|0\right\rangle 
\exp\left\{i\varpi\sum\limits_{i=1}^p l_i - i \frac{\Theta}{4}
\sum \limits_{i=1}^p l_i \xi_i\right\} 
\nonumber \\
&&\!\!\!\!\!\!\!\!\!\!\!\!\!\!\!\!\!\!
\cdot \left(\overline{\mu}\exp\left\{-g\left(\varsigma,\mu_1\right)- 
(a_1-a_2)\frac{\pi}{2}\left[(1+\sigma)^2+\left(\beta^2/4\pi\right)^2\rho^2\right]
\right\}\right)^{\left(\sum\limits_{i=1}^p l_i\right)^2(\pi/\beta^2)}
\nonumber \\
&&\!\!\!\!\!\!\!\!\!\!\!\!\!\!\!\!\!\!
\cdot \left(\overline{\mu}\exp\left\{-g\left(\varsigma,\mu_1\right)-
(a_1-a_2)\frac{\pi}{2}\left[(1-\sigma)^2+\left(4\pi/\beta^2\right)^2\rho^2\right]
\right\}\right)^{\left(\sum\limits_{i=1}^p l_i\xi_i\right)^2 (\beta^2/16\pi)}
\nonumber \\
&&\!\!\!\!\!\!\!\!\!\!\!\!\!\!\!\!\!\!
\cdot \exp\left\{-(a_1-a_2)\frac{\pi}{2}\rho\left[\cosh 2\eta+\sigma\sinh 2\eta \right]
\left(\sum\limits_{i=1}^p l_i\xi_i\right)\left(\sum\limits_{j=1}^p l_j\right)\right\}
\nonumber \\
&&\!\!\!\!\!\!\!\!\!\!\!\!\!\!\!\!\!\!
\cdot
\prod\limits_{i<k}^p \left\{e^{i \pi(\xi_i-\xi_k)}
\left[\frac{\displaystyle\sinh\left(\frac{\pi}{\varsigma}(x_i^{-}-x_k^{-}-i0)\right)}
{\displaystyle\sinh\left(\frac{\pi}{\varsigma}(x_i^{+}-x_k^{+}-i0)\right)}\right]
^{\xi_i+\xi_k}
\right.
\nonumber \\
&&\!\!\!\!\!\!\!\!\!\!\!\!\!\!\!\!\!\!
\left. 
\cdot \left[\left(\frac{i\varsigma}{\pi}\right)^2\sinh \left(\frac{\pi}{\varsigma}
(x_i^{-}-x_k^{-}-i0)\right)
\sinh \left(\frac{\pi}{\varsigma}(x_i^{+}-x_k^{+}-i0)\right)\right]
^{(4\pi/\beta^2)+\xi_i\xi_k(\beta^2/4\pi)} \right\}^{l_i l_k/4}. 
\label{VV_Tss} 
\end{eqnarray}
The changing of the factor $a_0\pi/4$ to $(a_1-a_2)\pi/2$ on comparing with 
(\ref{VV_rho}) corresponds to doubling of the number of degree of freedom. 
Only under both superselection rules (\ref{ssr_1}), (\ref{ssr_2}) the expression 
(\ref{VV_Tss}) simplifies to:
\begin{eqnarray}
&&\!\!\!\!\!\!\!\!\!\!\!\!\!\!\!\!\!\!
\left(\Lambda^{\overline{\beta}^2/4\pi}\sqrt{2\pi}\right)^p\!
\left\langle 0\left|\prod\limits_{i=1}^p
\Psi_{\xi_i}^{(l_i)}\left(x_i;[\pm]\varsigma\right)\right|0 \right\rangle = 
\delta_{\sum\limits_{i=1}^p l_i,0}\,\delta_{\sum\limits_{i=1}^p l_i\xi_i,0}
\nonumber  \\
&&\!\!\!\!\!\!\!\!\!\!\!\!\!\!\!\!\!\!\!\!
\nonumber \\
&&\!\!\!\!\!\!\!\!\!\!\!\!\!\!\!\!\!\!\!\!
\cdot\prod\limits_{i<k}^p\left\{e^{i\pi(\xi_i-\xi_k)}
\left[
\frac{\displaystyle\sinh\left(\frac{\pi}{\varsigma}(x_i^{-}-x_k^{-}-i0)\right)}
{\displaystyle\sinh\left(\frac{\pi}{\varsigma}(x_i^{+}-x_k^{+}-i0)\right)}
\right]^{\xi_i+\xi_k} 
\right.
\nonumber \\
&&\!\!\!\!\!\!\!\!\!\!\!\!\!\!\!\!\!\!\!\!
\left. 
\cdot \left[\left(\frac{i\varsigma}{\pi}\right)^2
\sinh\left(\frac{\pi}{\varsigma}(x_i^{-}-x_k^{-}-i0)\right)
\sinh\left(\frac{\pi}{\varsigma}(x_i^{+}-x_k^{+}-i0)\right)
\right]^{(4\pi/\beta^2)+\xi_i\xi_k(\beta^2/4\pi)}\right\}^{l_i l_k/4}. 
\label{VV_T} 
\end{eqnarray}
Thus, in accordance with \cite{abr}, only both zero temperature superselection rules 
(\ref{ssr_1}), (\ref{ssr_2}) assure again the elimination of all the old and new infrared 
divergences, regularized by parameters $\overline{\mu}$, $L$, $\overline{\mu}_1$, 
$a_{1,2}$, and  elimination of all dependencies on parameters $\sigma$, $\rho$, $\varpi$, 
$\Theta$, so that only the last two lines of Eqs. (\ref{VV_Tss}), or (\ref{VV_T}) 
survive again independently of the volume cut-off regularization function from the 
Table of Appendix C, and ultraviolet renormalization stays only necessary again. 
The zero temperature limit of Eq. (\ref{VV_T}) 
evidently gives the last lines of expressions (\ref{VV_rho}), or (\ref{Ps_p_3}), 
independently on type of the volume cut-off regularization was used. 
The mixed VEV of the HF with their tilde-partner reveal the similar properties. 

The Eqs. (\ref{M_2}), (\ref{exp_s_s}) easy show, that, for $a_1-a_2\Rightarrow 0$, with 
$L\to\infty$, e.g., for all continuous regularizations, both 
the old and new infrared divergences have one and the same character, given, as for zero 
temperature case, by the dynamical dimension ${\rm d}_{(\Psi)}$ (\ref{Sc_t}), 
(\ref{ID_ID}), and again, as for the above free case (\ref{u_u}), recasts the 
c- number spinor $w_\xi (\mu_1,\varsigma)$ (\ref{exp_s_s}) into its most simple 
``Oksak's'' form, which has a correct zero temperature behavior for Oksak case $a_0=0$ 
in Eq. (\ref{M_2}): 
\begin{eqnarray}
&&\!\!\!\!\!\!\!\!\!\!\!\!\!\!\!\!\!\!
w_\xi (\mu_1, \varsigma)\stackunder{L\to\infty}{\Longrightarrow} 
\left(\frac{\Lambda}{2\pi}\right)^{1/2}e^{i\varpi-i\xi\Theta/4} 
\exp\left\{\left(\frac{1}{2}+\frac{\overline{\beta}^2}{4\pi}\right)
\left[\ln\left(\frac{\overline{\mu}}{\Lambda}\right)-\,g(\varsigma,\mu_1)\right]\right\}
= w^{Ok}_\xi (\mu_1, \varsigma), 
\label{T_nweyi30_1} \\
&&\!\!\!\!\!\!\!\!\!\!\!\!\!\!\!\!\!\!
\lim_{\varsigma\to\infty} w^{Ok}_\xi\left(\mu_1,\varsigma\right)=v^{Ok}_\xi
\equiv v_\xi\bigr|_{a_0=0}.
\label{w_v} 
\end{eqnarray}
However for the usual box (see Appendix C) the spinor $w_\xi (\mu_1,\varsigma)$, as well 
as the free one (\ref{T_nblaie13}), due to the last exponential factor in 
(\ref{exp_s_s}) furnished by the relation (\ref{a1_a2}), also acquires additional 
multiplier of ``temperature induced 
anomalous dimension''. The multiplicatively renormalized field may be achieved now only 
for the field of Morchio et al. (\ref{MPS}), (\ref{v_Mor_r}) with discrete values 
of coupling constant, which follows from (\ref{s_l_i_n_i}), and 
recasts the last exponential $e^{\{...\}}$ of Eq. (\ref{exp_s_s}) to the common 
multiplier:  
\begin{eqnarray}
&&\!\!\!\!\!\!\!\!\!\!\!\!\!\!\!\!\!\!
-\sigma=(2\ell+1)\Rightarrow\coth 2\eta, \quad \frac g{\pi}=\sqrt{1+\frac 1{\ell}}-1,\;
\mbox { whence: }\;
 e^{\{...\}}\longmapsto\left(\frac{\pi}{\varsigma\Lambda}\right)^{{\cal M}(\ell,n)}, 
\label{g_n_0} \\
&&\!\!\!\!\!\!\!\!\!\!\!\!\!\!\!\!\!\!
\mbox {with: }\; 
{\cal M}(\ell,n)=\frac{(2\ell+1)}{16\sqrt{\ell(\ell+1)}}
\left[(2\ell+1)^2+(2n+1)^2-1\right].
\label{Anom_n_n}
\end{eqnarray}
As was discussed above, such a possibility of multiplicative renormalization is absent 
for the case of free field (\ref{T_nblaie12})--(\ref{T_nblaie13}), which formally 
corresponds here to limit $\ell\to\infty$. Thus, so obtained non-perturbative 
solution looks as artifact of the use of the usual box for the charge regularization 
\cite{sa_wp}.  

\section{Conclusion}

The main lesson of our work is very simple: the correct true HF should be only a fully 
normal ordered operator in the sense of DM onto irreducible physical fields. 
Only this form clarifies and assures correct renormalization, commutation and symmetry 
properties. It allows also a simple connections between different types of solutions with 
finite and zero temperature. 

Contrary to the recent works \cite{fab-iva}-\cite{fab-iva_2},\cite{sa_wp},
\cite{abr}-\cite{abr3}, we consider different types of charge's regularization and 
take into account all possible mutual commutation relations of bosonic thermofields and 
their charges. We reveal that their non-physical $(\widehat{x}^0-s)$ - dependence 
fixes the correct doubling of the number of degrees of freedom, and thus 
self-consistently eliminates from the normal forms of the free and Heisenberg fields 
operators and from the VEV of their products. 

The canonical transformations we found, mutually connect the different solutions by two 
additional parameters, which being arbitrary continuous for zero temperature solution, 
acquire only discrete nonzero values for the finite temperature solutions. 
The conditions (\ref{sgm_n_n}) obtained for these parameters provide the anticommutation, 
locality and kinematic independence relations for both the free and Thirring fields and 
their tilde partners simultaneously. 

We show that integration of HEqs by means of  the linearization procedure and
dynamical mapping onto the Schr\"odinger fields, - with generalized initial condition at 
$t=x^0=0$, $\lim\limits_{t\to 0}\Psi(x^1,t)\stackrel{\rm w}{=}\Upsilon[\psi_s(x^1,0)]$, 
is relevant also for the finite temperature case. 
The observed weak linearization (\ref{ns94m61}), (\ref{T_ns94m61}) of HEqs with 
so-generalized initial conditions in a weak sense allows to overcome the restrictions of 
Haag theorem, removing them into the representation construction of Schr\"odinger 
physical fields, at first, as reducible massless free Dirac fields:  
$\chi(x),\, \chi(x,\varsigma)$, and then, as irreducible massless (pseudo) scalar 
fields: $\phi(x),\, \phi(x,\varsigma)$. The latter ones arisen as Schr\"odinger 
physical fields, in fact play the role of asymptotic ones. Due to automatical elimination 
of zero mode's contributions, the chosen here representation space of free massless 
pseudoscalar field relaxes the problem of non-positivity of its inner product induced by 
Wightman functions. 

Within the thermofield dynamics formalism \cite{mtu,ojima} it is shown that 
for thus exactly linearizable and exactly solvable Thirring model at finite temperature 
the bosonization relations retain their operator sense at finite temperature only among 
the free fields operators. For the Heisenberg currents these bosonization rules are 
applicable only in a weak sense. 

The general solutions for HF (\ref{nweyi21}), (\ref{T_nweyi21}) keeps the Klaiber's 
normal form \cite{klaib}, but with distinct unitarily inequivalent representation of the 
free massless Dirac field sandwiched the simple dynamical factors and generated by 
distinct unitarily inequivalent representation of free massless (pseudo) scalar field. 
The zero temperature limit of thermal solution gives two-parametric generalization 
of the known Oksak solution \cite{blot,oksak}. 

The notion of ``hot'' and ``cold'' thermofields is found to be convenient to distinguish 
different thermofield representations giving the correct normal form of thermofield 
solution for finite temperature Thirring model with respect to different vacua. 
The new vacua always appear as infinite products of coherent states with respect to 
initial vacua of elementary field's oscillators. 

We show that very popular volume cut-off regularization  by usual box of length 
$L$, \cite{sa_wp}, which is crucial for the use of Bethe Ansatz method 
\cite{fujita,fujita_2}, when it is used for the charge definition, leads to 
non-physical properties of temperature solutions. 
While, for any continuous charge regularization function excluding the usual box, we 
found one and the same consistent properties of the field solutions for both zero and the 
finite temperature cases. Moreover, the latter case is independent of any type 
of continuous regularization at the corresponding thermodynamic limit $L\to\infty$. 
The non-mixed $n$ - point's VEV are independent of any regularization  at 
all, as well as on any non-physical parameters, if and only if the both superselection 
rules are fulfilled. And only for this case the thermodynamic limit $L\to\infty$ and the 
zero temperature limit $\varsigma\to\infty$ may be successfully interchanged for these 
Wightman functions with the one and the same result. 

We thank Y. Frishman, A.N. Vall, S.V. Lovtsov, V.M. Leviant, A.E. Kaloshin, 
N.V. Iljin and participants of seminar of the LTPh JINR for useful discussions, 
and D. Taychenachev for checking of some integrals.

This work was supported in part by the RFBR (project N
09-02-00749) and by the program ``Development of Scientific
Potential in Higher Schools'' (project N 2.2.1.1/1483,
2.1.1/1539).

\section{Appendix}

\subsection{Appendix A}

For arbitrary numbers $\lambda$, $u$, $v$, the following generalization of the results 
of \cite{kirj} may be obtained by means of the calculation of the values 
$\left(u\partial{\rm g}/\partial u\pm v\partial{\rm g}/\partial v\right){\rm g}^{-1}$, 
for ${\rm g}={\rm g}(u,v)$, with closed algebra of operators: 
\begin{eqnarray}
&&\!\!\!\!\!\!\!\!\!\!\!\!\!\!\!\!\!\!\!\!
[A,B] = 2\lambda C, \quad [C,A]=A,\quad [C,B]=-B,\quad 
{\rm g}(u,v)=\exp\left(uA+vB\right)\;:
\label{kir_1} \\
&&\!\!\!\!\!\!\!\!\!\!\!\!\!\!\!\!\!\!\!\!
{\rm g}(u,v)=
\exp\left\{\sqrt{\frac{u}{\lambda v}}\tanh\left(\sqrt{\lambda uv}\right)A\right\}
\exp\left\{-2\ln\left(\cosh\left(\sqrt{\lambda uv}\right)\right)C\right\}
\exp\left\{\sqrt{\frac{v}{\lambda u}}\tanh\left(\sqrt{\lambda uv}\right)B\right\}, 
\label{kir_2} 
\end{eqnarray}
where the main branches of analytic functions $\sqrt{z}$ and $\ln z$ are assumed. By 
using this formula, the operator $U^{-1}_\eta$ (\ref{intro_008}), (\ref{Fcc_009}) of 
Bogoliubov transformation transcribes with the help of operators ${\cal K}_{\pm,0}$ 
and/or $K_{\pm,0}(k^1)$, forming the algebras of the ``big'' and ``small'' groups 
$SU(1,1)$ (here the length of {\bf usual box} is $L$): 
\begin{eqnarray}
&&\!\!\!\!\!\!\!\!\!\!\!\!\!\!\!\!\!\!\!\!
{\cal K}_\pm=\int\limits^\infty_{-\infty}\frac{dk^1\theta(k^1)}{2\pi 2k^0}\left\{
\begin{array}{c} c^\dagger(k^1)c^\dagger(-k^1) \\ c(-k^1)c(k^1)\end{array}\right\}\equiv 
\frac{L}{2\pi}\int\limits^\infty_{-\infty} dk^1 K_\pm(k^1)
\stackunder{L\to\infty}{\Longleftarrow}\sum\limits^\infty_{k^1_n=-\infty}\!\!
K_\pm(k^1_n),\;\;\; k^1_n=\frac{2\pi n}L, 
\label{cK_pm} \\
&&\!\!\!\!\!\!\!\!\!\!\!\!\!\!\!\!\!\!\!\!
{\cal K}_0=\frac 12 \int\limits^\infty_{-\infty}\frac{dk^1\theta(k^1)}{2\pi 2k^0}\left(
c^\dagger(k^1)c(k^1)+c(-k^1)c^\dagger(-k^1)\right)\equiv
\frac{L}{2\pi}\int\limits^\infty_{-\infty} dk^1 K_0(k^1)
\stackunder{L\to\infty}{\Longleftarrow}\sum\limits^\infty_{k^1_n=-\infty}\!\!
K_0(k^1_n)=
\label{cK_0} \\
&&\!\!\!\!\!\!\!\!\!\!\!\!\!\!\!\!\!\!\!\!
=\frac 12 \int\limits^\infty_{-\infty}dk^1\theta(k^1)\left(
\frac{c^\dagger(k^1)c(k^1)+c^\dagger(-k^1)c(-k^1)}{2\pi 2k^0}+\delta(0)\right),\quad 
\delta(k^1-q^1)\Rightarrow\frac{L}{2\pi}\delta_{k^1,q^1},\quad 
\delta(0)\Rightarrow\frac{L}{2\pi},
\label{cK_00} \\
&&\!\!\!\!\!\!\!\!\!\!\!\!\!\!\!\!\!\!\!\!
\left[K_-(k^1),K_+(q^1)\right]=2K_0(k^1)\delta_{k^1,q^1}, \quad 
\left[K_0(k^1),K_\pm(q^1)\right]=\pm K_\pm(k^1)\delta_{k^1,q^1},\quad 
K_-(k^1)=\left\{K_+(k^1)\right\}^\dagger,
\label{crK_K_K} \\
&&\!\!\!\!\!\!\!\!\!\!\!\!\!\!\!\!\!\!\!\!
\left[{\cal K}_-,{\cal K}_+\right]=2{\cal K}_0,\quad 
\left[{\cal K}_0,{\cal K}_\pm\right]=\pm {\cal K}_\pm,\quad 
{\cal K}_-=\left\{{\cal K}_+\right\}^\dagger\!, \quad 
{\cal C}_2={\cal K}^2_0-{\cal K}_0-{\cal K}_+{\cal K}_-\Rightarrow
\ell(\ell-1)\widehat{I},\;\mbox{ as:}
\label{crKKK} \\
&&\!\!\!\!\!\!\!\!\!\!\!\!\!\!\!\!\!\!\!\!
U^{-1}_{\eta}=\exp\left\{-F_\eta\right\}\equiv
\exp\left\{\eta\left[{\cal K}_+-{\cal K}_-\right]\right\}
\stackunder{L\to\infty}{\Longleftarrow}
\prod\limits^\infty_{k^1_n=-\infty}\exp\left\{\eta
\left[K_+(k^1_n)-K_-(k^1_n)\right]\right\},\;\mbox{ for:}
\label{U_cK} \\
&&\!\!\!\!\!\!\!\!\!\!\!\!\!\!\!\!\!\!\!\!
|0\rangle\stackunder{L\to\infty}{\Longleftarrow}
\prod\limits^\infty_{k^1_n=-\infty}|0_{k^1_n}\rangle,\;\mbox{ or: }\;
U^{-1}_{\eta}=
\exp\left\{\tanh\eta\,{\cal K}_+\right\}
\exp\left\{-2\ln(\cosh\eta)\,{\cal K}_0\right\}
\exp\left\{-\tanh\eta\,{\cal K}_-\right\}.
\label{U_cKet} \\
&&\!\!\!\!\!\!\!\!\!\!\!\!\!\!\!\!\!\!\!\!
\mbox{With: }\;\widehat{\rm n}(k^1)=\frac{c^\dagger(k^1)c(k^1)}{2k^0},\quad 
K_0(k^1)=\frac{\theta(k^1)}{2}\left[1+
\frac{\widehat{\rm n}(k^1)+\widehat{\rm n}(-k^1)}{L}\right],\;
\mbox{ the Casimir operator are:}
\label{cCaz} \\
&&\!\!\!\!\!\!\!\!\!\!\!\!\!\!\!\!\!\!\!\!
C_2(k^1)= K^2_0(k^1)-K_0(k^1)-K_+(k^1)K_-(k^1)=
\frac{\theta(k^1)}{4}\left[-1+
\left(\frac{\widehat{\rm n}(k^1)-\widehat{\rm n}(-k^1)}{L}\right)^2\right]\Rightarrow
\kappa(\kappa-1)\widehat{I}.
\label{Caz} 
\end{eqnarray}
Here $\theta(k^1)\widehat{\rm n}(\pm k^1)/L$ are the density operators of right and 
left moving pseudoscalar particles with momentum $k^1$, $\widehat{I}$- unite operator. 
As a representation states of the ``small'' and ``big'' groups, the initial vacua for 
every oscillation mode $k^1_n$, $c(k^1_n)|0_{k^1_n}\rangle=0$ and the total vacuum 
(\ref{U_cKet}) have the quantum numbers $\kappa$, $\nu=\kappa+m$, $m$ - integer, and 
$\ell, N=\ell+m$, correspondingly, where for the ``small'' group: 
\begin{eqnarray}
&&\!\!\!\!\!\!\!\!\!\!\!\!\!\!\!\!\!\!\!\!
C_2(k^1)|\kappa,\nu\rangle=\kappa(\kappa-1)|\kappa,\nu\rangle, \quad 
K_0(k^1)|\kappa,\nu\rangle=\nu|\kappa,\nu\rangle, \quad 
|0_{k^1}\rangle\Rightarrow \left|\frac 12,\frac 12 \right\rangle, \quad 
\nu\Rightarrow\kappa=\frac 12, 
\label{vac_kn} \\
&&\!\!\!\!\!\!\!\!\!\!\!\!\!\!\!\!\!\!\!\!
\mbox{whereas for the ``big'' one: }\;
{\cal C}_2|\ell,N\rangle\!\rangle=\ell(\ell-1)|\ell,N\rangle\!\rangle,\quad 
{\cal K}_0|\ell,N\rangle\!\rangle=N |\ell,N\rangle\!\rangle,\quad 
|0\rangle\Rightarrow |\ell,\ell \rangle\!\rangle,
\label{vac_KN} \\
&&\!\!\!\!\!\!\!\!\!\!\!\!\!\!\!\!\!\!\!\!
\mbox{for: }\;N\Rightarrow \ell=\frac{L}{4\pi\rho},\;\mbox{ with: }\;
\frac 1{2\pi\rho}=\frac{\Lambda}{2\pi}=\int\limits^\Lambda_0\frac{dk^1}{2\pi}=
\int\limits^\infty_0\frac{dk^1}{2\pi}e^{-k^1/\Lambda},\;\mbox{ so that: }\;
|0\rangle\stackunder{\ell\to\infty}{\Longleftarrow}
\prod\limits^{2\ell}_{n=-2\ell}|0_{k^1_n}\rangle, 
\label{Lam_L} \\
&&\!\!\!\!\!\!\!\!\!\!\!\!\!\!\!\!\!\!\!\!
2\pi\rho\;\mbox{-- is the effective excitation volume for one mode, and }\;
2\ell=\frac{L}{2\pi\rho}\;\mbox{ is a number of excitations}
\label{L_ell}
\end{eqnarray}
that may be inserted in the system volume $L$ without overlap. The total new vacuum state 
$|\widehat{0}\rangle$ (\ref{N_vac}) is obviously a coherent state (\ref{U_cK}), 
(\ref{U_cKet}), \cite{perelom} for the discrete series representation of ``big'' group 
$SU(1,1)$ over the initial total vacuum $|0\rangle$, as infinite product (\ref{U_cK}),  
(\ref{Lam_L}) of coherent states over one mode vacua $|0_{k^1_n}\rangle$ for the 
``small'' groups for  $\ell\to\infty$:  
\begin{eqnarray}
&&\!\!\!\!\!\!\!\!\!\!\!\!\!\!\!\!\!\!\!\!
|\widehat{0}\rangle=U^{-1}_\eta|0\rangle\Longleftarrow (\cosh\eta)^{-2\ell}
\exp\left\{\tanh\eta\,{\cal K}_+\right\}|0\rangle, 
\label{Coher} \\
&&\!\!\!\!\!\!\!\!\!\!\!\!\!\!\!\!\!\!\!\!
|\widehat{0}\rangle=\left(1-\tanh^2\eta\right)^\ell\sum\limits^\infty_{m=0}
\left[\frac{\Gamma(m+2\ell)}{m!\Gamma(2\ell)}\right]^{1/2}\tanh^m\eta
|\ell,\ell+m \rangle\!\rangle,\quad \ell\to\infty, 
\label{Coher_big} \\
&&\!\!\!\!\!\!\!\!\!\!\!\!\!\!\!\!\!\!\!\!
|\ell,\ell+m \rangle\!\rangle=
\left[\frac{\Gamma(2\ell)}{m!\Gamma(m+2\ell)}\right]^{1/2} \left({\cal K}_+\right)^m
|\ell,\ell\rangle\!\rangle,
\label{discr_ser} \\
&&\!\!\!\!\!\!\!\!\!\!\!\!\!\!\!\!\!\!\!\!
\langle 0|\widehat{0}\rangle=\langle 0|U^{-1}_\eta|0\rangle=(\cosh\eta)^{-2\ell}=
\left(1-\tanh^2\eta\right)^\ell \stackunder{\ell\to\infty}{\longrightarrow}0, \;
\mbox{ for }\;\eta\neq 0, 
\label{vac_h_vac}
\end{eqnarray}
what means the orthogonality of these states for $\ell\to\infty$ and unitary 
inequivalence of corresponding representations. 
For the thermal Bogoliubov transformation (\ref{V_V_V}), (\ref{cosh_sinh}) the 
corresponding value reads: 
\begin{eqnarray}
&&\!\!\!\!\!\!\!\!\!\!\!\!\!\!\!\!\!\!\!\!
\langle 0\widetilde{0}|0(\varsigma)\rangle=
\langle 0\widetilde{0}|{\cal V}_{\vartheta(B)}^{-1}|0\widetilde{0}\rangle=
\exp\left\{-\,\delta(0)\int\limits_{-\infty}^{+\infty}
d k^1\ln\left(\cosh\vartheta(k^1,\varsigma)\right)\right\}\Longrightarrow
\label{V_V0} \\
&&\!\!\!\!\!\!\!\!\!\!\!\!\!\!\!\!\!\!\!\!
\Longrightarrow \exp\left\{-\,\frac{\delta(0)}{\varsigma}\frac{\pi^2}6\right\}
\Rightarrow \exp\left\{-\,\frac{L}{\varsigma}\frac{\pi}{12}\right\}
\stackunder{L\to\infty}{\longrightarrow}0,  
\label{V_L_vs}
\end{eqnarray}
also demonstrating the unitary inequivalence of corresponding representations 
\cite{mtu}. 

\subsection{Appendix B}

The Wightman functions for zero and nonzero temperature (\ref{nblaie16}) and 
(\ref{hot_DW}), (\ref{pm_DW}) admit the useful representations with different 
real parts, but with the same imaginary part, so that for: $\xi=\pm$, $k^0=|k^1|$, 
\begin{eqnarray}
&&\!\!\!\!\!\!\!\!\!\!\!\!\!\!\!\!\!\!
z^\xi=x^\xi-y^\xi, \quad z^\xi=z^0+\xi z^1,\quad 
z^+z^-=(z^0)^2-(z^1)^2\equiv z^2,\;\mbox{ and for any }\;A,B,{\cal F}(z)\; \mbox{ with}
\label{z_xi_x} \\
&&\!\!\!\!\!\!\!\!\!\!\!\!\!\!\!\!\!\!\!\!
\left({\cal F}(z^{-\xi})\right)^A\left({\cal F}(z^{\xi})\right)^B=
\biggl({\cal F}(z^-)\biggr)^{(A+B)/2+\xi(A-B)/2}
\biggl({\cal F}(z^+)\biggr)^{(A+B)/2-\xi(A-B)/2}=
\label{x_xi_AB} \\
&&\!\!\!\!\!\!\!\!\!\!\!\!\!\!\!\!\!\!\!\!
=\biggl({\cal F}(z^-){\cal F}(z^+)\biggr)^{(A+B)/2}
\left(\frac{{\cal F}(z^-)}{{\cal F}(z^+)}\right)^{\xi(A-B)/2},\;
\mbox{ one has it as following:}
\label{x_xi_AB_} \\
&&\!\!\!\!\!\!\!\!\!\!\!\!\!\!\!\!\!\!\!\!
\left[\varphi^{\xi(+)}(x^\xi),\varphi^{\xi'(-)}(y^{\xi'})\right]=
\frac{\delta_{\xi,\xi'}}{i} D^{(-)}(z^\xi),
\label{phi_x_phi_y} \\
&&\!\!\!\!\!\!\!\!\!\!\!\!\!\!\!\!\!\!\!\!
\frac{1}{i} D^{(-)}(z^\xi)=
\frac{1}{4\pi}\int\limits^\infty_\mu\frac{d\lambda}{\lambda}e^{-i\lambda(z^\xi-i0)}=
\frac{1}{4\pi} \int\limits^\infty_{i\mu(z^\xi-i0)}e^{-t}d(\ln t)=
-\frac{1}{4\pi}\ln\left(i\overline{\mu}\left\{z^\xi-i0\right\}\right)=
\label{D-_ln}\\
&&\!\!\!\!\!\!\!\!\!\!\!\!\!\!\!\!\!\!
=-\frac{1}{4\pi}\ln\left|\overline{\mu}z^\xi\right|-\frac i8 \varepsilon (z^\xi)=
 - \frac{1}{8\pi}\left[\ln|\overline{\mu}^2 z^2|
+ \xi \ln \left|\frac{z^{+}}{z^{-}}\right|\right]
- \frac{i}{8}\biggl[\varepsilon \left(z^0\right) \theta \left(z^2\right)
+ \xi \varepsilon \left(z^1\right) \theta \left(-z^2\right)\biggl],
\label{A_D_-} \\
&&\!\!\!\!\!\!\!\!\!\!\!\!\!\!\!\!\!\!
\left[\varphi^{\xi(+)}\left(x^\xi; [\pm]\varsigma\right), 
\varphi^{\xi'(-)}\left(y^{\xi'}; [\pm]\varsigma\right)\right]=
\frac{\delta_{\xi,\xi'}}{4\pi}\int\limits_{-\infty}^\infty\frac{d k^1}{k^0} 
\theta(-\xi k^1)\left[\cosh^2\vartheta\,e^{-i k^0z^\xi} + 
\sinh^2\vartheta\,e^{i k^0z^\xi}\right]=
\label{cD_z} \\
&&\!\!\!\!\!\!\!\!\!\!\!\!\!\!\!\!\!\!
=\frac{1}{4\pi}\delta_{\xi,\xi'}\left\{
\int\limits_\mu^\infty\frac{d k^1}{k^0}e^{-ik^0(z^\xi-i0)}+
\int\limits_0^\infty\frac{d k^1}{k^0}\,
\frac{2}{e^{\varsigma k^0}-1}\left[\cos(k^0 z^\xi)-1\right]+
\int\limits_{\mu_1}^\infty\frac{d k^1}{k^0}\,\frac{2}{e^{\varsigma k^0}-1}\right\}
\equiv
\label{cD_int} \\
&&\!\!\!\!\!\!\!\!\!\!\!\!\!\!\!\!\!\!
\equiv
\frac{\delta_{\xi,\xi'}}{i}{\cal D}^{(-)}(z^\xi,\varsigma;\mu_1)
\equiv -\,\frac{1}{4\pi} \delta_{\xi, \xi'} \left\{\ln \left(i\overline{\mu} 
\frac{\varsigma}{\pi}\sinh \left(\frac{\pi}{\varsigma}
(z^\xi-i0)\right)\right)-g \left(\varsigma, \mu_1\right)\right\}=
\label{cD_} \\
&&\!\!\!\!\!\!\!\!\!\!\!\!\!\!\!\!\!\!
=-\,\frac{1}{4\pi}\delta_{\xi, \xi'}
\left\{\ln\left|\overline{\mu}\frac{\varsigma}{\pi}\sinh \left(\frac{\pi}{\varsigma}
z^\xi\right)\right|+i\frac{\pi}2\varepsilon\left(z^\xi\right)-g(\varsigma,\mu_1)\right\},
\; \mbox{ or:}
\label{cD_ln_e} \\
&&\!\!\!\!\!\!\!\!\!\!\!\!\!\!\!\!\!\!
\frac{1}{i}{\cal D}^{(-)} (z^\xi;\varsigma;\mu_1)
= - \frac{1}{8\pi}\left[\ln \left|\left(\overline{\mu}\frac{\varsigma}{\pi}\right)^2
\sinh \left(\frac{\pi}{\varsigma}z^{+}\right)
\sinh \left(\frac{\pi}{\varsigma}z^{-}\right)\right|
+ \xi \ln \left|\frac{\sinh \left(\pi z^{+}/\varsigma\right)}
{\sinh\left(\pi z^{-}/\varsigma\right)}\right|
-2g \left(\varsigma, \mu_1\right)\right]-
\nonumber \\
&&\!\!\!\!\!\!\!\!\!\!\!\!\!\!\!\!\!\!
- \frac{i}{8} \biggl[\varepsilon \left(z^0\right) \theta \left(z^2\right)
+ \xi \varepsilon \left(z^1\right) \theta \left(-z^2\right)\biggl]\Longrightarrow 
-\frac{1}{4\pi}\left[\ln\left(i\overline{\mu}\left\{z^\xi-i0\right\}\right)
-g(\varsigma,\mu_1)\right],\;\mbox{ for: }\;z^\xi\rightarrow 0, 
\label{A_d_mu-} \\
&&\!\!\!\!\!\!\!\!\!\!\!\!\!\!\!\!\!\!
\mbox{where: }\;\varepsilon\left(z^\xi\right)=
\varepsilon\left(z^0\right)\theta\left(z^2\right)+
\xi\varepsilon\left(z^1\right)\theta\left(-z^2\right), \quad 
\overline{\mu}=\mu e^{C_\ni},\quad {C_\ni}=-\int\limits^\infty_{0}dt\,e^{-t} \ln t\, ,
\label{e_E_e} \\
&&\!\!\!\!\!\!\!\!\!\!\!\!\!\!\!\!\!\!
\mbox{and: }\;\cosh^2\vartheta=\frac{1}{1-e^{-\varsigma k^0}}=1+\sinh^2\vartheta, 
\quad \sinh^2\vartheta=\frac{1}{e^{\varsigma k^0}-1}, \quad 
\vartheta=\vartheta(k^1,\varsigma),\;\mbox{ so that:}
\label{cosh_sinh} \\
&&\!\!\!\!\!\!\!\!\!\!\!\!\!\!\!\!\!\!
\left[\varphi^{\xi(\pm)}(x^\xi;[\pm]\varsigma),
\widetilde{\varphi}^{\xi'(\mp)}(y^{\xi'};[\pm]\varsigma)\right]=(\pm 1)[\mp 1]
\frac{1}{2\pi}\delta_{\xi,\xi'}\int\limits_0^\infty\frac{d k^1}{k^0}
\cosh\vartheta\sinh\vartheta\cos(k^0 z^\xi)=
\label{phi_t_phi} \\
&&\!\!\!\!\!\!\!\!\!\!\!\!\!\!\!\!\!\!
=(\pm 1)[\pm 1]\frac{1}{4\pi}\delta_{\xi,\xi'}\left\{
\int\limits_0^\infty \frac{dk^1}{k^0}\,\frac{1}{\sinh\left(\varsigma k^0/2\right)} 
\left[1-\cos(k^0 z^\xi)\right]-
\int\limits_{\mu_2}^\infty \frac{dk^1}{k^0}\,\frac{1}{\sinh\left(\varsigma k^0/2\right)}
\right\}=
\label{ln_ch_ch} \\
&&\!\!\!\!\!\!\!\!\!\!\!\!\!\!\!\!\!\!
=(\pm 1)[\pm 1]\frac{1}{4\pi}\delta_{\xi,\xi'}
\left\{\ln\left(\cosh\left(\frac{\pi}{\varsigma}z^\xi\right)\right)
-f(\varsigma,\mu_2)\right\},
\label{ln_ch_f} \\
&&\!\!\!\!\!\!\!\!\!\!\!\!\!\!\!\!\!\!
\left[\varphi^{\xi(\pm)}(s;[\pm]\varsigma),Q^{\xi'(\mp)}([\pm]\varsigma)\right]=
\delta_{\xi,\xi'}\frac i2\!\!\int\limits^\infty_{-\infty}\!\!dk^1 
\theta(-\xi k^1)\left[e^{(\pm i)k^0(\widehat{x}^0-s)}\cosh^2\vartheta-
\right.
\nonumber  \\
&&\!\!\!\!\!\!\!\!\!\!\!\!\!\!\!\!\!\!
\left. 
-e^{(\mp i)k^0(\widehat{x}^0-s)}\sinh^2\vartheta\right]\delta_L(k^1)=
\delta_{\xi,\xi'}\frac i2\!\!\int\limits^\infty_{-\infty}\!\!dk^1 
\theta(-\xi k^1)\left[\cos\left(k^0(\widehat{x}^0-s)\right)\left(\cosh^2\vartheta-
\sinh^2\vartheta\right)+
\right.
\nonumber  \\
&&\!\!\!\!\!\!\!\!\!\!\!\!\!\!\!\!\!\!
\left. 
+(\pm 1)i \sin\left(k^0(\widehat{x}^0-s)\right)
\left(\cosh^2\vartheta+\sinh^2\vartheta\right)\right]\delta_L(k^1)=
\nonumber  \\
&&\!\!\!\!\!\!\!\!\!\!\!\!\!\!\!\!\!\!
=\delta_{\xi, \xi'}\frac{i}4\!\!\int\limits^\infty_{-\infty}\!\!dk^1
\left[\cos\left(k^0(\widehat{x}^0-s)\right)+(\pm 1)i\sin\left(k^0(\widehat{x}^0-s)\right)
\coth\left(\varsigma k^0/2\right)\right]\delta_L(k^1)
\stackunder{L\to\infty}{\Longrightarrow}
\nonumber  \\
&&\!\!\!\!\!\!\!\!\!\!\!\!\!\!\!\!\!\!
\stackunder{L\to\infty}{\Longrightarrow}
\delta_{\xi, \xi'}\left[\frac{i}{4}-
(\pm 1)\left(\frac{\widehat{x}^0-s}{2\varsigma}\right)\right],
\label{cc_phi_Q_1} \\
&&\!\!\!\!\!\!\!\!\!\!\!\!\!\!\!\!\!\!
\left[\varphi^{\xi(\pm)}(s;[\pm]\varsigma),
\widetilde{Q}^{\xi'(\mp)}([\pm]\varsigma)\right]=
(\pm 1)[\pm 1]\delta_{\xi, \xi'}\!\!\int\limits^\infty_{-\infty}\!\!
dk^1 \theta(-\xi k^1) 
\cosh\vartheta\sinh\vartheta\sin\left(k^0(\widehat{x}^0-s)\right)\delta_L(k^1)=
\label{cc_phi_tQ_0} \\
&&\!\!\!\!\!\!\!\!\!\!\!\!\!\!\!\!\!\!
=(\pm 1)[\pm 1]\frac{\delta_{\xi, \xi'}}2
\int\limits^\infty_0 dk^1\frac{\sin\left(k^0\left(\widehat{x}^0-s\right)\right)}
{\sinh\left(\varsigma k^0/2\right)}\,\delta_L(k^1)
\stackunder{L\to\infty}{\Longrightarrow}
(\pm 1)[\pm 1] \delta_{\xi, \xi'}\left(\frac{\widehat{x}^0-s}{2\varsigma}\right),
\label{cc_phi_tQ} 
\end{eqnarray}
with the same result for interchanged order of limit, $L\to\infty$, for example, 
for usual box: 
\begin{eqnarray}
&&\!\!\!\!\!\!\!\!\!\!\!\!\!\!\!\!\!\!
\int\limits^\infty_0\! dk^1\frac{\sin\left(k^1\left(\widehat{x}^0-s\right)\right)}
{\sinh\left(\varsigma k^1/2\right)}\frac{\sin k^1 L}{k^1}=\!\!
\int\limits^{\widehat{x}^0-s+L}_{\widehat{x}^0-s-L}\!\!\! d\rho 
\int\limits^\infty_0\frac{dk^1}2\frac{\sin(k^1\rho)}{\sinh\left(\varsigma k^1/2\right)}=
\frac 12\ln\!\left[\frac{\cosh\left(\pi(L+\widehat{x}^0-s)/\varsigma\right)}
{\cosh\left(\pi(L-\widehat{x}^0+s)/\varsigma\right)}\right]
\rightarrow\frac{\widehat{x}^0-s}{\varsigma}. 
\nonumber  
\end{eqnarray}
Here the formulas (\ref{tanh_coth})--(\ref{ln_cosh_ab}) are used (see also  
3.951.(18),(19) from \cite{gr}). 

The imaginary part of Wightman function (\ref{D_}), (\ref{DD_}) is defined by the 
commutative function (\ref{K_8}) \cite{blot}: 
\begin{eqnarray}
&&\!\!\!\!\!\!\!\!\!\!\!\!\!\!\!\!\!\!
\frac 1i{\rm D}^{(-)}(z)=\!\int\! \frac {d^2k}{2\pi}\theta(k^0)\delta(k^2)e^{-i(kz)}
\Rightarrow 
\frac {(-1)}{4\pi}\ln\left(-\overline{\mu}^2 z^2+i0\,\varepsilon(z^0)\right), 
\quad 
\frac 1i {\rm D}^{(-)}(z)-\left(\frac 1i {\rm D}^{(-)}(z)\right)^*\!\!=\quad 
\label{D_-0} \\
&&\!\!\!\!\!\!\!\!\!\!\!\!\!\!\!\!\!\!
=\frac 1i{\rm D}_0(z)=\int \frac {d^2k}{2\pi}\varepsilon(k^0)\delta(k^2)e^{-i(kz)}=
\frac{1}{2i}\varepsilon(z^0)\theta (z^2)=
\frac{1}{4i}\left[\varepsilon(z^\xi)+\varepsilon(z^{-\xi})\right]=
\frac{1}{2i}\left[\theta(z^\xi)-\theta(-z^{-\xi})\right],
\label{D_eps} \\
&&\!\!\!\!\!\!\!\!\!\!\!\!\!\!\!\!\!\! 
\mbox{with: }\;
\frac{\partial}{\partial z^\nu}\varepsilon (z^\xi )= 
2\left( \xi 1\right)^\nu\delta(z^\xi), \quad {\rm D}_0(0)=0, \quad 
\frac{\partial}{\partial z^0}{\rm D}_0(z)\biggr|_{z^0=0}=\delta(z^1).  
\label{D_0_eps} 
\end{eqnarray}
The relations (\ref{D_}), (\ref{DD_}) correspond to following expansion of the  
distribution \cite{wai}:
\begin{eqnarray}
&&\!\!\!\!\!\!\!\!\!\!\!\!\!\!\!\!\!\!
\theta(k^0)\delta(k^2)=\frac {\theta(k^\xi)}{k^\xi}\delta(k^{-\xi})+
\frac {\theta(k^{-\xi})}{k^{-\xi}}\delta(k^{\xi}), \quad k^{\xi}=k^0+\xi k^1.
\label{thet_delt} 
\end{eqnarray}

\subsection{Appendix C}

The asymptotic expansions (\ref{a_1_0}), (\ref{a_2_}) are obtained by use of the known 
series \cite{gr} with Bernoulli numbers $B_i$, and Bernoulli polynomials $B_i(x)$, 
that give the asymptotic expansion for corresponding integral till the moments 
$I^\Delta_n$ (\ref{a_0_I_n}) exist: 
\begin{eqnarray}
&&\!\!\!\!\!\!\!\!\!\!\!\!\!\!\!\!\!\!
\frac{z}{e^z-1}=
1-\frac z2+\sum\limits^\infty_{j=1}B_{2j}\frac{z^{2j}}{(2j)!},\quad B_0=1,\quad 
B_1=-\,\frac 12,\quad B_{2j+1}=0,\quad j\geq 1,\quad |z|<2\pi,\quad 
z=\frac{t\varsigma}{L}, 
\label{Bern_1} \\
&&\!\!\!\!\!\!\!\!\!\!\!\!\!\!\!\!\!\!
\frac{z/2}{\sinh(z/2)}=
1+\sum\limits^\infty_{j=1}B_{2j}\left(\frac 12\right)\frac{z^{2j}}{(2j)!},\quad 
B_{2j}\!\left(\frac 12\right)=\left(\frac 1{2^{2j-1}}-1\right)\!B_{2j},\quad 
B_{2j}=2(-1)^{j+1}\zeta(2j)\frac{(2j)!}{(2\pi)^{2j}}, \quad 
\label{Bern_2} 
\end{eqnarray}
where $\zeta(s)$ - is Riemann zeta-function. Various meanings of non-negative values of  
$I^\Delta_0, I^\Delta_1, I^\Delta_2$ depend on the choice of the volume cut-off 
regularization function $\Delta(y^1/L)$ and are depicted below for the some popular 
examples \cite{fab-iva,blot,i_z}: 

\noindent
\begin{tabular}{|c||c|c|c|c|c|}
\hline
$\Delta \left(x^1/L\right)$ & $ e^{-(x^1/L)^2}$ & $ e^{-|x^1/L|}$ &
$ (1-|x^1|/L) \theta (1-|x^1|/L)$ & $ \theta (1-|x^1|/L)$ & 
\cite{blot}: $\epsilon_r\to 0$, $r \rightarrow \infty $ \\
\hline \hline
$\overline{\Delta}(t)$ & $\displaystyle \frac{ e^{-(t/2)^2}}{2\sqrt{\pi}}$ &
$\displaystyle \frac{1}{\pi(1+t^2)}$ & 
$\displaystyle \frac{1}{2\pi}\,\frac{\sin^2 (t/2)}{(t/2)^2}$ &
$\displaystyle \frac{1}{\pi}\,\frac{\sin t}{t}$ & 
$\displaystyle {\cal T}^{\frac 12}_r(t)\,\frac{|t|^{\frac{1}{r}-1}}{2r}\theta (1-|t|)$ \\
\hline
$ 2\pi I_0^\Delta$ & $\sqrt{2\pi}/4$ & $1/2$ & $1/3$ & $1$ & 
$\displaystyle 0\leq\lim_{r\to\infty}\frac{\pi\ln 2}{2r^2\epsilon_r}\leq\infty $ \\
\hline            
$ \pi I_1^\Delta=a_0$ & $1/4$ & $1/2\pi$ & $\ln 2/\pi$ & 
$\infty$  & 0 \\
\hline
$\pi I_2^\Delta$ & $\sqrt{2\pi}/8$ & $1/4$ & $1/2$ & $\infty$ & 0 \\ 
\hline
\end{tabular}

\noindent
Note that the Oksak regularization \cite{blot} in the last column is essentially of 
another type, which does not implies the thermodynamic limit at all, since it implies:  
$L\mapsto L_0=$ const, $\lim_{r\to\infty}L_0\overline{\Delta}_r(k^1L_0)=\delta(k^1)$. 
Here $\sqrt{{\cal T}_r(t)}\simeq{\cal T}_r(t)$ is a smooth ``trapezium'' - like 
function: ${\cal T}_r(t)=1$, for $2\epsilon_r< t< 1-2\epsilon_r$, ${\cal T}_r(t)\neq 0$, 
for $\epsilon_r\leq t\leq 1-\epsilon_r$, so that 
$\epsilon_r{\cal T}^\prime_r(t)\simeq\pm 1$, when it is not equal to 0. The even  
regularization function $\Delta_r(x^1/L_0)$ may be found up to the terms of order 
$O(\epsilon_r)$, for $\delta_r=1/r$, as: 
\begin{eqnarray}
&&\!\!\!\!\!\!\!\!\!\!\!\!\!\!\!\!\!\!
\Delta_r(x^1/L_0)=\int\limits^\infty_{-\infty}dk^1 L_0\overline{\Delta}_r(k^1L_0) 
e^{ik^1x^1}\Longrightarrow
\int\limits^1_{0}d\left(t^{1/r}\right)\cos\left(tx^1/L_0\right)=
\label{Del_r_x_int}  \\
&&\!\!\!\!\!\!\!\!\!\!\!\!\!\!\!\!\!\!
=\Gamma(1+\delta_r)\left\{\cos \left(x^1/L_0\right)\sum^\infty_{n=0}
\frac{(-1)^n \left(x^1/L_0\right)^{2n}}{\Gamma(2n+1+\delta_r)}+
\sin \left(x^1/L_0\right) \sum^\infty_{n=0}\frac{(-1)^n
\left(x^1/L_0\right)^{2n+1}}{\Gamma(2n+2+\delta_r)}\right\}, 
\label{Del_r_x} 
\end{eqnarray}
which oscillates and very slowly decrease for any finite $r$, but uniformly for 
$x^1<\infty$ tends to 1 with $r\to\infty$: 
\begin{eqnarray}
&&\!\!\!\!\!\!\!\!\!\!\!\!\!\!\!\!\!\!
\Delta_r(x^1/L_0)\Longrightarrow \left(L_0/|x^1|\right)^{1/r}\Gamma(1+\delta_r)
\cos(\pi\delta_r/2),\;\mbox { for: }\;|x^1|\to\infty, \quad 
\Delta_r(x^1/L_0)\stackunder{r\to\infty}{\Longrightarrow} 1. 
\label{Del_r_x_1} 
\end{eqnarray}
Thus, the arbitrary fixed parameter $L_0$ has nothing to do with a thermodinamic 
parameter of effective box size. 
Nevertheless, the values of $a_{0,1,2}$ \cite{blot} can be defined by the same way 
(\ref{a_0_I_n})--(\ref{a_2_}) as above, with $a_0=a_1-a_2\Rightarrow 0$, for 
$r\to\infty$, because for $n>0$ all $I^\Delta_{n}=0$ with $r\to\infty$, independently on 
the view of ${\cal T}_r(t)$ and $\epsilon_r$. Only the value of $2\pi I_0^\Delta $ can be 
chosen as finite for $\epsilon_r=1/r^2$, as shown in the Table. 

The last but one column shows that the usual box needs a separate consideration, due to 
divergent values of integrals for $a_0$ (\ref{a_0_I_n}) and $a_1$ for this case. 
This may be done with the help of known elementary series and integrals \cite{gr}: 
\begin{eqnarray}
&&\!\!\!\!\!\!\!\!\!\!\!\!\!\!\!\!\!\!
\coth\pi y=\frac 1{\pi y}+\frac{2y}{\pi}\sum\limits^\infty_{n=0}\frac{1}{y^2+(n+1)^2}, 
\qquad \tanh\pi y=\frac{2y}{\pi}\sum\limits^\infty_{n=0}\frac{1}{y^2+(n+1/2)^2}, 
\label{tanh_coth} \\ 
&&\!\!\!\!\!\!\!\!\!\!\!\!\!\!\!\!\!\!
2\int\limits^\infty_0 dx\sin bx\,e^{-\lambda x}=\frac{2b}{b^2+\lambda^2},\;\mbox{ for }\;
\lambda>0 \quad 
\int\limits^b_a d\rho\sin\rho x=\frac{\cos ax-\cos bx}{x},\;\mbox{ that give:}
\label{sin_cos} \\ 
&&\!\!\!\!\!\!\!\!\!\!\!\!\!\!\!\!\!\!
2\int\limits^\infty_0 dx\frac{\sin bx}{e^{\lambda x}-1}=
\sum\limits^\infty_{n=0}\frac{2b}{b^2+\lambda^2(n+1)^2}=
\frac{\pi}{\lambda}\coth\left(\frac{\pi b}{\lambda}\right)-\frac 1b=
\frac{d}{db}\ln\left[\frac 1b\sinh\left(\frac{\pi b}{\lambda}\right)\right],
\label{ln_sinh} \\ 
&&\!\!\!\!\!\!\!\!\!\!\!\!\!\!\!\!\!\!
\int\limits^\infty_0 dx\frac{\sin bx}{\sinh(\lambda x/2)}=
\sum\limits^\infty_{n=0}\frac{2b}{b^2+\lambda^2(n+1/2)^2}=
\frac{\pi}{\lambda}\tanh\left(\frac{\pi b}{\lambda}\right)=
\frac{d}{db}\ln\left[\cosh\left(\frac{\pi b}{\lambda}\right)\right],
\label{ln_cosh} \\ 
&&\!\!\!\!\!\!\!\!\!\!\!\!\!\!\!\!\!\!
2\int\limits^\infty_0 \frac{dx}{x}\left(\frac{\cos ax-\cos bx}{e^{\lambda x}-1}\right)=
\ln\left[\frac 1b\sinh\left(\frac{\pi b}{\lambda}\right)\right]-
\ln\left[\frac 1a\sinh\left(\frac{\pi a}{\lambda}\right)\right],
\label{ln_sinh_ab} \\ 
&&\!\!\!\!\!\!\!\!\!\!\!\!\!\!\!\!\!\!
\int\limits^\infty_0\frac{dx}{x}\left(\frac{\cos ax-\cos bx}{\sinh(\lambda x/2)}\right)=
\ln\left[\cosh\left(\frac{\pi b}{\lambda}\right)\right]-
\ln\left[\cosh\left(\frac{\pi a}{\lambda}\right)\right].
\label{ln_cosh_ab} 
\end{eqnarray}
The $L$ - independence of $a_0$ implies the change of variable in Eq. (\ref{a_0_I_n}), 
which becomes impossible for divergent integral. Introducing again the ultraviolet 
cut-off $\Lambda=1/\rho$, following to (\ref{Lam_L}), the finite expressions for 
the values of $a_0$, $a_1$, $a_2$, by virtue of (\ref{sin_cos})--(\ref{ln_cosh_ab}), 
reads: 
\begin{eqnarray}
&&\!\!\!\!\!\!\!\!\!\!\!\!\!\!\!\!\!\!
a^{reg}_0(L) \Rightarrow \frac 1\pi\int\limits^\infty_0\frac{dk^1}{k^1}\sin^2(k^1L)e^{-k^1\rho}=
\int\limits^L_0 d\ell \frac{d a^{reg}_0(\ell)}{d\ell}=
\frac{1}{4\pi}\ln\left(\frac{(2L)^2+\rho^2}{\rho^2}\right)
\stackunder{L\to\infty}{\longrightarrow}
\frac{1}{2\pi}\ln\left(\frac{2L}{\rho}\right),
\label{a_0_reg} \\
&&\!\!\!\!\!\!\!\!\!\!\!\!\!\!\!\!\!\!
a^{reg}_1(L)\equiv 
\pi\int\limits^\infty_0 dk^1 k^1\left(L\overline{\Delta}(k^1 L)\right)^2
\coth\left(\frac{k^0\varsigma}{2}\right)\Rightarrow
\frac 1\pi\int\limits^\infty_0\frac{dk^1}{k^1}\sin^2(k^1L)e^{-k^1\rho}
\coth\left(\frac{k^0\varsigma}{2}\right)\Rightarrow 
\label{a_1_reg} \\
&&\!\!\!\!\!\!\!\!\!\!\!\!\!\!\!\!\!\!
\Rightarrow  a^{reg}_0(L)+\frac{1}{2\pi}
\ln\left[\frac{\varsigma}{2\pi L}\sinh\left(\frac{2\pi L}{\varsigma}\right)\right]
\stackunder{L\to\infty}{\longrightarrow}
\frac{L}{\varsigma}-\frac{1}{2\pi}\ln 2-
\frac{1}{2\pi}\ln\left(\frac{\pi\rho}{\varsigma}\right), 
\label{a_1_0reg} \\
&&\!\!\!\!\!\!\!\!\!\!\!\!\!\!\!\!\!\!
a_2(L)\equiv
\pi\int\limits^\infty_0 dk^1 k^1\frac{\left(L\overline{\Delta}(k^1 L)\right)^2}
{\sinh\left(k^0\varsigma/2\right)} \Rightarrow 
\frac 1\pi\int\limits^\infty_0\frac{dk^1}{k^1}\,\frac{\sin^2(k^1L)}
{\sinh\left(k^0\varsigma/2\right)}=\frac{1}{2\pi}
\ln\left[\cosh\left(\frac{2\pi L}{\varsigma}\right)\right]
\stackunder{L\to\infty}{\longrightarrow}
\frac{L}{\varsigma}-\frac{1}{2\pi}\ln 2,\quad 
\label{a_2_reg} \\
&&\!\!\!\!\!\!\!\!\!\!\!\!\!\!\!\!\!\!
\mbox{so that: }\;a^{reg}_1(L)-a_2(L)\stackunder{L\to\infty}{\longrightarrow}
-\,\frac{1}{2\pi}\ln\left(\frac{\pi\rho}{\varsigma}\right)=
\frac{1}{2\pi}\ln\left(\frac{\Lambda}{\pi{\rm k}_B T}\right). 
\label{a1_a2} 
\end{eqnarray}
Thus, all the $L$ -- dependence again is canceled exactly, as well as for the convergent 
case, and in accordance with the Table, the finite value of $2\pi I^\Delta_0=1$, while 
the remaining difference (\ref{a1_a2}) may be associated with zero only for high 
temperature $(\varsigma\to 0)$ by choosing the ultraviolet cut-off as 
$\Lambda=\pi{\rm k}_B T$.

The Table obviously demonstrate that asymptotic of Fourier image $\overline{\Delta}(t)$ 
in fact is defined by smoothness of the original $\Delta(x^1/L)$. For its discontinuous 
derivative of $n$-th order: $\overline{\Delta}(t)\sim t^{-1-n}$. Thus, $I^\Delta_{0,1,2}$ 
exist for $2n>1$, for example, for continuous function with first derivative 
discontinuous at finite number of points. 

\subsection{Appendix D}

From (\ref{T_intro_0_cc09}), by introducing the parameter $z$ for additional control, as: 
\begin{eqnarray}
&&\!\!\!\!\!\!\!\!\!\!\!\!\!\!\!\!\!\!
F_\eta ([\pm]\varsigma)
=\eta \int\limits_0^\infty \frac{d k^1}{4\pi k^0} \left[
c \left(k^1; [\pm]\varsigma\right) c \left(-k^1; [\pm]\varsigma\right)-
c^\dagger \left(k^1; [\pm]\varsigma\right) c^\dagger \left(-k^1; [\pm]
\varsigma\right)\right]+
\nonumber \\
&&\!\!\!\!\!\!\!\!\!\!\!\!\!\!\!\!\!\!
+z \eta \int\limits_0^\infty \frac{d k^1}{4\pi k^0} \left[
\widetilde{c} \left(k^1; [\pm]\varsigma\right)
\widetilde{c} \left(-k^1; [\pm]\varsigma\right)-
\widetilde{c}^\dagger \left(k^1; [\pm]\varsigma\right)
\widetilde{c}^\dagger \left(-k^1; [\pm]\varsigma\right)\right],
\nonumber 
\end{eqnarray}
one has the following nonzero contributions: 
\begin{eqnarray}
&&\!\!\!\!\!\!\!\!\!\!\!\!\!\!\!\!\!\!
\left[F_\eta([\pm]\varsigma), \varphi^{\xi(+)} \left(s; [\pm]\varsigma\right)\right]=
\nonumber \\
&&\!\!\!\!\!\!\!\!\!\!\!\!\!\!\!\!\!\!
=\eta \int\limits_0^\infty \frac{d k^1}{4\pi k^0}
\left(\frac{\xi}{2\pi}\right)\int\limits_{-\infty}^\infty \frac{d q^1}{2
q^0} \theta \left(- \xi q^1\right)
\cosh \vartheta e^{- i q^0 s}
\left[c^\dagger \left(k^1; [\pm]\varsigma\right) c^\dagger
\left(-k^1; [\pm]\varsigma\right),c \left(q^1\right)\right]-
\nonumber \\
&&\!\!\!\!\!\!\!\!\!\!\!\!\!\!\!\!\!\!
-z \eta \int\limits_0^\infty \frac{d k^1}{4\pi k^0}
\left(\frac{\xi}{2\pi}\right) \int\limits_{-\infty}^\infty \frac{d q^1}{2 q^0}
\theta\left(- \xi q^1\right) \cosh \vartheta e^{- i q^0 s}
\left[\widetilde{c} \left(k^1; [\pm]\varsigma\right) \widetilde{c}
\left(-k^1; [\pm]\varsigma\right),c \left(q^1\right)\right]+
\nonumber \\
&&\!\!\!\!\!\!\!\!\!\!\!\!\!\!\!\!\!\!
+\eta \int\limits_0^\infty \frac{d k^1}{4\pi k^0}
\left([\pm 1] \frac{\xi}{2\pi}\right) \int\limits_{-\infty}^\infty \frac{d
q^1}{2 q^0} \theta \left(- \xi q^1\right) \sinh \vartheta
e^{+ i q^0 s}
\left[c \left(k^1; [\pm]\varsigma\right) c \left(-k^1; [\pm]\varsigma\right),
\widetilde{c} \left(q^1\right)\right]-
\nonumber \\
&&\!\!\!\!\!\!\!\!\!\!\!\!\!\!\!\!\!\!
-z \eta \int\limits_0^\infty \frac{d k^1}{4\pi k^0}
\left([\pm 1] \frac{\xi}{2\pi}\right) \int\limits_{-\infty}^\infty \frac{d
q^1}{2 q^0} \theta \left(- \xi q^1\right) \sinh \vartheta
e^{i q^0 s} \left[\widetilde{c}^\dagger \left(k^1; [\pm]\varsigma\right)
\widetilde{c}^\dagger \left(-k^1; [\pm]\varsigma\right),
\widetilde{c} \left(q^1\right)\right].
\nonumber 
\end{eqnarray}
By virtue of (\ref{c_tc_V}) it is a simple matter to see that: 
\begin{eqnarray}
&&\!\!\!\!\!\!\!\!\!\!\!\!\!\!\!\!\!\!
\left[c^\dagger \left(k^1; [\pm]\varsigma\right),c \left(q^1\right)\right]=
\left[\widetilde{c}^\dagger \left(k^1; [\pm]\varsigma\right),
\widetilde{c} \left(q^1\right)\right]=
 - \left(2\pi\right) \left(2 k^0\right) \cosh \vartheta \delta \left(k^1-q^1\right),
\nonumber \\
&&\!\!\!\!\!\!\!\!\!\!\!\!\!\!\!\!\!\!
\left[\widetilde{c} \left(k^1; [\pm]\varsigma\right),c \left(q^1\right)\right]=
\left[c \left(k^1; [\pm]\varsigma\right),\widetilde{c} \left(q^1\right)\right]=
[\pm 1] \left(2\pi\right) \left(2 k^0\right)\sinh\vartheta\delta\left(k^1-q^1\right),
\nonumber 
\end{eqnarray}
what leads to: 
\begin{eqnarray}
&&\!\!\!\!\!\!\!\!\!\!\!\!\!\!\!\!\!\!
\left[c^\dagger(k^1; [\pm]\varsigma)c^\dagger(-k^1;[\pm]\varsigma),c(q^1)\right]
=-4\pi k^0\cosh\vartheta \left[\delta(k^1+q^1)c^\dagger(k^1;[\pm]\varsigma)
+ \delta(k^1-q^1)c^\dagger(-k^1;[\pm]\varsigma)\right],
\nonumber \\
&&\!\!\!\!\!\!\!\!\!\!\!\!\!\!\!\!\!\!
\left[\widetilde{c}(k^1;[\pm]\varsigma)\widetilde{c}(-k^1;[\pm]\varsigma),c(q^1)\right]
= [\pm 1]4\pi k^0\sinh\vartheta\left[\delta(k^1+q^1)\widetilde{c}(k^1;[\pm]\varsigma)
+\delta(k^1-q^1)\widetilde{c}(-k^1;[\pm]\varsigma)\right],
\nonumber \\
&&\!\!\!\!\!\!\!\!\!\!\!\!\!\!\!\!\!\!
\left[c(k^1;[\pm]\varsigma)c(-k^1;[\pm]\varsigma),\widetilde{c}(q^1)\right]
=[\pm 1]4\pi k^0\sinh\vartheta\left[\delta (k^1+q^1)c(k^1;[\pm]\varsigma)
+ \delta (k^1-q^1)c(-k^1;[\pm]\varsigma)\right],
\nonumber \\
&&\!\!\!\!\!\!\!\!\!\!\!\!\!\!\!\!\!\!
\left[\widetilde{c}^\dagger(k^1;[\pm]\varsigma)\widetilde{c}^\dagger(-k^1;[\pm]\varsigma),
\widetilde{c}(q^1)\right]=
-4\pi k^0\cosh\vartheta\left[\delta(k^1+q^1)\widetilde{c}^\dagger(k^1; [\pm]\varsigma)
+\delta(k^1-q^1)\widetilde{c}^\dagger(-k^1;[\pm]\varsigma)\right].
\nonumber 
\end{eqnarray}
Substitution of these expressions gives: 
\begin{eqnarray}
&&\!\!\!\!\!\!\!\!\!\!\!\!\!\!\!\!\!\!
\left[F_\eta([\pm]\varsigma), \varphi^{\xi(+)} \left(s; [\pm]\varsigma\right)\right]=
\nonumber \\
&&\!\!\!\!\!\!\!\!\!\!\!\!\!\!\!\!\!\!
= \frac{\xi\eta}{2\pi} \int\limits_{-\infty}^\infty \frac{d q^1}{2 q^0}
\theta \left(- \xi q^1\right) \cosh \vartheta e^{- i q^0 s}
\left(- \cosh \vartheta c^\dagger \left(-q^1; [\pm]\varsigma\right)
-[\pm 1] z \sinh \vartheta \widetilde{c} \left(-q^1; [\pm]\varsigma\right)
\right)+
\nonumber \\
&&\!\!\!\!\!\!\!\!\!\!\!\!\!\!\!\!\!\!
+ \frac{\xi\eta}{2\pi} \int\limits_{-\infty}^\infty \frac{d
q^1}{2 q^0}
\theta \left(- \xi q^1\right) \sinh \vartheta e^{+ i q^0 s}
\left( \sinh \vartheta c \left(-q^1; [\pm]\varsigma\right)
+[\pm 1] z \cosh \vartheta \widetilde{c}^\dagger \left(-q^1; [\pm]\varsigma\right)
\right),
\nonumber 
\end{eqnarray}
where: 
\begin{eqnarray}
&&\!\!\!\!\!\!\!\!\!\!\!\!\!\!\!\!\!\!
- \cosh \vartheta c^\dagger \left(-q^1; [\pm]\varsigma\right)
- [\pm 1] z \sinh \vartheta \widetilde{c} \left(-q^1;
[\pm]\varsigma\right)=
\nonumber  \\
&&\!\!\!\!\!\!\!\!\!\!\!\!\!\!\!\!\!\!
= - \left[\cosh^2 \vartheta
- z \sinh^2 \vartheta \right] c^\dagger \left(-q^1\right)
+[\pm 1] \left[1-z\right]\sinh\vartheta\cosh\vartheta\widetilde{c} \left(-q^1\right)
\Longrightarrow - c^\dagger \left(-q^1\right),
\nonumber \\
&&\!\!\!\!\!\!\!\!\!\!\!\!\!\!\!\!\!\!
\sinh \vartheta c \left(-q^1; [\pm]\varsigma\right)
+[\pm 1] z \cosh \vartheta \widetilde{c}^\dagger \left(-q^1;
[\pm]\varsigma\right)=
\nonumber  \\
&&\!\!\!\!\!\!\!\!\!\!\!\!\!\!\!\!\!\!
= - [\pm 1] \left[ \sinh^2 \vartheta
- z \cosh^2 \vartheta \right] \widetilde{c}^\dagger \left(-q^1\right)
+ \left[1-z\right] \sinh \vartheta \cosh \vartheta c \left(-q^1\right)
\Longrightarrow [\pm 1] \widetilde{c}^\dagger \left(-q^1\right), 
\nonumber 
\end{eqnarray}
for $z=1$. Thus, due to (\ref{phi_T_p}), (\ref{phi_T_m}): 
\begin{eqnarray}
&&\!\!\!\!\!\!\!\!\!\!\!\!\!\!\!\!\!\!
\left[\varphi^{\xi(+)} \left(s; [\pm]\varsigma\right),F_\eta([\pm]\varsigma)\right]
= \frac{\xi\eta}{2\pi} \int\limits_{-\infty}^\infty\frac{d q^1}{2 q^0}
\theta \left(- \xi q^1\right) \cosh \vartheta e^{- i q^0 s}
c^\dagger\left(-q^1\right)-
\nonumber \\
&&\!\!\!\!\!\!\!\!\!\!\!\!\!\!\!\!\!\!
-[\pm 1]\frac{\xi\eta}{2\pi}\int\limits_{-\infty}^\infty\frac{d q^1}{2 q^0}
\theta\left(-\xi q^1\right)\sinh\vartheta e^{+ i q^0 s}
\widetilde{c}^\dagger \left(-q^1\right)
\equiv \eta \varphi^{-\xi(-)} \left(-s; [\pm]\varsigma\right).
\nonumber 
\end{eqnarray}
The next commutator in (\ref{ph_T_F}) is obtained by the same way. 

\subsection{Appendix E}

The free fermionic annihilation/creation operators:  
$\{{\rm b}(p^1),{\rm b}^\dagger(q^1)\}=\{{\rm f}(p^1),{\rm f}^\dagger(q^1)\}=
\delta(p^1-q^1)$ with correct parity properties: 
${\cal P}{\rm b}^{\#}(p^1){\cal P}^{-1}={\rm b}^{\#}(-p^1)$,  
${\cal P}{\rm f}^{\#}(p^1){\cal P}^{-1}=-{\rm f}^{\#}(-p^1)$, for 
${\cal P}\chi(x^1, x^0){\cal P}^{-1}=\gamma^0\chi(-x^1, x^0)$,  
are defined in \cite{fab-iva,fab-iva-12} by following decompositions for the field 
$\chi(x)$, $\xi=\pm$:
\begin{eqnarray}
&&\!\!\!\!\!\!\!\!\!\!\!\!\!\!\!\!\!\!
\chi_\xi(x)=\int\limits^\infty_{-\infty}\frac{dp^1}{\sqrt{2\pi}}
\left[\theta(\xi p^1){\rm b}(p^1)e^{-i(px)}+
\xi\theta(\xi p^1){\rm f}^\dagger(p^1)e^{i(px)}\right]e^{i\varpi-i\xi\Theta/4}
\Rightarrow \chi_\xi(x^{-\xi}),
\label{chi_xi_0} 
\end{eqnarray}
where: $(px)\Rightarrow p^0x^{-\xi}$, for: $\xi p^1=p^0=|p^1|$, and initial 
overall and relative phases $\varpi$ and $\Theta$ are introduced;  
\begin{eqnarray}
&&\!\!\!\!\!\!\!\!\!\!\!\!\!\!\!\!\!\!
\left(\overline{\chi}(x)\right)_\xi=\left(\chi^\dagger(x)\gamma^0\right)_\xi=
\chi^\dagger_{-\xi}(x^\xi)=\!\!
\int\limits^\infty_{-\infty}\!\!\frac{dp^1}{\sqrt{2\pi}}
\left[\theta(-\xi p^1){\rm b}^\dagger(p^1)e^{i(px)}
-\xi\theta(-\xi p^1){\rm f}(p^1)e^{-i(px)}\right]e^{-i\varpi-i\xi\Theta/4},\quad 
\label{bar_chi_xi_0} 
\end{eqnarray}
where: $(px)\Rightarrow p^0x^{\xi}$, for: $-\xi p^1=p^0=|p^1|$. 
The corresponding conserved charges are defined by components of the vector 
current $:J_{(\chi)}^\nu (x):=:\overline{\chi}(x)\gamma^\nu\chi(x):=
:\chi^\dagger(x)\gamma^0\gamma^\nu\chi(x):$, as \cite{blot,shif,rub}: 
\begin{eqnarray}
&&\!\!\!\!\!\!\!\!\!\!\!\!\!\!\!\!\!\!
\frac{O_{(\chi)} } {\sqrt{\pi}} =\int\limits_{-\infty}^\infty d x^1 :J_{(\chi)}^0 (x):  
=\int\limits_{-\infty}^\infty d x^1 :\chi^\dagger(x)\chi(x):
=\int\limits_{-\infty}^\infty d x^1 \sum\limits_{\xi=\pm }
:\chi_{\xi}^\dagger (x^{-\xi})\chi_{\xi}(x^{-\xi}):\,, 
\label{E_1} \\
&&\!\!\!\!\!\!\!\!\!\!\!\!\!\!\!\!\!\!
\frac{O_{5(\chi)}}{\sqrt{\pi}} = \int\limits_{-\infty}^\infty d x^1 :J_{(\chi)}^1 (x): 
=\int\limits_{-\infty}^\infty d x^1 :\chi^\dagger(x)\gamma^5\chi(x):
=\int\limits_{-\infty}^\infty d x^1 \sum\limits_{\xi=\pm } \xi
:\chi_{\xi}^\dagger (x^{-\xi})\chi_{\xi}(x^{-\xi}):\,, \;\mbox{ where:}
\label{E_2} \\
&&\!\!\!\!\!\!\!\!\!\!\!\!\!\!\!\!\!\!
\frac{Q^{-\xi}}{\sqrt{\pi}}=
\frac{1}{2{\sqrt{\pi}}}\left[O_{(\chi)}+\xi O_{5(\chi)}\right]=
\int\limits_{-\infty}^\infty d x^1 :\chi_{\xi}^\dagger (x)\chi_{\xi} (x): 
= \int\limits_{-\infty}^\infty d p^1 \theta (\xi p^1) 
\left[{\rm b}^\dagger(p^1){\rm b}(p^1) - {\rm f}^\dagger(p^1){\rm f}(p^1)\right], 
\label{E_4} \\
&&\!\!\!\!\!\!\!\!\!\!\!\!\!\!\!\!\!\!
\mbox{recasts them into: }\;
\frac{O_{(\chi)} }{\sqrt{\pi}} = \int\limits_{-\infty}^\infty d p^1 
\left[{\rm b}^\dagger(p^1){\rm b}(p^1) - {\rm f}^\dagger(p^1){\rm f}(p^1)\right], \;
\mbox { with: } \;
\theta (\xi p^1)+\theta (-\xi p^1)=1, 
\label{E_5} \\
&&\!\!\!\!\!\!\!\!\!\!\!\!\!\!\!\!\!\!
\frac{O_{5(\chi)}}{\sqrt{\pi}}=\int\limits_{-\infty}^\infty d p^1\varepsilon (p^1) 
\left[{\rm b}^\dagger(p^1){\rm b}(p^1) - {\rm f}^\dagger(p^1){\rm f}(p^1)\right], \;
\mbox { with: } \;
\xi \left[\theta(\xi p^1)-\theta (-\xi p^1)\right]=\varepsilon (p^1).  
\label{E_6} \\
&&\!\!\!\!\!\!\!\!\!\!\!\!\!\!\!\!\!\!
\mbox{Thus: }\;
\frac{Q^{-}}{\sqrt{\pi}}={\rm Q}_{\rm R}=N^+_{\rm R}-N^-_{\rm R},\quad 
\frac{Q^{+}}{\sqrt{\pi}}={\rm Q}_{\rm L}=N^+_{\rm L}-N^-_{\rm L}, \quad 
N^\pm_{(F)}=N^\pm_{\rm R}+N^\pm_{\rm L}, 
\label{E_4_1} \\
&&\!\!\!\!\!\!\!\!\!\!\!\!\!\!\!\!\!\!
\frac{O_{(\chi)} }{\sqrt{\pi}}={\rm Q}_{(F)}={\rm Q}_{\rm R}+{\rm Q}_{\rm L}=
N^+_{(F)}-N^-_{(F)}, \quad 
\frac{O_{5(\chi)}}{\sqrt{\pi}}={\rm Q}_{5(F)}={\rm Q}_{\rm R}-{\rm Q}_{\rm L}.
\label{E_6_1}
\end{eqnarray} 
The normal form of free Hamiltonian (\ref{K_3}) reads: 
\begin{eqnarray}
&&\!\!\!\!\!\!\!\!\!\!\!\!\!\!\!\!\!\!
:H_{0[\chi]}(x^0):=\int\limits_{-\infty}^\infty
dx^1 :\chi^\dagger(x)\gamma^5(-i\partial_1)\chi(x): =
\int\limits_{-\infty}^\infty d p^1 |p^1|
\left[{\rm b}^\dagger(p^1){\rm b}(p^1) + {\rm f}^\dagger(p^1){\rm f}(p^1)\right].
\label{E_7}
\end{eqnarray}
The expressions (\ref{MuB_MuF})--(\ref{N_L_N_R}) are obtained following to 
\cite{blin}, but with $\eta_{(F)}=+1$, $\eta_{(B)}=-1$ and energy level density 
${\cal D}(\epsilon, L)\Rightarrow {\rm g}^{(\eta)}_{s}2L/(hc)$ for the spectrum 
like (\ref{E_7}), $\epsilon(p^1)=c|p^1|$ and spin degeneracy ${\rm g}^{(\eta)}_{s}$, 
from general definitions \cite{isih} with chemical potentials $\mu_\eta$ and  
$\gamma_\eta=\varsigma\mu_\eta=\mu_\eta/({\rm k}_B T)$: 
\begin{eqnarray}
&&\!\!\!\!\!\!\!\!\!\!\!\!\!\!\!\!\!\!
N_\eta=L\overline{n}_\eta=\int\limits^\infty_0 d\epsilon 
\left\langle\!\left\langle n_\eta(\epsilon,\mu)\right\rangle\!\right\rangle
{\cal D}(\epsilon, L) \Longrightarrow 
\frac {{\rm g}^{(\eta)}_{s}2L}{\varsigma hc}{\cal F}^{(\eta)}_0(\gamma_\eta)=
\frac {{\rm g}^{(\eta)}_{s}2L}{\varsigma hc} \eta\ln\left(1+\eta e^{\gamma_\eta}\right), 
\label{E_8}\\
&&\!\!\!\!\!\!\!\!\!\!\!\!\!\!\!\!\!\!
P_\eta L= \int\limits^\infty_0 d\epsilon 
\left\langle\!\left\langle n_\eta(\epsilon,\mu)\right\rangle\!\right\rangle
\int\limits^\epsilon_0 d\epsilon^\prime {\cal D}(\epsilon^\prime, L) \Longrightarrow 
\frac {{\rm g}^{(\eta)}_{s}2L}{\varsigma^2 hc}{\cal F}^{(\eta)}_1(\gamma_\eta), \quad 
 2{\cal F}^{(1)}_1(0)={\cal F}^{(-1)}_1(0)=\frac{\pi^2}6, 
\label{E_80}\\
&&\!\!\!\!\!\!\!\!\!\!\!\!\!\!\!\!\!\!
\mbox{with: }\;
\left\langle\!\left\langle n_\eta(\epsilon,\mu)\right\rangle\!\right\rangle=
\frac {1}{e^{\varsigma(\epsilon-\mu)}+\eta}, \quad 
{\cal F}^{(\eta)}_\lambda(\gamma)=
\int\limits^\infty_0 dx \frac {x^\lambda}{e^{x-\gamma}+\eta}, \quad 
\frac {d {\cal F}^{(\eta)}_\lambda(\gamma)}{d\gamma}=\lambda 
{\cal F}^{(\eta)}_{\lambda-1}(\gamma), 
\label{E_9}\\
&&\!\!\!\!\!\!\!\!\!\!\!\!\!\!\!\!\!\!
\frac {d}{d\gamma}\left({\cal F}^{(1)}_1(\gamma)+{\cal F}^{(1)}_1(-\gamma)\right)=
{\cal F}^{(1)}_0(\gamma)-{\cal F}^{(1)}_0(-\gamma)=\gamma, \quad 
{\cal F}^{(1)}_1(\gamma)+{\cal F}^{(1)}_1(-\gamma)=2{\cal F}^{(1)}_1(0)+\frac {\gamma^2}2. 
\label{E_90}
\end{eqnarray} 
Eqs. (\ref{MuB_MuF}) are given by Eq. (\ref{E_8}) for $N_\eta=N_{(B)}$ bose 
particles and for $N_\eta=N^\pm_{(F)}$ fermions and antifermions, pairing as 
$\chi^++\chi^- \rightleftharpoons\, $ to arbitrary number of these bosons. Since 
$\gamma_{(B)}=0$, then for equilibrium $\gamma^+_{(F)}+\gamma^-_{(F)}\Rightarrow 0$, 
what transcribes the Eqs. (\ref{E_80}), (\ref{E_90}) with 
${\rm g}^{(1,\pm)}_{s}={\rm g}^{(-1)}_{s}=1$, as Eq.(\ref{N_L_N_R}) for total 
charge, and Eq. (\ref{P_xi_P}) for total pressure $P_{(F)}=P^+_{(F)}+P^-_{(F)}$, with 
internal energy density: 
\begin{eqnarray}
&&\!\!\!\!\!\!\!\!\!\!\!\!\!\!\!\!\!\!
\frac{{\cal U}_{(F)}}L\equiv 
-\left[\left(\varsigma\frac{\partial P_{(F)}}{\partial\varsigma}\right)_{\gamma}+
P_{(F)}\right]=\frac 1{\varsigma^2 hc}\left(\frac{\pi^2}{3}+\gamma^2_{(F)}\right)=P_{(F)}. 
\label{E_10}
\end{eqnarray} 


\begin{thebibliography}{00}
\bibitem {thi}
  W.E. Thirring, A soluble relativistic field theory, {\it Annals of Physics}
  {\bf 3} (1958), 91--112.
\bibitem {Jon}
  K. Johnson, Solution of the equations for the Green's functions of
  a two-dimensional relativistic field theory, {\it Nuovo Cim.}
  {\bf 20} (1961), 773--790.
\bibitem {s_w}
  F. L. Scarf, J. Wess, A consistent operator solution for the Thirring model,
  {\it Nuovo Cim.} {\bf 26} (1962) 150--167.
\bibitem {Leut}
  H. Leutwyler, Symmetry breaking solutions of the Thirring model,
  {\it Helv. Phys. Acta} {\bf 38} (1965), 431--447.
\bibitem {klaib} 
   B. Klaiber, in ''Lectures in Theoretical Physics'', University of 
   Colorado (Boulder, 1967), edited by A. Barut and W. Brittin, Gordon and Breach, 
   New York, 1968, Vol. X, part A, pp. 141-176.
\bibitem {d_f_z}
  G.F. Dell'Antonio, Y. Frishman, D. Zwanziger, Thirring model in terms of
  currents: solution and light-cone expansions,
  {\it Phys. Rev.} {\bf D6} (1972), 988--1007.
\bibitem {wai}
  A.S. Wightman, Problems in relativistic dynamics of quantized fields
  (Nauka, Moscow, 1967).
\bibitem {col}
  S. Coleman, Quantum sine-Gordon equation as the massive Thirring model,
  {\it Phys. Rev.} {\bf D11} (1975), 2088--2097.
\bibitem {man}
  S. Mandelstam, Soliton operators for the quantized sine-Gordon equation,
  {\it Phys. Rev.} {\bf D11} (1975), 3026--3030.
\bibitem {nak}
  N. Nakanishi, Free massless scalar field in two-dimensional space-time: Revisited, 
  {Z. Phys.} {\bf C4} (1980), 17-25. 
  N. Nakanishi, Operator solutions in terms of asymptotic fields in the
  Thirring and Schwinger models, {\it Prog. Theor. Phys.} {\bf 57} (1977),
  580--592;  Operator solutions in terms of asymptotic fields in the
  Thirring and Schwinger models. II, {\it Prog. Theor. Phys.} 
  {\bf 57} (1977), 1025--1037. 
\bibitem {ot}
  A. Ogura, H. Takahashi, A new current regularization of the Thirring
  model, {\it Prog.Theor.Phys.} {\bf 105} (2001), 495--500.
\bibitem {ruij}
    S.N.M. Ruijsnaars, Integrable quantum fields theories and Bogoliubov transformations, 
    {\it Annals of Physics} {\bf 132} (1981), 328--382.
\bibitem {fab-iva}
  M. Faber, A.N. Ivanov, On the equivalence between Sine-Gordon model and
  Thirring model in the chirally broken phase of the Thirring model,
  {\it Eur. Phys. J.} {\bf C20} (2001) 723--757. {\it hep-th/0105057 v2}
  \bibitem {fab-iva-12}
  M. Faber, A.N. Ivanov, On the Solution of the Massless Thirring Model With
  Fermion Fields Quantized in the Chiral Symmetric Phase, {\it hep-th/0112183}, 2001.
\bibitem {fab-iva-03}
  M. Faber, A.N. Ivanov, On the Vacua in the Massless Thirring
  Model, {\it hep-th/0306229}, 2003.
\bibitem {fab-iva-04}
  M. Faber, A.N. Ivanov, Dynamical Breaking of Conformal Symmetry in the
  Massless Thirring Model, {\it hep-th/0305203}, 2003.
\bibitem {fab-iva-05}
  M. Faber, A.N. Ivanov, Goldstone Bosons in the Massless Thirring Model:
  Witten's Criterion, {\it hep-th/0305174}, 2003.
\bibitem {fab-iva-06}
  M. Faber, A.N. Ivanov, On the Ground State of a Free Massless (Pseudo-)
  Scalar Field in Two-Dimensions, {\it hep-th/0212226}, 2002.
\bibitem {fab-iva-07}
  M. Faber, A.N. Ivanov, Bosonic Vacuum Wave Functions
  From the BCS Type Wave Function of the Ground State of the
  Massless Thirring Model, {\it Phys.Lett.} {\bf B563} (2003),
  231--237.
\bibitem {fab-iva-08}
  M. Faber, A.N. Ivanov, Quantum Field Theory of a Free Massless
  (Pseudo-) Scalar Field in (1+1)-Dimensional Space-Time as a
  Test for the Massless Thirring Model, {\it hep-th/0206244}, 2002.
\bibitem {fab-iva-09}
  M. Faber, A.N. Ivanov, Massless Thirring Fermion Fields in the Boson Field
  Representation, {\it hep-th/0206034}, 2002.
\bibitem {fab-iva-11}
  M. Faber, A.N. Ivanov, On Spontaneous Breaking of Continuous Symmetry in
  1+1-Dimensional Space-Time, {\it hep-th/0204237}, 2002.
\bibitem {bip}
  H. Bozkaya, A.N. Ivanov, M. Pitschmann, On renormalizability of massless Thirring 
  model, {\it hep-th/0512286 v2}, 2006.
\bibitem {fab-iva_2}
  M. Faber, A.N. Ivanov, On Free Massless (Pseudo-) Scalar Quantum Field
  Theory in (1+1)-Dimensional Space-Time, {\it Eur.Phys.J.} {\bf C24}
  (2002), 653--663. (And References therein.)
\bibitem {raja}
  S.J. Chang, R. Rajaraman,
  Chiral vertex operators in off-conformal theory: the sine-Gordon example,
  {\it Phys. Rev.} {\bf D53} (1996), 2102--2114.
\bibitem {mps_1}
  G. Morchio, D. Pierotti, F. Strocchi, Infrared and vacuum structure in
  two-dimensional local quantum field theory models. The massless scalar field,
  {\it J. Math. Phys.} {\bf 31} (1990), 1467--1477.
\bibitem {mps_2}
  G. Morchio, D. Pierotti, F. Strocchi, Infrared and Vacuum Structure in
  Two-Dimensional Local Quantum Field Theory Models. Fermion Bosonization,
  {\it J. Math. Phys.} {\bf 33} (1992), 777--790.
\bibitem {fujita}
  T. Fujita, M. Hiramoto, T. Homma, H. Takahashi, New vacuum of Bethe
  ansatz solutions in Thirring model, {\it J. Phys. Soc. Jap.} {\bf 74} (2005),
  1143--1149. {\it arXiv: hep-th/0410221}
\bibitem {fujita_2}
  T. Fujita, M. Hiramoto, T. Homma, M. Matsumoto, H. Takahashi, Re-interpretation of 
  spontaneous symmetry breaking in quantum field theory and Goldstone theorem, 
  {\it arXiv: hep-th/0510151}.
\bibitem {gom_ste}
  A. G\'omez Nicola, D.A.Steer, Thermal Bosonization in the sine-Gordon and Massive
  Thirring Models, {\it Nucl. Phys.} {\bf B549} (1999), 409--449.
\bibitem {blin}
  S.I. Blinnikov, State equation of relativistic fermi-gas, 
  {\it Siberian phys. jour.} {\bf 1} (1993), 20-25.
\bibitem {alv-gom}
  R.F. Alvarez-Estrada, A. G\'omez Nicola, The Schwinger and Thirring models at finite
  chemical potential and temperature, {\it Phys.Rev.} {\bf D57} (1998), 3618--3633.
\bibitem {blot}
  N.N. Bogoliubov, A.A. Logunov, A.I. Oksak, I. T. Todorov,
  General principles of quantum field theory (Kluwer Academic Publishers, Boston, 1990).
\bibitem {oksak}
  A.I. Oksak, Non-Fock linear boson systems and their applications in
  two-dimentional models, {\it Teoret. Mat. Fiz.} {\bf 48} (1981), 297--318.
\bibitem {mtu}
  H. Umezawa, H. Matsumoto, M. Tachiki, Thermo-field dynamics and condensed states
  (North-Holland Publishing Company, Amsterdam, 1982).
\bibitem {green_1}
  O.W. Greenberg, Virtues of the Haag expansion in quantum field theory,
  Preprint UMD-PP-95-99, Saariselka, Finland, 1994.
\bibitem {green}
  O.W. Greenberg, Study of a model of quantum electrodynamics,
  {\it Found. Phys.} {\bf 30} (2000), 383--391.
\bibitem {fadd}
    L.D. Faddeev, On the separation of self-action and scattering effects in 
    perturbation theory, {\it Soviet physics Doklady} {\bf 8} (1964), 881--883. 
\bibitem {shir}
   A.V. Shebeko, M.I.Shirokov, Unitary Transformation in Quantum Field Theory and Bound 
   States, {\it Phys. of Part. and Nucl.} 
   {\bf 32} (2001), 31--93.  (And References therein).
\bibitem {hep}
  K. Hepp, Theorie de la renormalization (Springer, Berlin, 1969).
\bibitem {schwarz} 
       A.S. Schwarz, Mathematical foundation of quantum field theory (Atomizdat, Moscow, 
       1975)   
\bibitem {gldj}
  J. Glimm, A. Jaffe, Quantum physics: a functional integral point of view
  (Springer-Verlag, New York, 1981).
\bibitem {vklt}
  A.N. Vall, S.E. Korenblit, V.M. Leviant, A.B. Tanaev, A dynamical mapping
  method in non-relativistic models of quantum field theory,
  {\it J. Nonlin. Math. Phys.} {\bf 4} (1997), 492--502.
\bibitem {kt}
  S.E. Korenblit, A.B. Tanaev, Linearization of Heisenberg equations in
  four-fermion interaction model and bound state problem,
  Preprint BUDKERINP 2001-11, Novosibirsk, 2001.
\bibitem {ks}
  S.E. Korenblit, V.V. Semenov, Massless pseudoscalar fields and solution
  of the Federbush model, {\it J. Nonlin. Math. Phys.} {\bf 13} (2006), 271--284.
\bibitem {sok}
  V.V. Sokolov, Schwinger terms and the interaction Hamiltonian
  in the quantum electrodynamics, {\it Sov. J. of Nuclear Physics}
  {\bf 8} (1968), 559--570.
\bibitem {i_z} 
   C. Itzykson, J.-B. Zuber, Quantum field theory, vol. 1,2, (McGraw-Hill Inc., NY, 1980) 
\bibitem {ojima}
  I. Ojima, Gauge fields at finite temperatures: thermo field dynamics,
  KMS Condition and their extension to gauge theories,
  {\it Annals of Physics} {\bf 137} (1981), 1--32.
\bibitem {Vlad}
  A.A. Vladimirov, On the origin of the Schwinger anomaly,
  {\it J. Phys.} {\bf A23} (1990), 87--90.
\bibitem {sa_wp} 
    I. Sachs, A. Wipf, Generalized Thirring models, 
     {\it Annals of Physics} {\bf 249} (1996), 380-429.
\bibitem {abr}
  R.L.P.G. Amaral, L.V. Belvedere, K.D. Rothe, Two-dimensional thermofield bosonization, 
  {\it Annals of Physics} {\bf 320} (2005), 399--428.
\bibitem {abr2}
  R.L.P.G. Amaral, L.V. Belvedere, K.D. Rothe, Two-dimensional thermofield
  bosonization II: massive fermions, {\it Annals of Physics} {\bf 323} (2008), 2662--2684.
\bibitem {abr3}
  R.L.P.G. Amaral, L.V. Belvedere, K.D. Rothe, A.F. Rodrigues,
  Quantum electrodynamics in two dimensions at finite
  temperature: thermofield bosonization approach,
  {\it J.Phys.} {\bf A44} (2011), 025401.
  \bibitem {ks_tt}
  S.E. Korenblit, V.V. Semenov,  Massless Thirring model in canonical quantization
  scheme, {\it J. Nonlin. Math. Phys.} {\bf 18}, N 1, (2011), 65-74.  
  (arXiv: hep-th/1003.1439 v.2); S.E. Korenblit, V.V. Semenov, Integration of the 
  Thirring model equations,
  {\it Russian Physics Journal}, {\bf 53}, Is. 6 (2010), 630--638.
\bibitem {ks_ttt}
  S.E. Korenblit, V.V. Semenov, Integration of Quantum Thirring Model at Finite
  Temperature, Proceedings of the XI Baikal School on Fundamental Physics, (Irkutsk, 
  Russia, 2009), Edit. N.P. Perevalova, ISEP, Irkutsk, 2009, ISSN 0135-3748, pp. 344--348.
\bibitem {ks_tttt}
   S.E. Korenblit, V.V. Semenov, On fermionic tilde conjugation rules and thermal 
   bosonization. Hot and cold thermofields, {\it Phys. of Part. and Nucl. Lett.} 
   {\bf 8} No 7, (2011) pp. 1-7 (arXiv: hep-th/1108.5392); Proceedings of Baikal Summer 
   School 2010, pp. 369-380. (http://www.slac.stanford.edu/econf/C1007061).
   (http://astronu.jinr.ru/wiki/upload/2/23/BaikalProceedings\_Volume\_2010.pdf)
\bibitem {gshl} 
   I.M. Gel'fand, G.E. Shilov, Generalized functions and operations with them, 
   (Nauka, Moscow, 1959).
\bibitem {tvash}
  T. Vachaspati, Kinks and domain walls (Cambrige University
  Press, New York, 2006).
\bibitem {isih} 
  A.Isihara, Statistical Physics, (Academic Press, New York -- London, 1971). \\
  I.A. Kvasnikov, Statistical Physics, Vol. 1,2,3, (Moscow, URSS, 2002)
\bibitem {shif} 
   M.A. Shifman, Anomaly and low energy theorems of QCD, Usp. Fiz. Nauk, {\bf 157}, N 4, 
   (1989), 561-598.
\bibitem {rub} 
   V.A. Rubakov, Classical gauge fields. Theories with fermions. Noncommutative 
   theories. (Moscow, URSS, 2005)
\bibitem {perelom}
  A.M. Perelomov, Generalized coherent states and their applications 
  (Springer, Berlin, 1986).
\bibitem {lipkin}
  H.J. Lipkin, Quantum mechanics (North-Holland Publishing Company, Amsterdam, 1973).
\bibitem {kirj} 
     D.A. Kirjnits, Field's methods of Many Particles Theory, Moscow, Atomizdat, 1963. 
\bibitem {gr} 
   I.S. Gradshteyn, I.M. Ryzhik, Table of integrals, series, and products, 
   (7-s ed. by A.Jeffrey, D. Zwillinger) (Academic Press, 2007)
\end{thebibliography}
\end{document}